\documentclass[11pt]{article}

\usepackage{amsmath}
\usepackage{authblk}
\usepackage{enumitem}  
\usepackage[usenames, dvipsnames]{color} 
\usepackage{graphicx}     
\usepackage{subcaption}
\usepackage[utf8]{inputenc}
\usepackage{url}
\usepackage{cite}
\usepackage{amssymb}
\usepackage{mathrsfs}
\usepackage{comment}
\usepackage{amsthm}

\usepackage[
  top=3cm,
  bottom=3cm,
  left=3cm,
  right=3cm
]{geometry}

\usepackage{overpic}

 \setlist[description]{font=\normalfont\space}

\def\beq{\begin{equation}}
\def\eeq{\end{equation}}

\newcommand{\ben}{\begin{enumerate}}
\newcommand{\een}{\end{enumerate}}
\newcommand{\be}{\begin{equation}}
\newcommand{\ee}{\end{equation}}

\usepackage{tikz}
\usetikzlibrary{positioning}
\usetikzlibrary{intersections}
\usetikzlibrary{fadings} 
\usetikzlibrary{arrows.meta} 
\usetikzlibrary{arrows}

\tikzfading[name=fade out,
inner color=transparent!0,
outer color=transparent!100]

\definecolor{cherryblossompink}{rgb}{1.0, 0.72, 0.77}
\definecolor{lightblue}{rgb}{0.68, 0.85, 0.9}

\usetikzlibrary{decorations.pathmorphing}
\usetikzlibrary{decorations.pathreplacing,decorations.markings}

\usetikzlibrary{backgrounds,automata}

\usepackage{hyperref}

\hypersetup{
    bookmarks=true,         
    unicode=false,          
    pdftoolbar=true,        
    pdfmenubar=true,        
    pdffitwindow=false,     
    pdfstartview={FitH},    
    pdfauthor={Author},     
    pdfsubject={Subject},   
    pdfcreator={Creator},   
    pdfproducer={Producer}, 
    pdfkeywords={keyword1} {key2} {key3}, 
    pdfnewwindow=true,      
    linktocpage=true,
    linkcolor=blue,           
     linkbordercolor=blue,
    citecolor=blue,        
    filecolor=blue,      
    urlcolor=blue         
} 

\begin{document}
 
\numberwithin{equation}{section}
 
 \title{\vspace{-2cm}\bf\LARGE  Holographic   pressure and volume for black holes \\[9mm]
}

\author{\large Silvester G.A. Borsboom\thanks{silvester.borsboom@ru.nl} }
\author{\large Manus  R.  Visser\thanks{manus.visser@ru.nl}} 

\vspace{10mm}

\affil{\textit{Institute for Mathematics, Astrophysics and Particle Physics,}\\
\textit{and Radboud Center for Natural Philosophy,}\\
\textit{Radboud University, 6525 AJ Nijmegen, The Netherlands}}

 \date{ }
 
\maketitle

\begin{abstract}
\noindent
We advocate for a  holographic definition of thermodynamic pressure and volume for black holes based on quasi-local gravitational thermodynamics. When a black hole is enclosed by a finite timelike boundary, York's quasi-local first law includes a surface pressure conjugate to the boundary area. Assuming the existence of a holographically dual theory living on this boundary, these geometric quantities correspond to the pressure and volume of the dual thermal system. In this work we focus on static, spherically symmetric black holes, for which these quantities reduce to global thermodynamic variables. The holographic volume provides a notion of system size, allowing extensivity to be defined in standard thermodynamic terms, and it  yields a definition of the large-system limit. For the asymptotically flat case, we show that, in the canonical thermodynamic representation, small    Schwarzschild black holes are non-extensive, whereas   large black holes become extensive in the large-system limit. A similar conclusion applies to Anti-de Sitter Schwarzschild black holes, with the difference that the quasi-local energy of the large black hole also becomes extensive in the large-system limit. Before this limit, the  energy decomposes into subextensive and extensive contributions, and we derive an explicit expression for the extensive part as a function of the finite volume and entropy.

\end{abstract}

\thispagestyle{empty}

\newpage

 \tableofcontents

 \newpage

\section{Introduction} \label{sec:intro}

\paragraph{Pressure, volume, and extensivity in black hole thermodynamics.}
   Black hole thermodynamics is one of the most profound and conceptually puzzling subjects in modern gravitational physics (see, e.g., \cite{Wald:1999vt,Witten:2024upt} for a review). One particularly puzzling feature is the   absence of a pressure-volume work term in the black hole first law \cite{Dolan:2012jh}. For a Schwarzschild black hole, the first law relates variations of the mass $M$ to variations of the horizon area $A_{\mathrm h}$ according to
\begin{equation}
dM=\frac{\kappa}{8\pi G}\,dA_{\mathrm h}\,,
\end{equation}
where $\kappa$ denotes the surface gravity and $G$ is Newton's constant. Upon identifying the surface gravity with the Hawking temperature, $T_{\mathrm H}=\kappa/2\pi$ \cite{Hawking:1975vcx}, and the horizon area with the Bekenstein entropy, $S=A_{\mathrm h}/4G$ \cite{Bekenstein:1973ur}, this relation takes the thermodynamic form
\begin{equation} \label{blackholefirstlaw1}
dM=T_{\mathrm H}dS\,,
\end{equation}
with the black hole mass interpreted as the internal energy. Notably, this first law contains no   pressure-volume term, in contrast to ordinary thermodynamic systems.

A closely related and longstanding issue concerns the non-extensive character of black hole thermodynamic variables. In ordinary thermodynamics, extensivity is often loosely characterized by linear scaling with spatial volume, and it is frequently stated that black hole entropy is non-extensive because it scales with the horizon area rather than with a volume (see, for example, \cite{Tsallis_2013}). However, in the absence of a well-defined notion of thermodynamic volume for black holes, such arguments remain incomplete.

Two main proposals for defining a thermodynamic volume for black holes appear in the literature. 
The first, due to Padmanabhan \cite{Padmanabhan:2002sha}, applies to static, spherically symmetric spacetimes, while the second defines the volume as the quantity conjugate to the cosmological constant $\Lambda$ in an extended version of the black hole first law, in which $\Lambda$ is allowed to vary \cite{Kastor:2009wy,Dolan:2010ha,Cvetic:2010jb,Dolan:2011xt,Dolan:2012jh,Kubiznak:2014zwa} (see \cite{Kubiznak:2016qmn,Mann:2025xrb} for reviews). 
Neither proposal, however, yields a fully satisfactory notion of thermodynamic volume, as discussed   in Section~\ref{sec:previous}. In the asymptotically flat case, extended black hole thermodynamics does not even provide a notion of thermodynamic volume.
For (Anti-de~Sitter) Schwarzschild geometries, which form the primary focus of this work, the problem is especially acute: in both approaches the thermodynamic volume coincides with the naive Euclidean volume determined by the horizon radius. 
Consequently, the volume is not an independent thermodynamic variable but is instead completely fixed by the entropy. 
The resulting thermodynamic state space is therefore degenerate, and the issue of extensivity remains unresolved.

A more refined line of reasoning by Landsberg~\cite{Landsberg1984,Landsberg1992} concerning the non-extensivity of black hole entropy is based on the standard definition of extensivity in thermodynamics, according to which an extensive variable is a homogeneous function of degree one of its independent variables (see Eq. \eqref{eq:homogeneityoriginal} for a proper definition). For asymptotically flat, charged, rotating black holes in $d$ spacetime dimensions, the entropy satisfies the scaling relation \cite{smarr1973,Myers:1986un,Gauntlett:1998fz}
\begin{equation} \label{scalingtransformationkerrnewman1}
S(\lambda^{d-3}M,\lambda^{d-2}J,\lambda^{d-3}Q)=\lambda^{d-2}S(M,J,Q)\,,
\end{equation}
where $M$, $J$, and $Q$ denote the mass, angular momentum, and electric charge, respectively. Since the entropy $S(M,J,Q)$ is not a homogeneous function of degree one, i.e.  \ $S(\lambda M,\lambda J,\lambda Q)\neq \lambda S(M,J,Q)$, it is concluded that black hole entropy is non-extensive.

This conclusion, however, obscures an important conceptual distinction. The scaling transformation considered above does not correspond to a rescaling of the size of the black hole thermodynamic system, but rather to a change of the thermodynamic state at fixed system size. This is because the transformation \eqref{scalingtransformationkerrnewman1} only changes the conserved charges defined at a fixed boundary at infinity, and therefore does not alter the size of that boundary. Although the event horizon grows or shrinks under this transformation, the horizon size should not be interpreted as a measure of the black hole system's size but rather of its entropy. This is directly analogous to changing the energy or particle number of a gas at fixed volume, which alters the thermodynamic state of the gas rather than the size of the system. A parallel interpretation arises in the holographically dual conformal field theory (CFT) description of Anti-de Sitter (AdS) black holes, where rescaling the conserved charges (such as the energy) at fixed spatial volume changes the thermodynamic state of the CFT, understood here as a point in macroscopic state space,  without changing the size of the spatial manifold on which the theory is defined.

In conventional thermodynamics, extensivity is defined with respect to rescalings of a system’s size, encoded by one or more extensive ``system-size'' variables. While such variables need not be unique in general, the spatial volume plays this role in most thermodynamic systems. Without a corresponding notion of thermodynamic volume for black holes, the standard definition of extensivity cannot be meaningfully applied. 
A proper and physically motivated definition of black hole thermodynamic volume is therefore required in order to assess the extensivity of black hole variables in conventional terms. Such a definition should, in particular, apply to both asymptotically flat and AdS-Schwarzschild black holes, and not just to the latter (like in extended black hole thermodynamics \cite{Kastor:2009wy}).

\paragraph{Holographic pressure and volume for black holes.}
In this article we advocate the view that quasi-local gravitational thermodynamics, originally developed by York and collaborators \cite{york1986,Whiting:1988qr,Martinez:1989hn,Brown:1989fa,Braden:1990hw,Brown:1990fk,BrownYork1993,Brown:1992bq,BrownMann1994}, provides the right framework for defining pressure and volume in black hole thermodynamics. The central idea is to formulate thermodynamics for a gravitating system enclosed by a finite timelike boundary (the ``York boundary''), so that energy, temperature, and mechanical work are defined quasi-locally rather than at infinity. In this setting, gravitational thermodynamics acquires a genuine notion of system size, fixed by the area  of the boundary rather than by the area of the horizon.

For a Schwarzschild black hole enclosed by a finite boundary, York \cite{york1986} showed that the quasi-local first law  takes the form  
\begin{equation} \label{firstquasilocalfirstlawschw}
dE = T dS - s dA ,
\end{equation}
where all thermodynamic variables except for the Bekenstein-Hawking entropy $S$ are defined on the boundary: $E$ is the Brown-York quasi-local energy \cite{BrownYork1993}, $T$ is the Tolman temperature~\cite{Tolman:1930zza}, $A$ is the area of a spatial cross-section of the boundary, and $s$ is the associated surface pressure. The total thermodynamic state space is 5-dimensional in this case, with coordinates $(E,T,S,s,A)$, of which only two variables are independent. In the infinite-boundary limit one recovers the standard black hole first law \eqref{blackholefirstlaw1}, since $E \to M$, $T \to T_{\mathrm H}$, and $s \to 0$. In that limit the boundary area is infinite, but since the surface pressure vanishes the work term is nonetheless absent in \eqref{blackholefirstlaw1}.  Importantly, varying the quasi-local energy at fixed boundary area changes the thermodynamic state without changing the size of the gravitational system, clarifying why the scaling argument in \eqref{scalingtransformationkerrnewman1} (where the boundary is effectively at infinity) does not probe extensivity. 

Our main assumption is that a holographic duality exists between the gravitational theory in the bulk and a non-gravitational quantum theory living on the finite boundary \cite{tHooft:1993dmi,Susskind:1994vu}. Under this assumption, quasi-local gravitational thermodynamics should be dual to the thermodynamics of the boundary theory. Crucially, the cross-sectional area   becomes the spatial volume on the boundary, while the surface pressure corresponds to the ordinary thermodynamic pressure. This holographic identification was proposed previously by Banihashemi, Shaghoulian, and Shashi \cite{Banihashemi:2024yye}, who showed that the Brown-York quasi-local stress tensor  gives rise to a pressure-volume term in a holographic first law. 

Here we adopt the same identification and   develop  the holographic interpretation of quasi-local thermodynamics further   for both asymptotically flat and AdS gravity, and use this framework to study thermodynamic extensivity. Specifically, for static, spherically symmetric black holes we use the identification
\begin{equation}
P \equiv s \,, \qquad V \equiv A .
\end{equation}
With this identification, York’s quasi-local first law takes precisely the standard thermodynamic form $dE = T dS - P dV$ for the dual theory. A key advantage of this definition is that entropy and volume are independent variables even for Schwarzschild black holes, unlike in earlier proposals where the thermodynamic volume is fixed by the horizon radius \cite{Padmanabhan:2002sha,Kastor:2009wy}. 
The quasi-local framework and its holographic interpretation are reviewed in Section~\ref{sec:quasilocalholographic}.

\paragraph{Thermodynamic representations  and extensivity.}
Although York's original work~\cite{york1986} introduced a gravitational partition function in the presence of a boundary using the Euclidean path integral, in this paper we work solely at the level of equilibrium thermodynamics. 
At this macroscopic level one may choose different sets of independent thermodynamic variables, such as $(S,V)$ or $(T,V)$, together with a thermodynamic potential whose differential yields the first law. 
Such a choice defines a \emph{thermodynamic representation} \cite{callen1985thermodynamics}. 
In general, a representation need not be globally single valued: for fixed values of the independent variables there may exist multiple equilibrium configurations, giving rise to multiple branches of the thermodynamic potential. Legendre transformations, defined in Eq.~\eqref{generallegendretransf}, relate different thermodynamic representations locally by
 exchanging a variable for its conjugate, defined as a partial derivative of the chosen
 thermodynamic potential.

A thermodynamic representation should be distinguished from a statistical ensemble: whereas the former is a macroscopic description in terms of thermodynamic state variables, the latter is a microscopic probabilistic construction specified by which macroscopic quantities are held fixed (and which conjugate quantities are allowed to fluctuate), together with a probability measure on microstates. 
Throughout this work we do not assume the existence, uniqueness, or equivalence of any underlying microscopic ensembles in gravity; extensivity will be assessed entirely at the macroscopic level via the homogeneity properties of the relevant thermodynamic potentials.

We work primarily in two thermodynamic representations: the internal-energy representation $E(S,V)$, where the entropy $S$ and the volume $V$ are taken as independent variables, and the canonical (Helmholtz free energy) representation $F(T,V)$, where the independent variables are the temperature $T$ and the volume $V$.
In each case, we study whether the corresponding thermodynamic potential becomes extensive in an appropriate large-system limit. 
The large-system limit in the energy representation is defined by $V\to\infty$ at fixed entropy density $s\equiv S/V$, while in the canonical representation it is defined by $V\to\infty$ at fixed temperature~$T$. 
We define extensivity in the standard thermodynamic sense as homogeneity of degree one, namely
$E(\lambda S,\lambda V)=\lambda E(S,V)$ and $F(T,\lambda V)=\lambda F(T,V)$ for $\lambda>0$, where the temperature is not scaled. 
If either of these relations holds, then $(S,V)$ are extensive variables and their conjugates $(T,P)$ are intensive within that thermodynamic representation.
The thermodynamic potentials are extensive if the respective limits $\lim_{V\to\infty}E(sV,V)/V$ and $\lim_{V\to\infty}F(T,V)/V$ exist and are non-vanishing.

However, the existence of such extensive limits and the validity of the corresponding Euler relations $E=TS-PV$  and $F =-PV$ can be representation- and branch-dependent. 
Accordingly, we will say that a given equilibrium branch in a thermodynamic representation is extensive if  the thermodynamic potential of that representation scales homogeneously of degree one, so that the corresponding Euler relation is satisfied on that branch. General, i.e.\ representation- and branch-independent, extensivity of an entire thermodynamic system  holds only if the
thermodynamic potentials are extensive and non-vanishing on all equilibrium branches in all
thermodynamic representations.
A more detailed discussion of extensivity in thermodynamics is given in Section~\ref{sec:extensivityinthermo}.

\paragraph{Extensivity for asymptotically flat versus AdS black holes.}
Using these definitions, we analyze the realization of extensivity for (AdS-)Schwarzschild black holes in different thermodynamic representations. For asymptotically flat Schwarzschild black holes,
in the internal-energy representation, the quasi-local   energy $E(S,V)$ is non-extensive even in the large-system limit. 
In the canonical   representation, on the other hand,  equilibrium configurations appear as distinct small and large black hole branches \cite{York1985a,york1986} (see Section~\ref{sec:smalllarge}). 
The small black hole branch is non-extensive in the large-system limit. 
For the large black hole branch, however, the Helmholtz free energy $F(T,V)$ becomes extensive in the limit $V\to\infty$ at fixed temperature, hence this equilibrium branch is extensive in the canonical representation, contrary to a previous claim \cite{Banihashemi:2024yye}.  
Instead, the   internal energy $E(T,V)$ does not scale extensively in this representation, consistent with   the non-extensivity of $E(S,V)$.  
Consequently, in asymptotically flat gravity  extensivity is representation- and branch-dependent. 
Further details are given in Section~\ref{nonextensivityflatblackholes}.

For  AdS-Schwarzschild black holes the thermodynamic structure is qualitatively different. 
In the energy representation, the quasi-local energy at finite volume   separates into a subextensive contribution and an extensive part that dominates in the large-system limit. 
The extensive contribution is given, for any   $V$ and $S$ for which $S<V/4G$, by
\begin{align}
E_{\mathrm{ext}}(S,V)
=\frac{(d-2)V}{8\pi G L}
\left(1-\sqrt{1-\left(\frac{4G S}{V}\right)^{\frac{d-1}{d-2}}}\right)\,.
\end{align}
This function is homogeneous of degree one in $(S,V)$ and   can be used to define an intensive temperature and pressure, such that the Euler relation $E_\text{ext}=T_\text{int}S-P_\text{int}V$ is satisfied.
Solving for the entropy yields a closed-form expression $S(E_{\mathrm{ext}},V)$. 
Expanding $E_{\mathrm{ext}}$ at large volume and fixed entropy reproduces the extensive energy-entropy scaling of the dual conformal field theory,
$E_{\mathrm{CFT}} R \propto S^{\frac{d-1}{d-2}}$ for $(d-1)$-dimensional CFTs. 

In the canonical representation, the same equilibrium configurations again appear as small  and large black hole branches~\cite{Hawking:1982dh,BrownMann1994}. 
The small black hole branch remains non-extensive in the large-system limit, whereas for the large black hole   both the Helmholtz free energy and the internal energy are now governed by their extensive contributions as $V\to\infty$. 
Consequently, asymptotically AdS gravity admits a thermodynamic regime, much more general than the CFT limit, in which extensivity holds consistently for an equilibrium branch (the large black hole) across different thermodynamic representations (see Section~\ref{sec:extadsblackholes}).

\paragraph{Thermodynamic interpretation of the AdS Smarr formula.}
Our analysis of extensivity for AdS-Schwarzschild black holes also leads to a thermodynamic interpretation of the AdS Smarr formula that differs from previous approaches (see Section \ref{sec:adssmarrthermo} for a comparison). As a starting point, in Section \ref{sec:quasilocalsmarr}, we provide the first derivation of a quasi-local Smarr relation for AdS-Schwarzschild black holes,
\begin{equation} \label{firstquasilocalsmarrintro}
(d-3)E = (d-2)(TS - PV) - \frac{\Lambda\tilde V_\xi}{4\pi GN_B}\,,
\end{equation}
where $\Lambda$ is the cosmological constant, $\tilde V_\xi$ is the background-subtracted Killing volume defined in Eqs.~\eqref{backkillingvolumeads} and \eqref{defquasilocalall}, and $N_B$ is the lapse function evaluated at the boundary.   The final term is absent for asymptotically flat black holes, for which $\Lambda=0$.

In extended black hole thermodynamics \cite{Kastor:2009wy,Dolan:2010ha,Cvetic:2010jb,Dolan:2011xt,Dolan:2012jh,Kubiznak:2014zwa}, the cosmological constant is   interpreted as a bulk pressure,
$P_\Lambda=-\Lambda/(8\pi G)$, with an associated thermodynamic volume. Notably, the quantity $\tilde V_\xi$
appearing above is not identical to the thermodynamic volume conjugate to $P_\Lambda$, and we discuss the
relation between the two in detail in Section~\ref{sec:adssmarrthermo}. In the present quasi-local framework,
however, a thermodynamic pressure is already present, and introducing an additional pressure is neither
necessary nor natural. Instead, we interpret the final term in the quasi-local Smarr relation as signaling a failure of exact (quasi-)homogeneity of the thermodynamic variables. Importantly, this term is not associated with a new pair of independent variables, but is instead a function of the existing thermodynamic variables.

This interpretation connects   with the holographic conformal field theory description of AdS black holes.
 Passing to the CFT variables by multiplying both sides of Eq. \eqref{firstquasilocalsmarrintro} by the Weyl factor $r_B/R$, where $r_B$ is the bulk boundary radius and $R$ is the radius of the sphere on which the CFT resides, and then taking the limit $r_B\to\infty$, we obtain in Eq.~\eqref{cftsubextensiveeuler}:
\begin{equation} \label{cftsmarrrr}
E_{\text{CFT}} - T_{\text{CFT}} S + P_{\text{CFT}} V_{\text{CFT}}
= 2C
\left(\frac{\Omega_{d-2}}{V_{\text{CFT}}}\right)^{\frac{1}{d-2}}
\left(\frac{S}{4\pi C}\right)^{\frac{d-3}{d-2}} \,,
\end{equation}
where $C$ is the central charge and $V_{\text{CFT}}=\Omega_{d-2}R^{d-2}$ is the CFT spatial volume. The relevant holographic dictionary is summarized in Section~\ref{sec:cftextensivity}. Dividing both sides by the volume and then taking the large-system limit $V_{\text{CFT}}\to\infty$ at fixed entropy density $s=S/V_{\text{CFT}}$, both the energy density and entropy density remain finite, while the right-hand side vanishes as
\begin{equation}
s^{\frac{d-3}{d-2}}\,
V_{\text{CFT}}^{-\frac{2}{d-2}} \to 0 .
\end{equation}
Thus, the homogeneous Euler relation and hence extensivity are recovered in the large-system limit, and the right-hand side of \eqref{cftsmarrrr} is identified as a finite-size, subextensive correction. In the CFT, this contribution has an interpretation as a thermal Casimir energy \cite{Verlinde:2000wg}. From this perspective, the AdS Smarr formula encodes the failure of extensivity at finite volume, which is restored in the large-system limit.

\section{Previous proposals for black hole pressure and volume} \label{sec:previous}
 
The concept of volume plays a central role in ordinary thermodynamics, but its extension to
black hole thermodynamics is subtle. In general relativity, diffeomorphism invariance
eliminates any preferred slicing of spacetime into space and time, so spatial volumes are
inherently foliation-dependent. This ambiguity is partially alleviated in static spacetimes, which admit a hypersurface-orthogonal   Killing vector field that is timelike in some region, since one may then define spatial hypersurfaces as
orthogonal to this vector field. However, inside the event horizon such Killing vector
fields become spacelike, so that the orthogonal hypersurfaces are timelike, and no natural
notion of spatial volume can be defined on these surfaces inside the horizon.

A further obstruction arises from the absence of a local, covariant notion of gravitational energy
density in general relativity. As a result, conserved quantities such as energy are not defined
locally, but only globally or  quasi-locally, that is, as integrals over finite closed
surfaces rather than as integrals of a local density. Familiar examples include the ADM mass
defined at infinity and the Brown-York energy associated with a finite timelike boundary.
Since thermodynamic variables in gravitational systems are necessarily tied to such
quasi-local constructions, it is difficult to define a notion of a bulk  volume that
functions as an independent thermodynamic variable for black holes in close analogy with ordinary
thermodynamics.

Despite these challenges, several proposals aim to define a volume that contributes to a black hole first law. Below we critically assess two prominent approaches~\cite{Padmanabhan:2002sha,Kastor:2009wy}, both of which assign thermodynamic significance to a volume conjugate to some sort of pressure. 
Other volume concepts found in the literature~\cite{Parikh:2005qs,DiNunno:2009cuq,Ballik:2010rx,Ballik:2013uia,Finch:2012vli,Christodoulou:2014yia,Stanford:2014jda,Couch:2016exn,Couch:2018phr} were not constructed to enter a black hole first law as thermodynamic variables, and will not be considered. 

For instance, Parikh \cite{Parikh:2005qs} introduced a volume for stationary spacetimes defined as the derivative of spacetime volume with respect to Killing time. This volume is time-independent,   invariant under changes of stationary slicing and yields  the   Euclidean volume for static black holes, but is not obtained from a first law or as a quantity conjugate to pressure.
Further, Christodoulou and Rovelli \cite{Christodoulou:2014yia} defined the interior volume as   the maximal volume of a spacelike hypersurface bounded by a given round two-sphere on the horizon.  
Although this definition is geometrically well defined for spherically symmetric black holes, the resulting volume grows linearly with advanced time,   reflecting  the non-staticity of the interior of a black hole rather than equilibrium properties.
For AdS black holes, Stanford and Susskind \cite{Stanford:2014jda} defined the volume as that of the maximal spacelike codimension-one surface anchored on the two asymptotic boundaries of an eternal black hole, and conjectured that it is dual to the computational complexity of the boundary state.  
While these  notions of black hole volume may be geometrically or holographically well motivated, they   do not arise as variables conjugate to pressure in an equilibrium first law, and hence they are not relevant for   black hole thermodynamics.

\subsection{Padmanabhan's proposal for static, spherically symmetric black holes}

We begin by reviewing the proposal by Padmanabhan \cite{Padmanabhan:2002sha}, who defines the thermodynamic volume for static, spherically symmetric black hole solutions to Einstein gravity as the naive Euclidean volume enclosed by the horizon,
\begin{align} \label{euclvolume}
    V_{\text{h}}(r_h) =  \Omega_{d-2} r_h^{d-1} /(d-1)\,,
\end{align}
where \( \Omega_{d-2} \) is the surface area of the unit \( (d-2) \)-sphere, \( r_h \) is the horizon radius, and $d$ is the number of spacetime dimensions. The conjugate pressure is taken to be the radial component of the energy-momentum tensor evaluated at the horizon, \( P_{\text{h}} = T^r{}_r(r_h) \), which vanishes for Schwarzschild but is nonzero for instance for Reissner-Nordstr\"{o}m black holes. These identifications are motivated by rewriting the Einstein equation at the horizon in the form of a thermodynamic identity. For static, spherically symmetric line elements of the form \begin{align} \label{lineelement1}
ds^2 = -f(r)\,dt^2 + f(r)^{-1}\,dr^2 + r^2 d\Omega_{d-2}^2\,,
\end{align} where $d \Omega_{d-2}$ is the line element on a unit $(d-2)$-sphere, the radial Einstein equation ${G^r}_r = 8\pi G {T^r}_r$ evaluated at \( r = r_h \) (the largest positive root of $f(r_h)=0$) becomes
\begin{equation} \label{padmana}
    P_{\text{h}} = \frac{T_\text{H}}{4G} \frac{d-2}{r_h} - \frac{(d-2)(d-3)}{16\pi G r_h^2}\,,
\end{equation}
with Hawking temperature \( T_\text{H} = f'(r_h)/(4\pi) \). Since  by assumption \( V_{\text{h}} \sim r_h^{d-1} \), Padmanabhan interprets this as an equation of state \( P_{\text{h}} = P_{\text{h}}(V_{\text{h}}, T_\text{H}) \), even though $V_\text{h}$ and $T_{\text{H}}$ are not independent. Multiplying \eqref{padmana} with a virtual volume displacement $dV_{\text{h}}$ yields a horizon ``first law''
\begin{equation} \label{padmana2}
    P_{\text{h}} dV_{\text{h}} = T_\text{H} dS - dE_{\mathrm{MS}}\,,
\end{equation}
where $S=\Omega_{d-2}r_h^{d-2}/4G$ is the Bekenstein-Hawking entropy, and  \( E_{\mathrm{MS}} \) is the Misner-Sharp energy  \cite{MisnerSharp}, a quasi-local energy defined for spherically symmetric spacetimes with line element $ds^2 = h_{ab} dx^a dx^b + r(x)^2 d \Omega_{d-2}^2$ as
\begin{equation}
E_{\mathrm{MS}} (r)= \frac{(d-2)\Omega_{d-2}}{16 \pi G} r^{d-3} \left ( 1- h^{ab} \nabla_a r \nabla_b r\right)\,.
\end{equation} For the  line element  \eqref{lineelement1}  we have $h^{ab} \nabla_a r \nabla_b r = f(r)$, which vanishes at the horizon, hence the Misner-Sharp energy at the horizon is
\begin{equation}
E_{\mathrm{MS}} (r_h)= \frac{(d-2)\Omega_{d-2}}{16 \pi G} r_h^{d-3}  \,.
\end{equation}
This is the energy that appears in the horizon first-law-like equation \eqref{padmana2}.

However, although \eqref{padmana2}   formally resembles a thermodynamic first law, this   interpretation is not justified. Since both entropy and volume  are   functions of the single parameter \( r_h \), they are not independent thermodynamic variables, as also emphasized in \cite{Hansen:2016wdg,Hansen:2016gud}. 
As a result, in \eqref{padmana2} the heat term \( T_\text{H} dS \) and the work term \( P_{\text{h}} dV_{\text{h}} \)  are functionally dependent, and cannot define a proper thermodynamic differential structure. Moreover, the fundamental equation for the energy function $E_{\text{MS}}(S,V_{\text{h}})$ is ill defined, since $S$ and $V_{\text{h}}$ are not independent coordinates. This leads to a degenerate state space, violating one of the basic requirements of thermodynamic consistency: namely  that the extensive variables (such as entropy and volume) must be independently variable so that their conjugates (temperature and pressure, respectively) are unambiguously defined as partial derivatives of energy.

\subsection{Extended black hole thermodynamics}
\label{sec:extendedthermo}
In the  second proposal,   called  ``extended black hole thermodynamics'' or ``black hole chemistry'' \cite{Kastor:2009wy,Dolan:2010ha,Cvetic:2010jb,Dolan:2011xt,Dolan:2012jh,Kubiznak:2014zwa} (see \cite{Kubiznak:2016qmn,Mann:2025xrb} for reviews),   the cosmological constant \( \Lambda \) is promoted to a thermodynamic   pressure,
\begin{equation} \label{pressureextended}
    P_\Lambda = -\frac{\Lambda}{8\pi G}\,.
\end{equation}
  This prescription stems from the observation that $\Lambda$ can be formally varied in the action and that it appears in the Einstein equations as a perfect fluid energy-momentum source with pressure \eqref{pressureextended}. 
  Allowing for variations of the cosmological constant, the  extended first law of AdS-Schwarzschild  black holes (with $\Lambda <0$)  takes the form
\begin{equation}\label{extendedadsfirstlaw1}
dM = \frac{\kappa}{8\pi G} d A_{\text h} + \frac{\bar V_\xi}{8\pi G} d \Lambda\,,
\end{equation}
where $M$ is the black hole mass, $\kappa$ is the surface gravity of the horizon, and $A_{\text{h}}$ is the area of the event horizon. The  quantity $\bar V_\xi$ (usually denoted by $\Theta$ in the literature) conjugate to $\Lambda$ in the first law can be defined as \cite{Jacobson:2018ahi}\begin{align}\label{backgroundsubvolume}
\bar V_\xi = V_\xi^{\text{bh}} - V_\xi^{\text{AdS}}= \int_{\Sigma_{{\text{bh}}}} \xi_{\text{bh}} \cdot \epsilon_{\text{bh}} - \int_{\Sigma_{\text{AdS}}} \xi_{\text{AdS}} \cdot \epsilon_{\text{AdS}}\,,\end{align}  where $\xi_\text{BH}$ is the horizon generating Killing vector field, $\xi_{\text{AdS}}$ is the stationary Killing field of  AdS spacetime, $\epsilon$ is the spacetime volume form, and a subtraction with respect to empty AdS is implemented to regulate the divergence arising from the infinite spatial volume. Here, the domain of integration $\Sigma_{\text{bh}}$ extends from the horizon to infinity, while $\Sigma_{\text{AdS}}$ is a complete spacelike hypersurface of   AdS.
In extended black hole thermodynamics $\bar V_\xi$ is interpreted as minus the ``thermodynamic volume'',  $ \bar V_\xi= -V_{\Lambda}  $, since it is conjugate to the    pressure \eqref{pressureextended} associated to the cosmological constant. For AdS-Schwarzschild   this volume coincides with  
the Euclidean volume $  V_\Lambda=   \Omega_{d-2} r_h^{d-1}/ (d-1)$ excluded by the black hole.

Upon identifying $\Lambda$ with the pressure, $\bar V_\xi$ with the thermodynamic volume, $\kappa$ with the temperature and $A_{\text h}$ with the entropy,   the extended first law becomes \cite{Kastor:2009wy}
\begin{equation} \label{extendedadsfirstlaw2}
dM = T_{\text{H}}dS + V_{\Lambda}dP_\Lambda\,.
\end{equation}
Here $T_\text{H}=\kappa/2\pi$ is the Hawking temperature and $S=A_{\text{h}}/4G$ is the Bekenstein entropy.  This relation is formally identical to the thermodynamic first law provided the black hole mass is interpreted not as the internal energy but as   enthalpy,
 $M=  E + P_\Lambda V_\Lambda\equiv H$. With this enthalpy interpretation the extended first law indeed turns into the thermodynamic first law for  enthalpy, $dH = TdS + V_\Lambda dP_\Lambda $. For asymptotically flat black holes the pressure vanishes, since $\Lambda=0$, hence   no pressure-volume conjugate pair arises.

Although this framework yields a consistent differential first law \eqref{extendedadsfirstlaw2} for the enthalpic mass of AdS black holes and has led to intriguing analogies with classical thermodynamics (such as Van der Waals critical behavior \cite{Kubiznak:2012wp}), it also suffers from conceptual difficulties. For AdS-Schwarzschild black holes the degeneracy problem persists: the thermodynamic volume coincides with the naive Euclidean volume, so that both the entropy and the volume depend on a single parameter, the horizon radius. As a result, 
 the fundamental equation for the internal energy  $E(S,V_\Lambda)$ is not well defined. This means   there is no consistent first law for the internal energy of the form $dE = T_{\text H} dS - P_\Lambda dV_\Lambda.$  

Only when the black hole solution depends on additional conserved parameters, such as angular momentum, does the state space acquire sufficient dimensionality to support a non-degenerate thermodynamic structure. For Kerr black holes, for example, the internal energy 
$E = M - P_\Lambda V_\Lambda$ depends   on three independent variables $E = E(S,V_\Lambda, J)$ \cite{Caldarelli:1999xj,Cvetic:2010jb,Dolan:2011xt,Dolan:2012jh}, but when $J=0$ the entropy and volume are again mutually dependent. By contrast, for AdS-Reissner-Nordstr\"{o}m black holes both the entropy and the volume remain functions of the horizon radius alone, and not of the electric charge, hence the state space is degenerate.

More fundamentally, Mancilla \cite{Mancilla:2024spp} has recently argued that treating the cosmological constant as a thermodynamic state variable is conceptually problematic. The cosmological constant is a coupling constant of the gravitational theory, and varying 
$\Lambda$ therefore changes the theory itself rather than the equilibrium state of a fixed system. Thermodynamic state variables, by contrast, characterize equilibrium configurations within a given theory. From this perspective, interpreting 
$\Lambda$ as a thermodynamic variable conflates a flow in theory space with motion in the space of thermodynamic states.

In the holographic context, in particular the Anti-de Sitter/Conformal Field Theory duality, this issue becomes even more poignant, since changing $\Lambda$ also corresponds to changing the field content of the CFT. Before we can explain   this issue, we need a holographic interpretation of extended black hole thermodynamics.
The bulk pressure 
$P_\Lambda$ and its conjugate thermodynamic volume 
$V_\Lambda$ have no clear interpretation in the dual conformal field theory. In particular, they are not dual to the pressure and volume of the boundary theory \cite{Johnson:2014yja}. Moreover, the black hole mass corresponds holographically to the internal energy of the CFT, not to its enthalpy. The bulk pressure-volume term therefore lacks a   boundary counterpart.

Instead,   the cosmological constant  controls the central charge 
\( C \propto L^{d-2}/G \) of the dual CFT, where $L \propto \sqrt{-\Lambda}$ is the AdS curvature scale. Allowing 
$\Lambda$ to vary thus corresponds to varying the number of field degrees of freedom of the CFT \cite{Johnson:2014yja,Dolan:2014cja,Karch:2015rpa}. In   previous work \cite{Karch:2015rpa,Visser:2021eqk,Cong:2021jgb,Ahmed:2023snm} it was suggested that extended  thermodynamics of AdS black holes corresponds to an extended version of CFT thermodynamics where the central charge is allowed to vary, with an associated chemical potential $\mu$ conjugate to 
$C$. In this framework   the extended first law \eqref{extendedadsfirstlaw1} for AdS-Schwarzschild black holes is dual to an extended   first law of CFT thermodynamics:
\begin{equation}
dE_{\text{CFT}} = T_{\text{CFT}}dS -P_{\text{CFT}}dV_{\text{CFT}} + \mu dC\,.
\end{equation} Moreover,   the Smarr relation for AdS black holes is dual to an Euler relation   that is valid at large $C$ for high-energy states: \begin{equation} \label{largeceuler}
E_{\text{CFT}} = T_{\text{CFT}} S + \mu C\,.
\end{equation} Crucially, there is no pressure-volume pair in this Euler relation. 

Even though such an extension of CFT thermodynamics can be formally defined, Mancilla's critique implies for the boundary theory that varying the central charge corresponds to a flow in the space of quantum field theories, rather than to a thermodynamic process within a fixed theory. Since $C$ is not a function of spacetime coordinates, it cannot be interpreted as a standard thermodynamic state variable. It is conceivable that $C$ could instead play the role of a thermodynamic variable in a grand canonical ensemble over CFTs with different central charges. Symmetric product CFTs or D-brane constructions might allow one to vary the number of copies or branes, and hence the central charge of the CFT.
In particular, in~\cite{DeLange:2018wbz} a grand canonical ensemble was formulated for symmetric product CFTs, with a chemical potential conjugate to the number of copies. It would be interesting to clarify the precise relation between this construction and extended AdS black hole thermodynamics. In this paper, however, we will not pursue this direction and instead restrict attention to standard black hole thermodynamics at fixed $\Lambda$. Accordingly, we keep the central charge fixed and focus on a pressure-volume term in the black hole first law that admits a genuine thermodynamic interpretation.

\section{Black hole pressure and volume from quasi-local thermodynamics}
\label{sec:quasilocalholographic}

We take a different approach to the thermodynamic volume and pressure for black holes, based on combining York’s \cite{york1986} quasi-local formulation of gravitational thermodynamics with the holographic principle. A similar   perspective was  described in recent work~\cite{Banihashemi:2024yye} by Banihashemi, Shaghoulian, and Shashi, who   identify thermodynamic pressure   with the spatial (diagonal) components of the quasi-local stress tensor and thermodynamic volume with the area of a finite timelike boundary, and who   recover a pressure-volume term in the first law.
 While their primary focus is on conformal boundary conditions \cite{York:1972sj,anderson2008boundary,Witten:2018lgb,An:2021fcq,Odak:2021axr,Anninos:2023epi} and the resulting thermodynamics of flat-space gravity at finite cutoff, they also briefly discuss York's original quasi-local thermodynamics subject to Dirichlet boundary conditions. We analyze  quasi-local thermodynamics with Dirichlet boundary conditions  and its holographic interpretation in more detail, both for flat-space  and AdS gravity, and study the issue of thermodynamic extensivity from a different perspective. 

 \subsection{The surface  character of gravitational thermodynamics}

The general philosophy is that thermodynamic variables in gravitational systems must be associated with boundaries \cite{york1986,Martinez:1996vy}. This follows from diffeomorphism invariance: in any diffeomorphism-invariant theory of gravity, the bulk Hamiltonian is an integral over the constraints and therefore vanishes on shell,  while the
boundary Hamiltonian is a codimension-2 surface integral that encodes the conserved charges associated with diffeomorphisms \cite{Arnowitt:1962hi,Regge:1974zd}. 
Likewise in black hole thermodynamics, diffeomorphism invariance dictates that black hole entropy is given by a local geometric expression on a  horizon  cross-section \cite{wald1993entropy,iyer1994noether}, while mass and angular momentum are defined by asymptotic boundary integrals. From this perspective, one would  expect that any meaningful notion of thermodynamic pressure and volume for black holes should likewise be defined  at a boundary.

This expectation is realized concretely in York’s formulation of quasi-local black hole thermodynamics. When a Schwarzschild black hole is enclosed by a finite timelike boundary, the quasi-local first law for general relativity extends the thermodynamic state space by introducing a surface pressure $s$ conjugate to the area $A$ of a spatial spherical cross-section of the boundary. Surface pressure and area are  quasi-local quantities, like the quasi-local energy that also appears in York's first law, in the sense that they are associated to   codimension-2 surfaces (round spheres for Schwarzschild geometry).   

The pair 
 $(s,A)$ enters the first law as a work term, but differs from the usual pressure-volume contribution in that both quantities are intrinsically lower-dimensional: 
$s$ is a surface pressure and 
$A$ is an area rather than a bulk volume. Such surface work terms are familiar in physical systems with interfaces, including systems with surface tension (such as soap bubbles) and elastic media with boundary stresses, where interfacial stresses are conjugate to changes in area rather than volume.  In the gravitational setting, however, the appearance of such a surface work term raises the question of how these quasi-local variables should be interpreted thermodynamically.

The crucial additional input is provided by the holographic principle \cite{tHooft:1993dmi,Susskind:1994vu}. If the gravitational system in the interior of a York boundary admits a dual description in terms of a non-gravitational theory living on that boundary, then   quasi-local gravitational thermodynamics should be dual to the thermodynamics of the boundary theory. Since the dual theory lives in one fewer spatial dimension, the boundary area plays the role of a thermodynamic volume, while the conjugate surface pressure is   interpreted as the thermodynamic pressure in the dual theory. In \cite{Creighton:1995au} they also called the boundary area ``the volume of the system'', but they did not invoke holography.

We therefore define the thermodynamic pressure and volume for static,
spherically symmetric black holes holographically by identifying
\begin{equation} \label{dictionarypressurearea}
P \equiv s\,, \qquad V \equiv A \, .
\end{equation}
Below we also briefly discuss quasi-local thermodynamics for static but
non-spherically symmetric black holes, for which the quasi-local first
law instead involves the full spatial stress tensor on the boundary.
Although a corresponding quasi-local first law can be formulated for
stationary but non-static, non-spherically symmetric black holes
\cite{Brown:1990fk,Brown:1992bq,Creighton:1995au}, such cases require more
general boundary data and will not be considered here. In the absence of spherical symmetry, the surface pressure is locally defined and conjugate to the local area element, leading to a continuum description analogous to hydrodynamics, formulated in terms of local densities rather than global variables.

The identification \eqref{dictionarypressurearea} is precisely the one
that arises in finite-cutoff generalizations of AdS/CFT, most notably
in holography based on $T\bar{T}$ deformations \cite{McGough:2016lol}.
More generally, however, the holographic dictionary
\eqref{dictionarypressurearea} applies beyond AdS, for example to
cutoff surfaces in flat-space gravity.

Moreover, we note that the present identification is consistent with,
and conceptually inspired by, earlier effective descriptions of black
hole dynamics, in particular the membrane paradigm \cite{Thorne} and the
fluid/gravity correspondence
\cite{Bhattacharyya:2007vjd,Bredberg:2011jq,Compere:2011dx}. In the
membrane paradigm the stretched horizon supports an effective fluid
description, with local pressure and transport properties capturing
aspects of horizon dynamics. The Brown-York formalism provides a
  quasi-local realization of this idea on a timelike
hypersurface, while subsequent developments have shown that the
associated stress tensor admits a hydrodynamic interpretation at
arbitrary cutoff radius within a long-wavelength expansion. The
standard AdS fluid/gravity correspondence may then be viewed as an
infinite-volume realization of this general framework. The identification \eqref{dictionarypressurearea} fits
well within  these  approaches.

In the remainder of this Section, we review the relevant elements of quasi-local gravitational thermodynamics. 
We then derive the quasi-local first law for static solutions to  general relativity using the covariant phase space formalism. Finally, we derive a new quasi-local Smarr relation for AdS-Schwarzschild  black  holes within the same formalism.

\subsection{Surface pressure and quasi-local energy}

We consider a gravitational system with an event horizon enclosed by a finite timelike
York boundary $B$, equipped with Dirichlet boundary conditions that fix the induced
metric $\gamma_{ij}$. The spacetime region inside the boundary, extending from the
horizon up to~$B$, constitutes the thermodynamic system, while the exterior plays the role of an environment   with which the system may exchange energy.
We assume the system is in thermal equilibrium with this exterior region and
so restrict attention to stationary spacetimes admitting a Killing horizon generated by a Killing vector  field $\xi$, such
as a black hole or cosmological Killing horizon.

For a spherical cross-section $\mathcal S$ of $B$ in a spherically symmetric spacetime,
fixing the induced metric $\gamma_{ij}$ implies that the area $A$ of $\mathcal S$ and the
lapse function $N_B=\sqrt{-\xi \cdot \xi}\big|_B$ at the boundary are fixed. In equilibrium, the latter determines the proper boundary
temperature. This proper boundary temperature $T$ is related to the asymptotic Hawking
temperature $T_{\rm H}$ by the Tolman relation~\cite{Tolman:1930zza}
\begin{equation} \label{tolman}
T N_B =  T_{\rm H} \,.
\end{equation}
Thus,   Dirichlet boundary conditions   restrict the allowed variations to those that keep the boundary temperature and area fixed, thereby ensuring that the appropriate thermodynamic potential, the Helmholtz free energy, is stationary. The corresponding thermodynamic representation is the  canonical  representation, in which equilibrium states are parametrized by the independent variables~\((T,A)\).

The central object in quasi-local thermodynamics is the Brown-York stress tensor~\cite{BrownYork1993}, a covariant rank-two tensor defined on $B$, whose normal and tangential projections onto a spatial cross-section $\mathcal S$ yield the quasi-local energy surface density, momentum surface density, and spatial stress. For the Dirichlet variational principle, the Brown-York quasi-local stress tensor is
defined as the functional derivative of the on-shell gravitational action with respect
to the induced boundary metric $\gamma_{ij}$. For general relativity, this requires
supplementing the Einstein-Hilbert action with the Gibbons-Hawking-York boundary
term. The quasi-local stress tensor then takes the form

\begin{equation}
\tau^{ij} \equiv \frac{2}{\sqrt{-\gamma}}\frac{\delta S_{\text{grav}}}{\delta \gamma_{ij}} = \frac{1}{8\pi G} \left( K^{ij} - K \gamma^{ij} \right),
\end{equation}
where \(K_{ij}\) is the extrinsic curvature of  \(B\) embedded in the bulk spacetime, and its trace is \(K \equiv \gamma^{ij} K_{ij}\). The quasi-local energy \(E\) is the integral over \(\mathcal S\) of the energy surface density $\varepsilon$ measured by observers with four-velocity \(u^{a} \), i.e,  the future pointing timelike unit normal to a  spacelike surface  $\Sigma$ that is orthogonal to $B$,
\begin{equation} \label{BYenergy}
E \equiv \oint_{\mathcal S} d^{d-2} x \sqrt{\sigma} \, \varepsilon, \quad \varepsilon \equiv u^{i} u^{j} \tau_{ij}\,,
\end{equation}
where $\sigma$ is the determinant of the induced metric \(\sigma_{ab}\) on   $\mathcal S$.
For general relativity we have $\varepsilon =-k/(8\pi G) $, where $k$ is the extrinsic trace of the $(d-2)$-surface $\mathcal S= \Sigma \cap B$ as embedded in~$\Sigma.$
The spatial components of the stress-energy tensor define the spatial stress tensor, which for general relativity is given by 
\begin{equation} \label{spatialstresstensor}
    s^{ab}\equiv\sigma^a_i \sigma^b_j \tau^{ij} = \frac{1}{8\pi G} \left [ - k^{ab} +(n \cdot a + k ) \sigma^{ab}\right]\,.
\end{equation}
Here, $a \equiv \sqrt{a_b a^b}$ and $a^b \equiv u^a \nabla_a u^b$ is the acceleration vector. Next, the trace of the spatial stress is proportional to the surface pressure  
\begin{equation} \label{surfacepressure}
s \equiv \frac{1}{d-2} \sigma_{ab} s^{ab} = \frac{1}{8\pi G} \left[ \frac{d-3}{d-2} k + n \cdot a \right].
\end{equation}
  In our spherically symmetric setup the Tolman temperature, quasi-local energy and  surface pressure are constant over $\mathcal S$. For a static, spherically symmetric solution to the Einstein equation with line element 
\begin{align}\label{metric}
    ds^2=-N^2(r)d t ^2+h^2(r) dr^2 +r^2d\Omega_{d-2}^2\,,
\end{align}
the quasi-local energy and surface pressure take the form \cite{BrownYork1993} 
\begin{align}
    E&= - \frac{1}{8\pi G} \frac{d-2}{r_B h(r_B)} A(r_B)\,,\\
    s&= \frac{1}{8\pi G} \left [ \frac{d-3}{r_B h(r_B)} + \frac{N'(r_B)}{ h(r_B) N (r_B) } \right],
\end{align}
where  $A= \Omega_{d-2}r_B^{d-2}$ is the area  (``volume'') of a round $(d-2)$-sphere with radius $r_B$, and the lapse at the boundary is $N_B=N(r_B) $ in this case.
For (AdS-)Schwarzschild black holes   $h(r)=N^{-1}(r)$, and $E$ and $s$ diverge as $r_B \to \infty$, so we adopt the background subtraction method to regulate these divergences, following the original literature~\cite{york1986,BrownYork1993,BrownMann1994}. That is, we define the quasi-local energy and   pressure of   (AdS-)Schwarzschild black holes such that they vanish if the mass $M$ of the black hole is zero, i.e. for Minkowski  and AdS spacetime, 
\begin{align}
E &\equiv E(M)- E(M=0)\,, \label{backgroundsubenergy}\\
 s &\equiv s(M)- s(M=0)\,. \label{backgroundsubpressure}
 \end{align}
 In the canonical representation the quasi-local first law for these black holes is
 \begin{equation}
     dF = - S dT - sdA\,.
 \end{equation}
 Clearly, the Helmholtz free energy $F$ is stationary if we impose the Dirichlet boundary condition that fixes $T$ and $A$.
In the internal energy representation instead, where the entropy and boundary area are the independent variables, the quasi-local first law is given by Eq.~\eqref{firstquasilocalfirstlawschw}.
In the holographically dual boundary theory, upon making the identification~\eqref{dictionarypressurearea}, the quasi-local first law takes the form \begin{equation} \label{boundaryfirstlaw}
dE = T dS - P dV \,,
\end{equation}
where the pressure 
$P$ is dual to the gravitational surface pressure 
$s$, and the volume 
$V$ corresponds to the boundary area 
$A$.  

In the holographic description, the pressure $P$ should be understood as the spatial stress of the boundary theory, defined thermodynamically as the response of its energy to variations of the metric (in particular, the volume) of the space on which it lives. No bulk interpretation is required: $P$ does not represent a force pushing in an ambient volume, but rather encodes how the energy of the boundary theory responds to geometric deformations of its spatial manifold. With the standard thermodynamic convention adopted here, $PdV$ represents work done by the boundary theory under an infinitesimal increase of its spatial volume; consequently, the contribution $-PdV$ in the first law reflects the   decrease of the system's internal energy.

\subsection{Quasi-local first law from covariant phase space}

The quasi-local gravitational first law \eqref{boundaryfirstlaw} can be checked explicitly using the expressions for the quasi-local quantities in Eqs. \eqref{eq:schwarzschildquantities} and \eqref{eq:AdSthermodynamicquantities} for (AdS-)Schwarzschild black holes, but we now review a more generic derivation using the covariant phase space (CPS) formalism~\cite{Lee:1990nz,wald1993entropy,iyer1994noether}. Our derivation closely follows the quasi-local analysis of Appendix~A in
\cite{Banihashemi:2022htw}, which was carried out in the context of de Sitter space,
but applies more generally to static black hole backgrounds with a Killing horizon. Extending the covariant phase space derivation to genuinely stationary but non-static spacetimes 
would require additional momentum and rotational contributions, see e.g. \cite{Brown:1990fk,BrownYork1993,Brown:1992bq}, and we leave this
to future work.
See also \cite{Svesko:2022txo} for a similar derivation of the quasi-local first law for   two-dimensional de Sitter space in Jackiw-Teitelboim gravity.

Recall that, in the CPS formalism, one starts from a Lagrangian $(d,0)$-form\footnote{Here the notion of a $(p,q)$-form refers to the double grading with respect to the horizontal (spacetime) and vertical (field-space) derivatives, denoted $d$ and $\delta$ respectively, in the variational bicomplex \cite{andersonbicomplex}. } $\mathfrak{L}[\phi]$, which satisfies the fundamental variational identity: 
\begin{align}\label{CPSfundidentity}
    \delta\mathfrak{L}=\mathfrak{E}(\mathfrak{L})+d\theta,
\end{align}
where $\mathfrak{E}(\mathfrak{L})$ is the Euler Lagrange $(d,1)$-form and $\theta$ is a $(d-1,1)$-form known as the presympletic potential density. The presymplectic $(d-1,2)$-form is then given by $\omega=\delta\theta$.

Now consider a diffeomorphism-invariant Lagrangian $\mathfrak L$, meaning that
for any spacetime diffeomorphism $\psi\colon\mathcal M\to\mathcal M$ one has
\begin{equation}
\mathfrak L[\psi^*\phi]=\psi^*\mathfrak{L}[\phi]\,.
\end{equation}
This invariance implies the existence of a conserved Noether current $(d-1,0)$-form
$J_\chi$ associated with an arbitrary vector field
$\chi\in\mathfrak X(\mathcal M)$ generating an infinitesimal diffeomorphism. The
induced variation of the dynamical fields is $\delta_\chi\phi=\mathcal L_\chi\phi$,
and the corresponding variation of the Lagrangian is also given by the Lie
derivative,
\begin{equation}
\delta_\chi\mathfrak L=\mathcal L_\chi\mathfrak L=d(\iota_\chi\mathfrak L)\,,
\end{equation}
where the last equality follows from Cartan's magic formula and the fact that $\mathfrak{L}$ is a top-form with respect to $d$. On the other hand, the
fundamental variational identity Eq.~\eqref{CPSfundidentity} yields
\begin{equation}
\delta_\chi\mathfrak L
=
\mathcal E(\mathfrak L)\cdot\mathcal L_\chi\phi
+
d\bigl(\theta(\mathcal L_\chi\phi)\bigr)\,.
\end{equation}
Comparing these expressions, one defines the Noether current
associated with $\chi$ as
\begin{equation}
J_\chi\equiv\theta(\mathcal L_\chi\phi)-\iota_\chi\mathfrak L\,,
\end{equation}
which satisfies $dJ_\chi=-\mathcal E(\mathfrak L)\cdot\mathcal L_\chi\phi$ and is
therefore conserved on shell. By the Poincar\'{e} lemma, this implies that on shell, the current is exact and can be expressed as the exterior derivative of a Noether charge \((d-2,0)\)-form \(Q_\chi\),
\begin{equation} \label{onshellidentitynoether}
J_\chi =dQ_\chi\,.
\end{equation}
By considering variations of the Noether current with respect to the diffeomorphisms generated by a vector field $\chi$, one can derive the on-shell variational identity \cite{wald1993entropy,iyer1994noether}
\begin{equation}\label{fundamental_identity}
\omega(\cdot, \mathcal{L}_\chi \phi) = d\left(\delta Q_\chi(\cdot) - \iota_\chi \theta(\cdot)\right)\,,
\end{equation}
where the open slot denotes a place where a field variation can be inserted. For simplicity we assume that the vector field $\chi$ is independent of the dynamical
fields $\phi$, so that its field-space variation vanishes.
The generalization to field-dependent vector fields, for which an additional contribution
appears in \eqref{fundamental_identity}, has been discussed extensively in the literature;
see e.g.\ \cite{Barnich:2001jy,Gao:2003ys,Compere:2006my,Freidel:2021cjp,Chandrasekaran:2021vyu,Ashtekar:2024stm}.

Next we derive the quasi-local first law from the on-shell identity
\eqref{fundamental_identity}, applied to a static black hole background solution
with a bifurcate Killing horizon, with $(d-2)$-dimensional bifurcation surface $\mathcal B$.
We specialize to the case where the vector field $\chi$ is chosen to be the timelike
Killing field $\xi$ of the static background spacetime, which   vanishes at $\mathcal B$
and has an arbitrary normalization
$N_B = \sqrt{-\xi \cdot \xi}\big|_B$ at the York boundary.

Because the symplectic current $\omega$ is   bilinear in its arguments, and $\xi$   generates a symmetry of the background solution, 
$\mathcal{L}_\xi \phi = 0$, it follows that
$\omega(\cdot,\mathcal{L}_\xi\phi)=0$. The fundamental identity therefore
reduces to the conservation law
\begin{equation}
\label{conservation_eq}
0 = d \bigl(\delta Q_\xi - \iota_\xi \theta \bigr)\,.
\end{equation}
To derive the quasi-local first law, we integrate this exact form over a $(d-1)$-dimensional spacelike hypersurface~\(\Sigma\) that is everywhere   orthogonal to   $\xi$. The surface \(\Sigma\) extends from the bifurcation surface \(\mathcal{B}\) of the Killing horizon   to the spatial cross-section  \(\mathcal{S}\) of the finite York boundary. 
By Stokes' theorem, the integral of the exact form reduces to two boundary terms:
\begin{equation} \label{orientationmatters}
0=\int_\Sigma d(\delta Q_\xi-\iota_\xi \theta)=\oint_{\partial\Sigma}(\delta Q_\xi-\iota_\xi \theta) = \oint_{\mathcal{B}}\delta Q_\xi + \oint_{\mathcal{S}}(\delta Q_\xi-\iota_\xi \theta) \,,
\end{equation}
where we used $\xi |_{\mathcal B}=0$ in the last equality.
The boundary components are oriented as follows: the surface
$\mathcal B$ is equipped with the inward-pointing orientation (toward smaller radius), while the surface $\mathcal S$ is equipped
with the outward-pointing orientation (toward larger radius),   
consistent with the orientation induced from $\Sigma$. Note that the orientation of
$\mathcal B$ is opposite to the convention adopted in \cite{wald1993entropy}.

Rearranging the terms, we obtain a relation equating the variation of quantities at the horizon to those at the York boundary:
\begin{equation} \label{flux_equality}
-\oint_{\mathcal{B}}\delta Q_\xi  = \oint_{\mathcal{S}}(\delta Q_\xi-\iota_\xi \theta)\,.
\end{equation}
We first evaluate the horizon contribution. Although the Killing field $\xi$ vanishes on
the bifurcation surface $\mathcal B$, its derivative does not. In particular, in the
binormal directions one has
$ 
\nabla_a \xi_b\big|_{\mathcal B} = \kappa\,\epsilon_{ab},
$
which defines the surface gravity $\kappa$. As shown by Wald~\cite{wald1993entropy}, the
integral of the Noether charge variation over the bifurcation surface is then
\begin{equation} \label{Bterm_entropy}
- \oint_{\mathcal B} \delta Q_\xi
= T_{\text{H}}\delta S\,.
\end{equation}
  For general relativity, the Noether charge entropy  reduces to the Bekenstein  entropy, so that
\begin{equation}
- \oint_{\mathcal B}\delta Q_\xi=\frac{\kappa}{ 8\pi G}\delta A_{\mathcal B}\,.
\end{equation}
This expression holds for arbitrary perturbations satisfying the linearized equations of
motion, regardless of whether the perturbation is stationary. For generic cross-sections
of the horizon and non-stationary perturbations, an additional dynamical correction term   appears in the entropy \cite{Rignon-Bret:2023fjq,Hollands:2024vbe,Visser:2024pwz}. In this
work we restrict attention to the bifurcation surface $\mathcal B$, where such corrections
vanish, and the     entropy is simply proportional to the area of the event horizon.

We now turn to the boundary term at the York surface $\mathcal S$.
Using the explicit form of the Noether charge and presymplectic potential for vacuum
general relativity, and rewriting the result in terms of Brown-York data on
$\mathcal S$, the boundary integrand decomposes as~\cite{Banihashemi:2022htw}
\begin{equation}
\left(\delta Q_\xi - \iota_\xi\theta\right)\Big|_{\mathcal S}
= -\frac{N_B}{8\pi G}\delta(k\epsilon_{\mathcal S})
+ N_B\epsilon_{\mathcal S}\frac12s^{ab}\delta\sigma_{ab}\,.
\label{QL_A30}
\end{equation}
Here $k$ is the trace of the extrinsic curvature of
$\mathcal S$ in $\Sigma$, $\sigma_{ab}$ is the induced metric on $\mathcal S$,
$\epsilon_{\mathcal S}$ its volume form, and $s^{ab}$ the Brown-York spatial stress
tensor~\eqref{spatialstresstensor} on $\mathcal S$.  

Returning to the integrated identity \eqref{flux_equality}, we thus find for vacuum general
relativity
\begin{equation}
-\frac{1}{8\pi G}\oint_{\mathcal S} N_B\delta\!\left(k\,\epsilon_{\mathcal S}\right)
= \frac{\kappa}{8\pi G}\delta A_{\mathcal B}
-\oint_{\mathcal S} N_B\epsilon_{\mathcal S}\frac12s^{ab}\delta\sigma_{ab}\,.
\label{QL_A31}
\end{equation}
This is the quasi-local first law for   static spacetimes with a Killing horizon,  in agreement with~\cite{Brown:1990fk,Brown:1992bq}.  In the presence of matter, additional contributions involving the
matter stress-energy tensor appear on the right-hand side, as derived in~\cite{Banihashemi:2022htw}. 
The boundary lapse $N_B$ is arbitrary on $\mathcal S$ and need not be constant. On the left side,  the combination $\varepsilon_{\text{loc}}\equiv - k\,\epsilon_{\mathcal S}/(8\pi G)$
represents the  local energy density on $\mathcal S$.
Multiplying \eqref{QL_A31} by $2\pi/\kappa$ shows that the left-hand side takes the
form $\oint_{\mathcal S}\beta \delta \varepsilon_{\text{loc}}$, where
$\beta=(2\pi/\kappa)N_B$ is the local inverse temperature at the boundary.
On the right side, the final term   involves the full spatial stress tensor $s^{ab}$ and represents
the quasi-local work associated with deformations of the induced metric
 on $\mathcal S$. 

Furthermore, the spatial stress tensor on $\mathcal S$ can be decomposed into its trace and trace-free parts as 
\begin{equation}
s^{ab}=\hat s^{ab}+s\,\sigma^{ab}, \qquad \sigma_{ab}\hat s^{ab}=0 \,,
\end{equation}
where $s$ denotes the isotropic surface pressure in Eq.~\eqref{surfacepressure} and $\hat s^{ab}$ the trace-free, anisotropic surface stress \cite{Brown:1994su}.
Employing $\delta\epsilon_{\mathcal S}=\epsilon_{\mathcal S}\tfrac12\sigma^{ab}\delta\sigma_{ab}$,
the work term in \eqref{QL_A31} splits as
\begin{equation} \label{reflocalsurfacepressure}
-\oint_{\mathcal S} N_B\epsilon_{\mathcal S}\frac12 s^{ab}\delta\sigma_{ab}
=
-\oint_{\mathcal S} N_B\left(
\frac12\,\epsilon_{\mathcal S}\hat s^{ab}\delta\sigma_{ab}
+s\,\delta\epsilon_{\mathcal S}
\right).
\end{equation}
This identifies $s$ as a local surface pressure conjugate to the area element
$\epsilon_{\mathcal S}$, while $\hat s^{ab}$ encodes shear stresses associated
with shape deformations of the induced metric. Hence, in the absence of spherical symmetry, there still is a (local) notion of
surface pressure.

As a final simplification, assume $\mathcal S$ is a round $(d-2)$-sphere in a
spherically symmetric spacetime. Then $k^{ab}=\frac{1}{d-2}k\,\sigma^{ab}$ and hence
$s^{ab}=s\sigma^{ab}$, i.e. $\hat s^{ab}=0$.
Moreover, spherical symmetry implies that the lapse $N_B$ and surface pressure $s$ are constant over~$\mathcal S$, and therefore may be pulled out of the integrals. Writing $A_{\mathcal S}\equiv\oint_{\mathcal S}\epsilon_{\mathcal S}$, the identity
\eqref{QL_A31} reduces to  
\begin{equation}
\delta E
= \frac{\kappa}{8\pi G\,N_B}\delta A_{\mathcal B}
- s\delta A_{\mathcal S}\,,
\label{QL_A32}
\end{equation}
where $E$ is the Brown-York quasi-local energy defined in \eqref{BYenergy}. In terms of the Bekenstein entropy and the Tolman temperature  \eqref{tolman} the first law reads
\begin{equation} \label{firstlawquasithree}
\delta E = T\delta S - s\delta A \,,
\end{equation}
where $A \equiv A_{\mathcal S}$. 
Importantly, the quasi-local first law~\eqref{firstlawquasithree} is a variational identity on the solution space, 
on which no boundary conditions are imposed; only the vacuum Einstein equations and their linearizations are satisfied.
Boundary conditions enter only at a later stage, when one
restricts the allowed variations such that the appropriate thermodynamic potential is
stationary in equilibrium.

\subsection{Quasi-local Smarr relation from covariant phase space}
\label{sec:quasilocalsmarr}

We now derive a Smarr-like relation \cite{Smarr:1972kt} for the quasi-local thermodynamic variables of AdS-Schwarzschild black holes. The derivation closely follows the analysis of~\cite{Banihashemi:2022htw} for de~Sitter spacetimes, with the essential difference that, in asymptotically AdS spacetimes, quasi-local quantities are typically defined relative to a reference background, since otherwise they diverge at   infinity. In particular, the quasi-local energy, surface pressure, and the quantity conjugate to the cosmological constant are defined here via background subtraction with respect to  AdS. An alternative would be to use boundary covariant counterterms, which have been implemented within the covariant phase space formalism and used to derive the thermodynamics of asymptotically locally AdS black holes  in \cite{Papadimitriou:2005ii}, but we leave this for future work.

We work in vacuum general relativity with negative cosmological constant 
 $\Lambda$ and Einstein-Hilbert Lagrangian form $\mathfrak L = \frac{\epsilon}{16 \pi G} (R-  2 \Lambda)$.
 Integrating the on-shell identity \eqref{onshellidentitynoether} over a spacelike hypersurface $\Sigma$ with
boundary $\partial\Sigma=\mathcal S\cup\mathcal B$ and applying Stokes' theorem yields
\begin{equation}
\int_{\Sigma} J_\xi  
=
\oint_{\mathcal B} Q_\xi
+
\oint_{\mathcal S} Q_\xi \,.
\label{Smarr_A15}
\end{equation}
Here the orientation of $\mathcal B$ and $\mathcal S$ is the same as in \eqref{orientationmatters}. We have omitted the Gibbons-Hawking-York exact form from the   Lagrangian. 
If it were included, the Lagrangian would be \(\mathfrak L' = \mathfrak L + d\mu\), which shifts 
the Noether current and charge as 
\(J'_\xi = J_\xi + d(\xi \cdot \mu)\) and 
\(Q'_\xi = Q_\xi + \xi \cdot \mu\), respectively, 
so that these contributions cancel in the identity above.

First, the bulk contribution in \eqref{Smarr_A15} can be evaluated using the equations of motion and the stationarity condition. Recall that the Noether current is $J_\xi = \theta (\mathcal L_\xi \phi) - i_\xi \mathfrak L$. Since  $\theta (\mathcal L_\xi \phi)$ depends linearly on $\mathcal L_\xi \phi$, this quantity vanishes if $\xi$ is a Killing vector field. Further, the trace of the vacuum Einstein equation  yields 
$R-2\Lambda=4 \Lambda / (d-2)$, hence one obtains
\begin{equation}
\int_\Sigma J_\xi
=
-\frac{ \Lambda}{(d-2) 4 \pi G}
\int_\Sigma \xi\!\cdot\!\epsilon
=
-\frac{ \Lambda}{(d-2)4\pi G}\,V_\xi \,,
\end{equation}
where $V_\xi\equiv \int_\Sigma \xi \cdot \epsilon $ is the Killing volume \cite{Jacobson:2018ahi} and $\epsilon$ is the volume $d$-form. 

Second, the Noether charge contribution from the bifurcation surface $\mathcal B$ is proportional to the surface
gravity   and the area of $\mathcal B$,  
\begin{equation}
\oint_{\mathcal B} Q_\xi
=
-\frac{\kappa}{8\pi G}A_{\mathcal B} \,,
\label{Smarr_A16}
\end{equation}
whose variational version we saw in \eqref{Bterm_entropy}.
Third, evaluating the Noether charge on the York surface $\mathcal S$ and expressing the result
in terms of Brown-York data yields a contribution from the quasi-local energy density and from the
surface stress tensor $s^{ab}$. In particular, one finds that the pullback of the Noether charge to $\mathcal S$ for general relativity is \cite{Banihashemi:2022htw}
\begin{equation}
Q_\xi\big|_{\mathcal S}
=
-\frac{d-3}{d-2}\frac{N_Bk\epsilon_{\mathcal S}}{8\pi G}
+
\frac{N_B\epsilon_{\mathcal S}}{d-2}s^{ab}\sigma_{ab} \,.
\label{Smarr_A19}
\end{equation}
Combining Eqs.~\eqref{Smarr_A15}--\eqref{Smarr_A19},   using spherical symmetry to pull the lapse
$N_B$ and $s^{ab}\sigma_{ab}=(d-2)s$ out of the integral over $\mathcal S$, and dividing both sides of the integral equation by $N_B$, one arrives at
the quasi-local  relation 
\begin{equation}
    - \frac{d-3}{d-2}\frac{1}{8\pi G} \oint_{\mathcal S} k_{\text{bh}} \epsilon_{\mathcal S}= \frac{\kappa}{8\pi G N_B} A_{\mathcal B}  -  s_{\text{bh}} A_{\mathcal S} - \frac{ \Lambda}{(d-2)4\pi G} \frac{V_\xi^{\text{bh}}}{N_{B}}\,.
\end{equation}
The issue with this relation for AdS-Schwarzschild black holes is that both sides diverge in the infinite boundary limit.  However, these divergences are the same as
those that arise for pure AdS, for which the quasi-local relation takes the form 
\begin{equation}
       - \frac{d-3}{d-2}\frac{1}{8\pi G} \oint_{\mathcal S} k_{\text{AdS}} \epsilon_{\mathcal S}=  -   s_{\text{AdS}} A_{\mathcal S} - \frac{ \Lambda}{(d-2)4\pi G } \frac{\tilde V_\xi^{\text{AdS}}}{N_{B}}\,,
\end{equation}
where the horizon term is absent because there is no inner boundary in pure AdS. The Killing volume in   AdS space is defined as $\tilde V_\xi^{\text{AdS}} = \int_{\Sigma_{\rm AdS}} \tilde \xi_{\text{AdS}} \cdot \epsilon_{\text{AdS}}.$  

In performing the AdS background subtraction we require the York boundary data to match, which ensures that the   canonical   thermodynamic representation  coincides in the black hole spacetime and the reference AdS background. 
Equivalently, the induced metric on the timelike boundary $B$ is identified between pure AdS and AdS-Schwarzschild, so that the area $A_{\mathcal S}$ and the boundary lapse
$
N_B 
$
are the same in the two spacetimes. 
This fixes the normalization of the stationary Killing field in AdS by requiring that it generates the same proper-time translations on $B$.\footnote{
In Euclidean AdS gravity, the same matching condition is imposed by fixing
the proper boundary temperature, as required by the canonical ensemble.
This is implemented by adjusting the periodicity of Euclidean time so that the
thermal circle has the same proper length at the radial cutoff in the black hole
and background AdS geometry 
\cite{Witten:1998zw}.
} 

In particular, in static coordinates~\eqref{metric}, if $\xi_{\text{bh}}=\partial_{t_{\rm bh}}$ in the black hole geometry, then correspondingly in pure AdS the representation-matched Killing field is
$
\tilde \xi_{\text{AdS}}
=
\partial_{\tilde t},$
where $
\tilde t
=
t_{\rm AdS} N_{\rm AdS}(r_B) /N_{\rm bh}(r_B),
$
so that in both spacetimes the stationary Killing field satisfies \begin{equation}\sqrt{-\xi_{\text{bh}}\cdot\xi_{\text{bh}}}\bigg|_B=\sqrt{-\tilde \xi_{\text{AdS}}\cdot \tilde \xi_{\text{AdS}}}\bigg|_B=N_{\rm bh}(r_B)\,.\end{equation}  This implies 
\begin{equation}\label{backkillingvolumeads}\tilde V_\xi^{\text{AdS}} = \int_{\Sigma_{\rm AdS}} \tilde \xi_{\text{AdS}} \cdot \epsilon_{\text{AdS}} =  \frac{N_{\rm bh}(r_B) }{N_{\rm AdS}(r_B)}\int_{\Sigma_{\rm AdS}}  \xi_{\text{AdS}} \cdot \epsilon_{\text{AdS}}\,,\end{equation}
where $\xi_{\text{AdS}} = \partial_{t_{\text{AdS}}}$, $N_{\text{bh}}(r_B)$ is given in \eqref{lapseadsschw} and $N_{\text{AdS}}(r_B) = \lim_{r_h \to 0} N_{\text{bh}}(r_B)$.

Subtracting the quasi-local Smarr relation for pure AdS from that of AdS-Schwarzschild then removes the shared boundary divergences and yields a finite relation:

\begin{equation}  \label{quasilocaladslaw}
(d-3)  E 
=
(d-2)\left[
\frac{\kappa}{8\pi G N_B}A_{\mathcal B}
-
sA_{\mathcal S}
\right]
-
\frac{\Lambda \tilde{V}_\xi}{4\pi G N_B}
,
\end{equation}
with 
\begin{equation} \label{defquasilocalall}
    E \equiv E_{\text{bh}} - E_{\text{AdS}}\,, \qquad s \equiv s_{\text{bh}} - s_{\text{AdS}}, \qquad \tilde V_\xi \equiv V_\xi^{\text{bh}}- \tilde V_\xi^{\text{AdS}}\,,
\end{equation}
where $E$ is the background subtracted quasi-local energy   \eqref{backgroundsubenergy}, $s$ is the background subtracted surface pressure \eqref{backgroundsubpressure} and $\tilde{V}_\xi$ is the background subtracted Killing volume. We note that $V_\xi^{\text{bh}}$ is an integral from the horizon to $\mathcal S$, whereas $\tilde{V}_\xi^{\text{AdS}}$ is an integral over a ball-shaped region in AdS with boundary $\mathcal S.$
Further, for $\Lambda=0$ the final term in \eqref{quasilocaladslaw}  vanishes, so we recover the quasi-local Smarr relation for asymptotically flat Schwarzschild black holes  derived by  York   \cite{york1986}. 
We discuss the thermodynamic interpretation of the quasi-local AdS Smarr formula in Section \ref{sec:adssmarrthermo}, in particular the thermodynamic meaning of the Killing volume term.

\section{Small and large (AdS-)Schwarzschild black holes }
\label{sec:smalllarge}

In this Section we provide explicit coordinate expressions for the quasi-local thermodynamic quantities for Schwarzschild and AdS-Schwarzschild black holes. Moreover, we review York's result \cite{York1985a,york1986} that there are two  black hole solutions for a given boundary area and temperature: the so-called small and large black holes. This holds both for asymptotically flat and   AdS-Schwarzschild black holes in any number of spacetime dimensions $d\ge 4$  \cite{BrownMann1994,Andre:2020czm,Andre:2021ctu}. We then finish this section by taking the infinite-boundary limit of   the quasi-local AdS Smarr formula, and discussing its thermodynamic interpretation. 

From now on we denote the holographic pressure and volume for spherically symmetric black holes by $P$ and $V$, respectively, instead of $s$ and $A.$

\subsection{Schwarzschild black holes}\label{Yorksolutionsappendix}

For a Schwarzschild black hole  in $d$ spacetime dimensions, a.k.a.  Tangherlini black holes \cite{Tangherlini:1963bw}, the static, spherically symmetric line element is given by Eq.~\eqref{metric}, with $h(r)=N^{-1}(r)$ and lapse function
\begin{equation}
N(r)=\sqrt{1-\frac{r_h^{\,d-3}}{r^{\,d-3}}}\, .
\end{equation}
Here $r_h$ denotes the horizon radius, which is related to the total mass $M$ of the black hole by
\begin{equation}
M=\frac{(d-2)\Omega_{d-2}}{16\pi G}\, r_h^{\,d-3}\, .
\end{equation}
We consider the thermodynamics of this spacetime when enclosed by a spherical York boundary of radius $r_B> r_h$. The associated  entropy, volume, energy, temperature, and pressure are
\begin{align}\label{eq:schwarzschildquantities}
S(r_h)&=\frac{\Omega_{d-2} r_h^{\,d-2}}{4G}\,, 
\qquad 
V(r_B)=\Omega_{d-2} r_B^{\,d-2}\,, \nonumber\\
E(r_B,r_h)&=\frac{(d-2)\Omega_{d-2} r_B^{\,d-3}}{8\pi G}
\left[1-\left(1-\frac{r_h^{\,d-3}}{r_B^{\,d-3}}\right)^{1/2}\right], \nonumber\\
T(r_B,r_h)&=\frac{d-3}{4\pi r_h}
\left(1-\frac{r_h^{\,d-3}}{r_B^{\,d-3}}\right)^{-1/2},\\
P(r_B,r_h)&=\frac{d-3}{8\pi G r_B}
\left[
\left(1-\frac{r_h^{\,d-3}}{2r_B^{\,d-3}}\right)
\left(1-\frac{r_h^{\,d-3}}{r_B^{\,d-3}}\right)^{-1/2}
-1
\right]. \nonumber
\end{align}
A flat-space background subtraction is employed to regulate the energy and pressure, fixing them to vanish in flat spacetime. These quantities  satisfy the quasi-local Smarr relation \cite{york1986,Andre:2021ctu}
\begin{equation} \label{eq:quasilocalsmarrflat}
(d-3)E=(d-2)(TS-PV)\, .
\end{equation}
In the limit $r_B\to\infty$ we have $E\to M$ and $T\to T_{\rm H}$.\footnote{This is because $r_B^{d-3} \left [1-\left(1-\frac{r_h^{\,d-3}}{r_B^{\,d-3}}\right)^{1/2} \right]\to   \frac{r_h^{d-3}}{2}$ and $\left(1-\frac{r_h^{\,d-3}}{r_B^{\,d-3}}\right)^{-1/2}\to 1$, as $r_B \to \infty$.} Moreover,
$V \sim r_B^{d-2}$, while to leading order $P \sim r_B^{-(2d-5)}$,\footnote{This follows from the expansion $(1-x/2)(1-x)^{-1/2}\approx 1+x^2/8$ for small $x$.} so that
$PV\sim r_B^{-(d-3)}\to 0$ (for $d>3$). We therefore recover the standard Smarr
formula for  Schwarzschild black holes,
$ 
(d-3)M=(d-2)T_{\rm H} S.
$

The thermodynamic variables in \eqref{eq:schwarzschildquantities} are expressed in terms of
the horizon radius $r_h$ and the cavity radius $r_B$, which in turn are proportional to the
$(d-2)$-th roots of the entropy and volume, respectively. In this $(S,V)$ representation,
Eqs.~\eqref{eq:schwarzschildquantities} therefore define the energy, temperature, and
pressure as functions $E(S,V)$, $T(S,V)$, and $P(S,V)$. At the level of equilibrium thermodynamics one may use different thermodynamic representations, related by Legendre transformations, which provide   equivalent descriptions of equilibrium configurations.

In the canonical
thermodynamic representation the independent variables are $(T,V)$, and the horizon radius is
determined by inverting the Tolman temperature,
\(
T = T(r_B,r_h) ,
\)
to obtain $r_h=r_h(r_B,T)$. In general this inversion must be carried out implicitly. In
four-dimensional~\cite{york1986} and five-dimensional~\cite{Andre:2020czm} asymptotically
flat spacetime, however, the resulting equation can be solved explicitly.

At fixed temperature and cavity size, the temperature relation generically admits more than one solution for the horizon radius. In four dimensions, York showed that for values $r_BT \geq \sqrt{27}/(8\pi)$ of the dimensionless combination $r_BT$, there exist two  black hole solutions: the small and large black hole. More generally, in $d$ spacetime dimensions the condition for the existence of two real solutions is \cite{Andre:2021ctu}
\begin{equation} \label{boundgeneraldimensionsrT}
r_BT \geq  \left(\frac{d-1}{2}\right)^{\!\frac{1}{d-3}}\, \frac{\sqrt{(d-3)(d-1)}}{4\pi }.
\end{equation}
Specializing to $d=4$, the Tolman temperature reads
\begin{equation} \label{tolmanflat}
T(r_B,r_h)=\frac{1}{4\pi r_h}\left(1-\frac{r_h}{r_B}\right)^{-1/2}.
\end{equation}
Rearranging and squaring yields the cubic equation
\begin{equation}\label{eq:cubicrh}
16\pi^2 T^2 r_h^2\left(1-\frac{r_h}{r_B}\right)-1=0\,,
\end{equation}
which determines the horizon radius at fixed $(T,r_B)$.

\begin{figure}[t]
    \centering
    \begin{subfigure}[b]{0.45\linewidth}
        \centering
        \includegraphics[width=\linewidth]{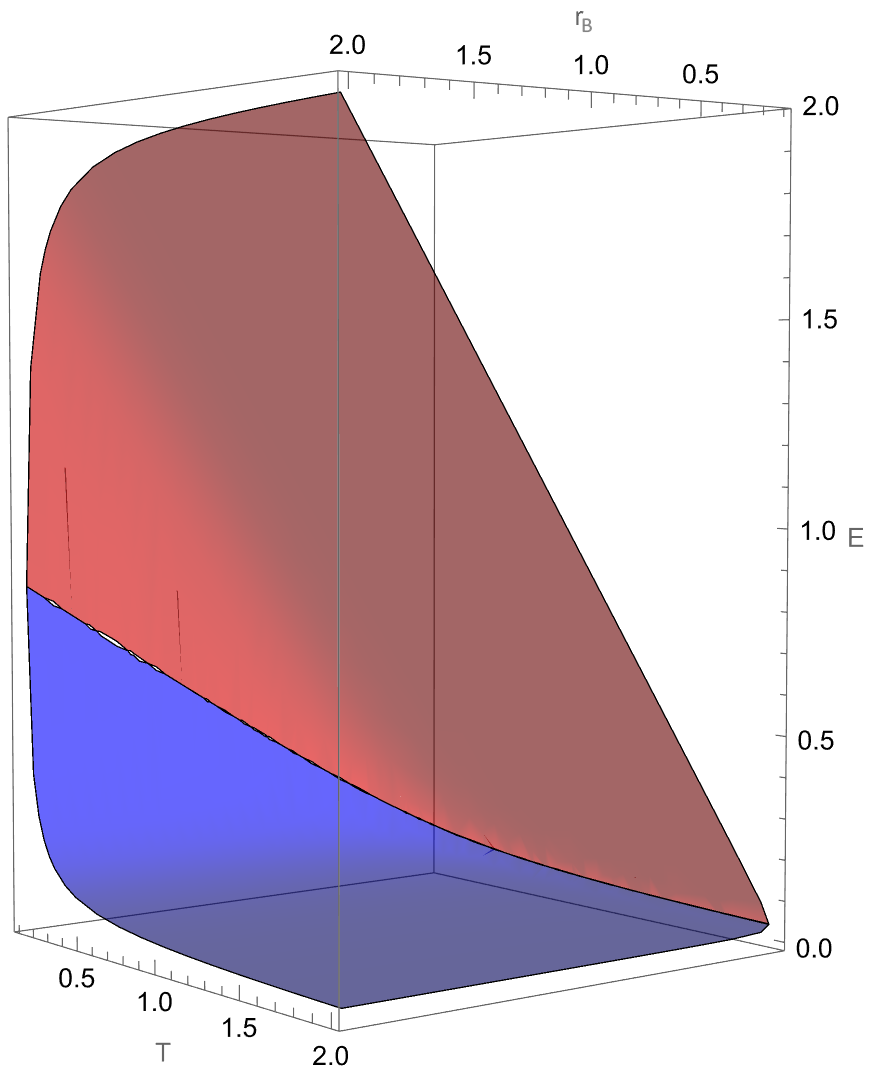}
    \end{subfigure}
    \hfill
    \begin{subfigure}[b]{0.48\linewidth}
        \centering
        \includegraphics[width=\linewidth]{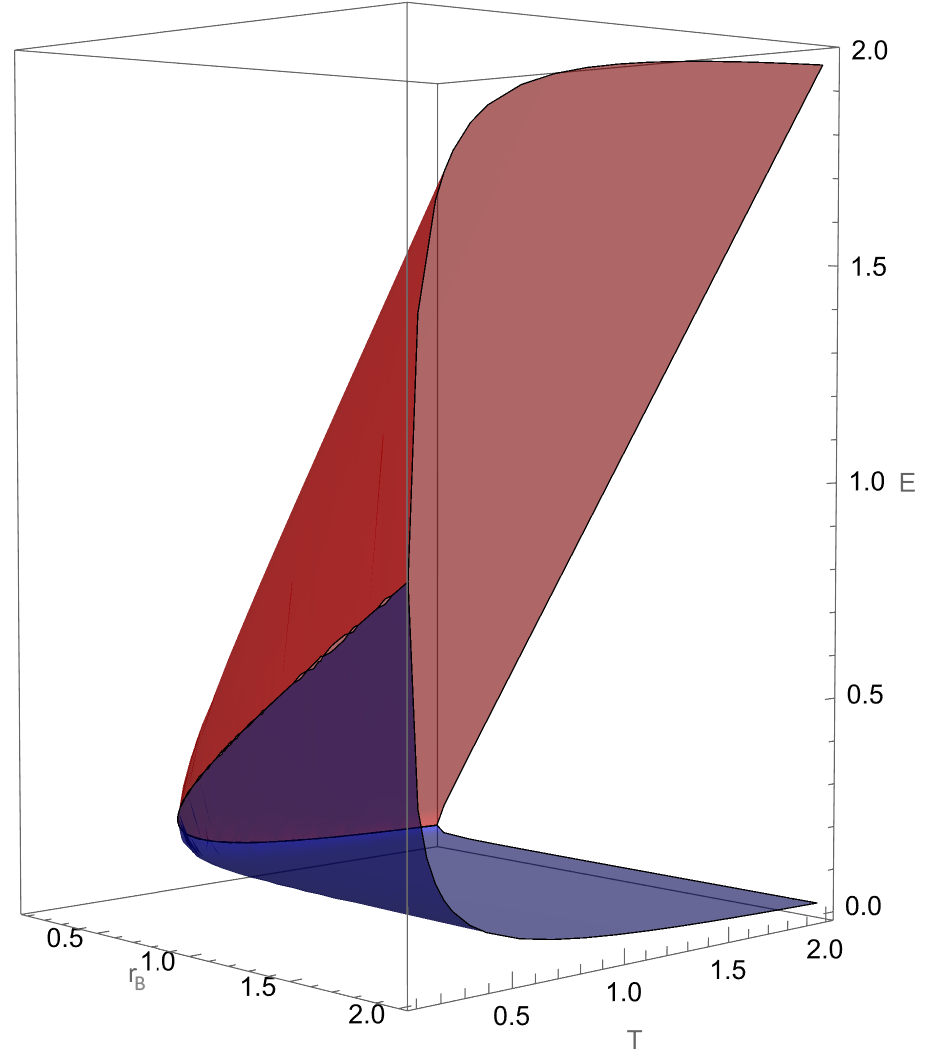}
    \end{subfigure}
    \caption{Quasi-local energy $E$ of an asymptotically flat Schwarzschild black hole in 3+1 dimensions as a function of the boundary radius $r_B$ and the Tolman temperature $T$, seen from two different angles. The red corresponds to the large black hole solution, whereas the blue part corresponds to the small black hole solution.}
    \label{fig:schwarzschildPrT}
\end{figure}

The structure of Eq.~\eqref{eq:cubicrh} makes the origin of the multiplicity transparent. For fixed cavity radius $r_B$, the left-hand side starts at $-1$ for $r_h=0$, increases as the positive quadratic term dominates and crosses zero once, but then decreases at larger $r_h$ due to the dominance of the negative cubic term, crossing zero a second time. Consequently, whenever the bound $r_BT \geq \sqrt{27}/(8\pi)$ is satisfied, the equation admits two real solutions. When the inequality holds these solutions are distinct, while at saturation they merge at
\(
2r_B = 3r_h,
\)
corresponding to the photon sphere. These two solutions can be visualized by plotting a thermodynamic quantity as a function of $T$ and $r$, for instance the energy, as in Figure \ref{fig:schwarzschildPrT}.

For $d=4$ the small and large black hole solutions can be written, respectively,   as \cite{york1986}
\begin{align} \label{smallbhsch}
r_h^{(1)}&=\frac{r_B}{3}\left[1-2\cos\!\left(\frac{\alpha+\pi}{3}\right)\right],\\ \label{largebhsch}
r_h^{(2)}&=\frac{r_B}{3}\left[1+2\cos\!\left(\frac{\alpha}{3}\right)\right],
\end{align}
where
\begin{equation} 
\cos\alpha = 1-\frac{54}{(8\pi r_BT)^2}\,,
\qquad 0\leq\alpha\leq\pi\,.
\end{equation}
Expanding these solutions around $  Tr_B =\infty$ yields
\begin{align} \label{smalllarger}
r_h^{(1)} &= \frac{1}{4\pi T}\left(1+\frac{1}{8\pi r_BT}
+\frac{5}{128\,\pi^2 (r_BT)^2} + \mathcal O((r_BT)^{-3})\right),\\
r_h^{(2)} &= r_B\left(1-\frac{1}{(4\pi r_BT)^2}
-\frac{2}{(4\pi r_BT)^4}+ \mathcal O((r_BT)^{-5})\right)\,. \label{largelarger}
\end{align}
At leading order, the horizon radius of the small black hole is independent of the cavity size, while the horizon radius of the large black hole approaches the cavity wall.

\begin{figure}[t]
    \centering
    \includegraphics[width=0.5\linewidth]{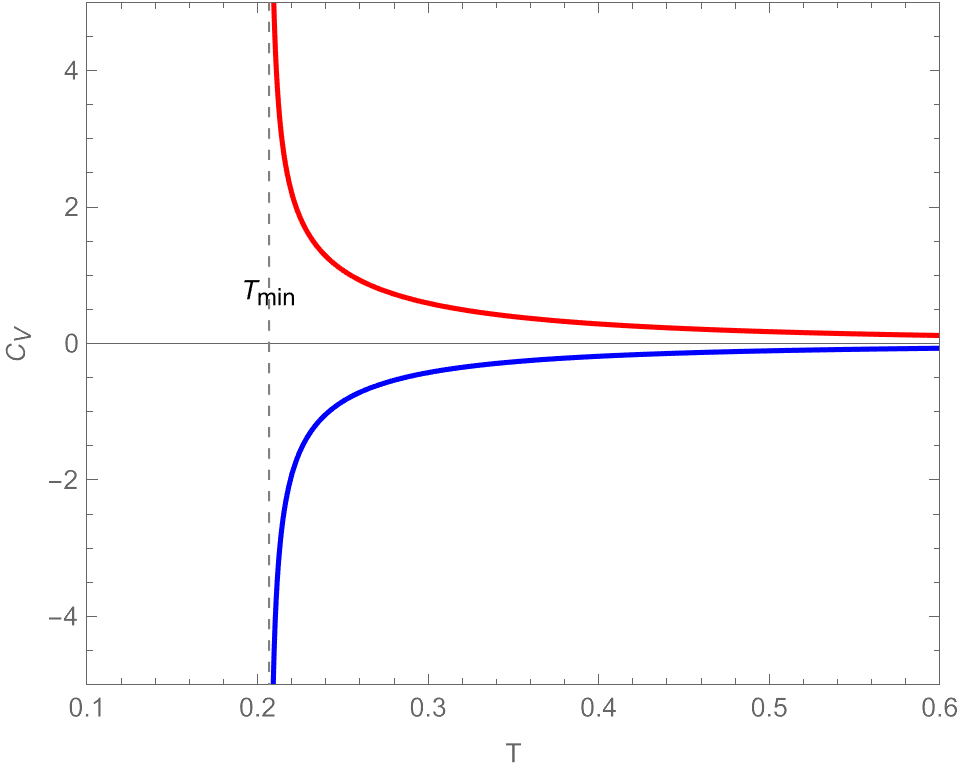}
    \caption{Heat capacity $C_V$   as a function of the temperature $T$ for a Schwarzschild black hole in $d=4$ at fixed $r_B=1$, with $G=1$. The small black hole (blue line) has a negative heat capacity, whereas the large black hole (red line) has a positive heat capacity.}
    \label{fig:schwarzschildCV}
\end{figure}

Thermal stability in the canonical representation is governed by the heat capacity at fixed volume $V$  
\begin{equation} \label{heatcapacitydef}
C_V \equiv T\left(\frac{\partial S}{\partial T}\right)_{V}
=\left(\frac{\partial E}{\partial T}\right)_{V}.
\end{equation}
Using $S=S(r_h)$ together with the Tolman temperature  in \eqref{eq:schwarzschildquantities}, one obtains~\cite{york1986,Andre:2021ctu}
\begin{equation}\label{eq:CA_general_d}
C_V
=\frac{(d-2)\Omega_{d-2}r_h^{d-2}}{4G}\,
\left(1-\frac{r_h^{d-3}}{r_B^{d-3}}\right)
\left(\frac{d-1}{2}\frac{r_h^{d-3}}{r_B^{d-3}}-1\right)^{-1}\,.
\end{equation}
Since $r_B>r_h$ implies the numerator is positive, the sign of $C_V$ is controlled by the denominator. The heat capacity diverges when
\begin{equation}
\frac{r_h^{d-3}}{r_*^{d-3}}=\frac{2}{d-1}
\qquad\Longleftrightarrow\qquad
r_*=\left(\frac{d-1}{2}\right)^{\!\frac{1}{d-3}}\,r_h\,,
\end{equation}
which coincides with the photon-sphere radius and marks the cusp where the two branches of solutions merge at the minimal temperature  $T_\text{min}=\sqrt{(d-3)(d-1)}/(4\pi r_h) $. Consequently, in the two-solution regime the larger-horizon branch $r_h < r_B < r_*$ (the ``large'' black hole) has $C_V>0$ and is   thermodynamically stable, whereas the smaller-horizon branch $r_B > r_*$ (the ``small'' black hole) has $C_V<0$ and is thermodynamically unstable. We plotted the heat capacity in four dimensions in Figure \ref{fig:schwarzschildCV}.

\subsection{AdS-Schwarzschild black holes}\label{AdSsection}

For an AdS-Schwarzschild black hole the line element is again given by  \eqref{metric}, with $h(r)=N^{-1}(r)$, where now the lapse function is
\begin{equation} \label{lapseadsschw}
    N(r)= \sqrt{1+\frac{r^2}{L^2}-\frac{r_h^{d-3}}{r^{d-3}}\left(1+\frac{r_h^2}{L^2}\right)}\,.
\end{equation}
The mass $M$ of the black hole is related to the horizon radius $r_h$ by 
\begin{equation}
    \frac{16 \pi G M}{(d-2) \Omega_{d-2}} =   r_h^{d-3} \left(1+\frac{r_h^2}{L^2}\right)\,.
\end{equation}
The quasi-local thermodynamic variables for an AdS-Schwarzschild black hole enclosed by a spherical  cavity of radius $r_B$ are \cite{BrownMann1994}
\begin{align}\label{eq:AdSthermodynamicquantities}
       S(r_h)&= \frac{\Omega_{d-2} r_h^{d-2}}{4 G}\,, \quad V(r_B) = \Omega_{d-2} r_B^{d-2}\,, \nonumber\\
       E (r_B, r_h)&= \frac{(d-2)\Omega_{d-2}r_B^{d-3}}{8\pi G}\left(\sqrt{1+\frac{r_B^2}{L^2}}-\sqrt{1+\frac{r_B^2}{L^2}-\frac{r_h^{d-3}}{r_B^{d-3}}\left(1+\frac{r_h^2}{L^2}\right)}\right), \! \nonumber\\
    T (r_B,r_h)&= \frac{1}{4\pi r_h }  \left(d-3 + (d-1)\frac{r_h^2}{L^2}\right) \left ( 1+\frac{r_B^2}{L^2}-\frac{r_h^{d-3}}{r_B^{d-3}}\left(1+\frac{r_h^2}{L^2}\right) \right)^{-1/2}\,,\\
  P(r_B,r_h)&=\frac{1}{8\pi G\,r_B}\left[
\frac{(d-3)+\frac{(d-2)r_B^{2}}{L^{2}}
-\frac{(d-3)}{2}\frac{r_h^{d-3}}{r_B^{d-3}}\!\left(1+\frac{r_h^{2}}{L^{2}}\right)}
{\sqrt{\,1+\frac{r_B^{2}}{L^{2}}
- \frac{r_h^{d-3}}{r_B^{d-3}} \!\left(1+\frac{r_h^{2}}{L^{2}}\right)\,}}  \;-\;
\frac{(d-3)+\frac{(d-2)r_B^{2}}{L^{2}}}
{\sqrt{\,1+\frac{r_B^{2}}{L^{2}}\,}}
\right]\,. \nonumber
\end{align}
The energy and pressure are defined with an AdS background subtraction, fixing them to vanish in empty AdS spacetime.

These quantities satisfy the quasi-local Smarr relation 
\eqref{quasilocaladslaw}
\begin{equation}
\label{eq:AdSsmarrrelation}
    (d-3) E =(d-2)\left(T S-P V \right)-\frac{ \Lambda \tilde V_\xi}{4 \pi G N_{\text{bh}}(r_B)} \,,
\end{equation}
where the background subtracted Killing volume \eqref{defquasilocalall} is,    in static coordinates, 
\begin{equation}\begin{aligned} \label{killingvolumeadsfinite}
    \tilde V_\xi &= \Omega_{d-2} \left ( \int_{r_h}^{r_B}\!\! \!\! dr r^{d-2} -  \frac{N_{\text{bh}}(r_B )}{N_{\text{AdS}}(r_B)} \int_{0}^{r_B} \!\! \!\!dr r^{d-2}\right)\, \\
    &=\frac{\Omega_{d-2}}{d-1} \left (  r_B^{d-1} - r_h^{d-1} - r_B^{d-1} \frac{N_{\text{bh}}(r_B )}{N_{\text{AdS}}(r_B)}  \right)\,.
\end{aligned}
\end{equation}
As in the asymptotically flat case, passing from the $(S,V)$ entropy representation to the $(T,V)$ Helmholtz representation    requires solving the Tolman temperature relation $T=T(r_B,r_h)$ for the horizon radius. In contrast to the flat-space Schwarzschild case, this inversion does not admit a simple closed form for $d \ge 4$. We now check that, for sufficiently large cavity radius $r_B$, the temperature relation again admits two real solutions for $r_h$ if $d \ge 4$ \cite{BrownMann1994}. This qualitative structure is consistent with the existence of small and large black hole branches for AdS-Schwarzschild first identified by Hawking and Page \cite{Hawking:1982dh}.

Multiplying the temperature relation by $4\pi r_h N(r_B)$
and squaring yields
\begin{equation}\label{eq:AdS_rh_poly_start}
16\pi^2T^2 r_h^2\left(1+\frac{r_B^2}{L^2}-\frac{r_h^{d-3}}{r_B^{d-3}}\left(1+\frac{r_h^2}{L^2}\right)\right)
=\left(d-3+ (d-1)\frac{r_h^2}{L^2}\right)^2.
\end{equation}
Already for $d=4$ this becomes a quintic equation in $r_h$, which in principle does not have an exact solution, while in higher dimensions the degree increases further. Writing Eq.~\eqref{eq:AdS_rh_poly_start} as a polynomial in $r_h$ gives
\begin{align}\label{eq:AdS_rh_poly}
&
-\frac{16\pi^2T^2}{r_B^{d-3}L^2}\,r_h^{d+1}
-\frac{16\pi^2T^2}{r_B^{d-3}}\,r_h^{d-1}
-\frac{(d-1)^2}{L^4}\,r_h^4 \nonumber\\
&+\left(16\pi^2T^2\left(1+\frac{r_B^2}{L^2}\right)-\frac{2}{L^2}(d-1)(d-3)\right)r_h^2
-(d-3)^2=0\,.
\end{align}
This ordering is only correct for $d \ge 5,$ since for $d=4$ the second term is cubic, and for $d=3$ it becomes quadratic.
From this expression it is straightforward to determine the number of real solutions in each spacetime dimension.

For $d=3$ the polynomial admits only a single real, positive solution. In this case the constant term vanishes, and the polynomial contains a quadratic term with a positive coefficient (in the regime of sufficiently large 
$r_B$) together with a quartic term with a negative coefficient. Consequently, the function initially rises from zero, but is eventually dominated by the negative quartic term and decreases again, crossing the horizontal axis only once. The corresponding horizon radius of a BTZ black hole in a cavity can be determined explicitly:
\begin{equation}
r_h(r_B,T)=\frac{2\pi L T\,r_B}{\sqrt{1+4\pi^2L^2T^2}}\,.
\end{equation}
For $d\ge4$, the left side of \eqref{eq:AdS_rh_poly} has a negative constant term, a quadratic term whose coefficient is positive in the appropriate regime, and higher-order terms with negative coefficients. Thus, for sufficiently large $r_B$, the left-hand side starts negative at $r_h=0$, rises and crosses zero, and eventually becomes negative again at large $r_h$, implying a second crossing. This corresponds to the familiar small  and large black hole solutions.

\begin{figure}[t]
    \centering
    \includegraphics[width=0.5\linewidth]{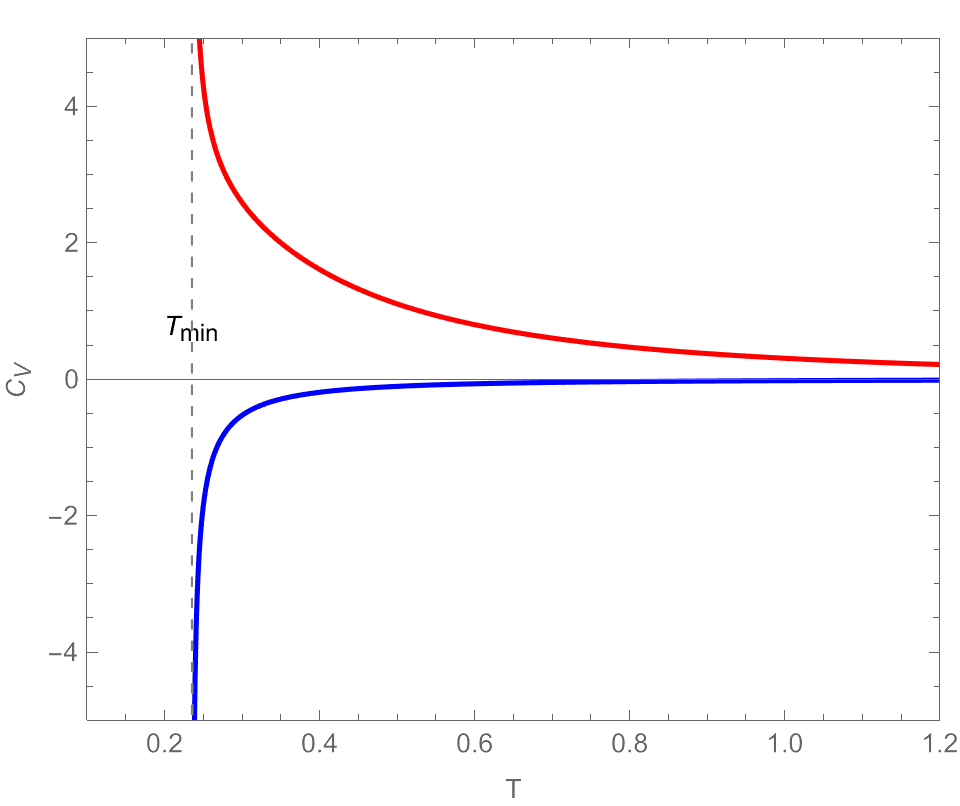}
    \caption{Heat capacity $C_V$ as a function of temperature $T$ at fixed $r_B=1$ for AdS-Schwarzschild in $d=4$, with $G=L=1$. $T_\text{min}$ is the temperature at which the two solutions merge. The small black hole (blue line) has a negative heat capacity, whereas the large black hole (red line) has a positive heat capacity.}
    \label{fig:AdSschwarzschildCV}
\end{figure}

To estimate these two real solutions, consider the large-cavity regime, in which the \textcolor{black}{$r_h^{d+1}$ term is suppressed on the small branch and the $r_h^{d-1}$ term is suppressed on both branches by powers of $r_B^{-(d-3)}$}. Dropping these terms gives an effective equation,
\begin{align}\label{eq:AdS_quartic_approx}
&\textcolor{black}{-\frac{16\pi^2T^2}{r_B^{d-3}L^2}\,r_h^{d+1}}-\frac{(d-1)^2}{L^4}r_h^4
\\
&+\left(16\pi^2T^2\left(1+\frac{r_B^2}{L^2}\right)-\frac{2}{L^2}(d-1)(d-3)\right)r_h^2
-(d-3)^2
\approx 0\,,
\end{align}
\textcolor{black}{where the first term is suppressed for the small branch.} When $r_BT\gg 1$, the dominant contribution in the coefficient of $r_h^2$ is $16\pi^2T^2 r_B^2/L^2$, and \eqref{eq:AdS_quartic_approx} further simplifies to
\begin{equation}
\textcolor{black}{-\frac{16\pi^2T^2}{r_B^{d-3}L^2}\,r_h^{d+1}}-\frac{(d-1)^2}{L^4}r_h^4+\frac{16\pi^2T^2 r_B^2}{L^2}r_h^2-(d-3)^2\approx 0.
\end{equation}
The two real roots then correspond to the small and large black hole branches, which to leading order in $r_B$ are given by
\begin{equation}\label{AdSrhapprox}
r_h^{(1)}\approx \frac{L(d-3)}{4\pi Tr_B}\,,
\qquad
r_h^{(2)}\approx \textcolor{black}{cr_B}\,,
\qquad
\textcolor{black}{\frac{c}{\sqrt{1-c^{d-1}}}=\frac{4\pi LT}{d-1}.}
\end{equation}
Finally, the heat capacity \eqref{heatcapacitydef} at fixed cavity volume is given by
\begin{align}\label{eq:CV_AdS_general_d}
C_V
&=T\left(\frac{\partial S}{\partial T}\right)_V=T\left(\frac{\partial S}{\partial r_h}\right)_V\left(\frac{\partial T}{\partial r_h}\right)_V^{-1}
\\
&=\frac{(d-2)\Omega_{d-2}r_h^{d-2}}{4G}\left[
\frac{2(d-1)\,r_h^2}{\,L^2(d-3)+(d-1)r_h^2\,}
+\frac{d-3+(d-1)\frac{r_h^2}{L^2}}{2 N^2_{\text{bh}}(r_B)}
 \frac{r_h^{d-3}}{r_B^{d-3}} 
-1\right]^{-1}. \nonumber
\end{align}
The heat capacity diverges at the minimal temperature at fixed $r_B$ where the two branches merge. Away from this point, the branch with the larger horizon radius has $C_V>0$ and is   thermodynamically stable, whereas the branch with the smaller horizon radius has $C_V<0$ and is thermally unstable. This is visualized in Figure \ref{fig:AdSschwarzschildCV}.

\subsection{Thermodynamic interpretation of the AdS Smarr relation}
\label{sec:adssmarrthermo}
Using the expressions for the quasi-local thermodynamic quantities presented in the previous Section, we can take the infinite-boundary limit of the quasi-local AdS Smarr relation \eqref{quasilocaladslaw}, and compare it to the standard Smarr relation  for AdS black holes. This comparison will aid us in interpreting the quasi-local quantities thermodynamically.

If one multiplies both sides of the quasi-local Smarr relation for AdS-Schwarzschild  by~$N_B$ and takes the infinite boundary limit, then 
 surprisingly all the terms remain finite and non-zero.   In particular, using the expressions for the quasi-local variables of AdS-Schwarzschild in static coordinates in  Eq.~\eqref{eq:AdSthermodynamicquantities}, in the limit   $r_B\to \infty$  one finds
 \begin{equation}
    E N_B \to M\,,\qquad (d-2)  s A_{\mathcal S} N_B \to  M\,. 
\end{equation}
 where $M$ is the mass of the black hole.
 Hence, in this limit the Smarr relation becomes 
 \begin{equation} \label{inflimitsmarr}
     (d-3) M = (d-2)   \frac{\kappa A_{\mathcal B}}{8\pi G} - M - \frac{\Lambda \tilde V_\xi}{4 \pi G}\,.
 \end{equation}
 Adding  $M$ on both sides of this equation yields a Smarr relation where the dimension-dependent factors in front of the mass and horizon area term are the same, i.e.
 \begin{equation} \label{smarrlimitnodim}
     M = \frac{\kappa A_{\mathcal B}}{8\pi G} - \frac{\Lambda \tilde V_\xi}{(d-2)4\pi G}\,.
 \end{equation}
 This differs from the standard   Smarr formula for AdS-Schwarzschild black holes \cite{Kastor:2009wy}. We can recover the standard formula by taking the $r_B\to \infty$ limit of the Killing volume term, using Eq. \eqref{killingvolumeadsfinite}, and rewriting it
\begin{align}
       & \frac{\Lambda \tilde V_\xi}{4\pi G }\to \frac{(d-2)\Omega_{d-2}}{16 \pi G} r_h^{d-3}\left ( \frac{r_h^2}{L^2} - 1 \right) = - M + \frac{\Lambda \bar V_\xi}{4\pi G} \,,\label{inflimitkillingvol}
\end{align}
with $\bar V_\xi = -\Omega_{d-2} r_h^{d-1}/(d-1)$ and $\Lambda = - (d-1)(d-2)/(2L^2)$. Inserting  this limit into \eqref{inflimitsmarr} yields the standard AdS Smarr formula \cite{Kastor:2009wy}
\begin{equation} \label{standardadsasymptsmarr}
    (d-3) M = (d-2)  \frac{\kappa A_{\mathcal B}}{8\pi G}  - \frac{\Lambda \bar V_\xi}{4 \pi G}\,,
\end{equation}
where $\bar V_\xi$ is another background subtracted Killing volume defined in \cite{Jacobson:2018ahi}, see Eq.~\eqref{backgroundsubvolume}, which does not coincide with $\tilde V_\xi.$
The difference lies in the fact that in $\tilde V_\xi$ the norm of the AdS Killing field is matched to the norm of the black hole Killing field at the boundary, whereas for $\bar V_\xi$ this matching condition is absent. Explicitly, in static coordinates for AdS-Schwarzschild geometry the former is\footnote{This is obtained by expanding $ N_\text{bh}/N_\text{AdS}$ around $r_B = \infty$. At an intermediate step, the Killing volume takes the form $\Omega_{d-2} \lim\limits_{r_B \to \infty}\left((r_B^{d-1}-r_h^{d-1})/(d-1)-\left(1-r_h^{d-3}(r_h^2+L^2)/(2r_B^{d-1})+\mathcal{O}\left(r_B^{-(d+1)}\right)\right)r_B^{d-1}/(d-1)\right)$, which in turn yields $\frac{\Omega_{d-2}}{d-1}  \lim_{r_B \to \infty}\left(-r_h^{d-1}+r_h^{d-3}(r_h^2+L^2)/2+\mathcal{O}\left(r_B^{-2}\right)\right)$, resulting in the final expression.}
 \begin{align}
     \tilde V_\xi &= \int_{\Sigma_{\text{bh}}} \!\! \xi_{\text{bh}} \cdot \epsilon_{\text{bh}} -  \int_{\Sigma_{\text{AdS}}}\tilde \xi_{\text{AdS}} \cdot \epsilon_{\text{AdS}}
     \\
     &= \Omega_{d-2}  \lim_{r_B \to \infty}\left ( \int_{r_h}^{r_B}\!\!dr r^{d-2} -  \frac{N_{\text{bh}}(r_B )}{N_{\text{AdS}}(r_B)} \int_{0}^{r_B} \!\!dr r^{d-2}\right) \nonumber 
     = - \frac{\Omega_{d-2}}{2 (d-1)} r_h^{d-3} (r_h^2 - L^2)\,,
 \end{align}
whereas the latter is
 \begin{align}
     \bar   V_\xi &= \int_{\Sigma_{\text{bh}}} \!\! \!\! \xi_{\text{bh}} \cdot \epsilon_{\text{bh}} -  \int_{\Sigma_{\text{AdS}}} \!\! \!\!\xi_{\text{AdS}} \cdot \epsilon_{\text{AdS}}
     = \Omega_{d-2}  \left ( \int_{r_h}^{\infty}\!\! \!\! dr r^{d-2} -  \int_{0}^{\infty} \!\! \!\!dr r^{d-2}\right) = - \frac{\Omega_{d-2}r_h^{d-1}}{d-1}\,.
 \end{align}
The thermodynamic interpretation of the background-subtracted Killing volumes remains unclear. In the following, we discuss four possible interpretations and argue that the final one is the most compelling.\\

\noindent\textbf{1. Extended black hole thermodynamics.}
We begin with the interpretation of the cosmological constant as a bulk pressure, together with
an associated thermodynamic volume (see Section~\ref{sec:extendedthermo}) \cite{Kastor:2009wy,Dolan:2010ha,Cvetic:2010jb,Dolan:2011xt,Dolan:2012jh,Kubiznak:2014zwa}.
 In the present context, however, two distinct quantities appear to be conjugate to the cosmological constant, namely $\tilde V_\xi$ and $\bar V_\xi$, and they enter the extended first law for AdS-Schwarzschild black holes in qualitatively different ways. Expressed in terms of $\bar V_\xi$, the first law takes the form
\begin{equation} \label{somefirstlawkillingvolume}
dM = \frac{\kappa}{8\pi G}\, dA_{\mathcal B} + \frac{\bar V_\xi}{8\pi G}\, d\Lambda\, .
\end{equation}
By contrast, when written in terms of $\tilde V_\xi$ one finds
\begin{equation}
dM = \frac{\kappa}{8\pi G}\, dA_{\mathcal B} + M\,\frac{d\Lambda}{2\Lambda} + \frac{\tilde V_\xi}{8\pi G}\, d\Lambda ,
\end{equation}
or, equivalently, in terms of the AdS radius $L$, using $d\Lambda/\Lambda = -2\, dL/L$,
\begin{equation}\label{newfirstlawcft}
dM = \frac{\kappa}{8\pi G}\, dA_{\mathcal B}
 - \frac{M}{d-2}\, \frac{dL^{d-2}}{L^{d-2}}
 + \frac{\tilde V_\xi}{8\pi G}\, d\Lambda \,.
\end{equation}
As emphasized in~\cite{Visser:2021eqk}, the second term on the right-hand side becomes a  pressure-volume term in the first law of the dual CFT. Indeed, if $M$ is identified with the CFT energy and the CFT volume scales as $V_{\text{CFT}} \propto L^{d-2}$, then its conjugate  pressure is proportional to  $
P_{\text{CFT}}  \propto M  /((d-2)L^{d-2})
$, which is the equation of state
for a CFT in $d-1$ spacetime dimensions. Since a pressure-volume term is already present at the boundary, it would be redundant and conceptually problematic to interpret the final term in~\eqref{newfirstlawcft} as an additional bulk pressure-volume contribution. As we discuss below, this term instead admits a   holographic interpretation.

From this perspective, one might attempt to identify $\bar V_\xi$ in \eqref{somefirstlawkillingvolume} with the bulk thermodynamic volume, $V_\Lambda = - \bar V_\xi$, as is standard in extended black hole thermodynamics. However, this identification  treats $\Lambda$ as a thermodynamic state variable, which is conceptually questionable.\\

\noindent\textbf{2. Holographic extended thermodynamics.}
Within an extended version of the AdS/CFT correspondence \cite{Karch:2015rpa,Visser:2021eqk,Ahmed:2023snm}, $\tilde V_\xi$ admits a  
holographic interpretation.
Specifically, comparing the Smarr relation~\eqref{smarrlimitnodim} to the large-$C$ Euler relation \eqref{largeceuler}, and using the standard holographic dictionary 
 $ 
E_{\text{CFT}} = M  
$
and $ T_{\text{CFT}} = \kappa/ 2\pi  $
which holds if $L$ is the radius of the sphere on which the CFT lives, corresponding to $R=L$ in Eq.~\eqref{cftmetricroverl},  we are led to identify the term involving $\tilde V_\xi$ with the product of the chemical potential $\mu$ and the central charge $C$. This yields
\begin{equation}\label{chemicalpotentialvolume1}
\mu C = - \frac{\Lambda \tilde V_\xi}{(d-2)\,4\pi G} \, .
\end{equation}
Fixing the normalization of the central charge as
\begin{equation}
C = \frac{\Omega_{d-2} L^{d-2}}{16\pi G} \, ,
\end{equation}
the corresponding holographic dictionary for the chemical potential becomes
\begin{equation}
\mu = \frac{2(d-1)\, \tilde V_\xi}{\Omega_{d-2} L^{d}} \, .
\end{equation}
The alternative form~\eqref{inflimitsmarr} of the AdS Smarr relation also admits a consistent holographic interpretation. By invoking the dictionary~\eqref{chemicalpotentialvolume1} together with the CFT equation of state
 $ 
E_{\text{CFT}} = (d-2 ) P_{\text{CFT}} V_{\text{CFT}}  ,
$
  the Smarr relation~\eqref{inflimitsmarr} is dual to
\begin{equation}
(d-3) E_{\text{CFT}} = (d-2)\, (T_{\text{CFT}}S - P_{\text{CFT}}V_{\text{CFT}} + \mu C) \, .
\end{equation}
Moreover, the extended bulk first law~\eqref{newfirstlawcft} maps precisely to the extended CFT first law
\begin{equation}
dE_{\text{CFT}} = T_{\text{CFT}} dS - P_{\text{CFT}} dV_{\text{CFT}} + \mu dC \, .
\end{equation}
Within this framework, the background-subtracted Killing volume $\tilde V_\xi$ is   dual to the chemical potential associated with the central charge. However, this interpretation relies on treating the central charge as a   state variable, an assumption that lies outside the standard formulation of AdS/CFT, where $C$ is fixed by the microscopic definition of the theory. \\

\noindent \textbf{3. Generalized Euler equation.} 
 Mancilla \cite{Mancilla:2024spp} has recently proposed an alternative thermodynamic interpretation of the
standard AdS Smarr formula based on an effective action formulation of perfect fluids. The starting point of his analysis is to rewrite the AdS Smarr relation
in a form that closely resembles the usual Euler relation, but with an additional correction
term that is attributed to the presence of a geometric scale of the thermal CFT system. Concretely, by adding and subtracting $M/ (d-2)$ on the right side in Eq. \eqref{smarrlimitnodim} and using the conformal equation of state, 
the AdS-Schwarzschild Smarr formula can be recast into the suggestive ``Euler-like'' form
\begin{equation}\label{eq:mancilla_eulerlike}
E_{\text{CFT}} = T_{\text{CFT}}S - P_{\text{CFT}}V_{\text{CFT}} + \frac{\Omega_{d-2}}{8\pi G}\, r_h^{\,d-3}\,.
\end{equation}
Here the final term arises from
\( 
\frac{1}{d-2}\!\left(M - \frac{\Lambda \tilde V_\xi}{4\pi G}\right)
= \frac{\Omega_{d-2}}{8\pi G}\, r_h^{\,d-3}.
\)
For different horizon topologies, this contribution is multiplied by a factor $k = 1,0,-1$ for spherical, planar, and hyperbolic cases, respectively. Holographically, the parameter $k$ characterizes the constant-curvature spatial geometry on which the dual CFT is defined.

Mancilla then argues that the correction term compared to the standard extensive Euler relation can be understood systematically within the effective action formulation of relativistic perfect fluids. When the dual CFT is defined
on a curved or compact spatial manifold, the background geometry introduces an additional
length scale.
At the level of the zeroth-order fluid action, according to Mancilla, this
introduces an auxiliary scalar variable \(y\), which parametrizes the response of local
thermodynamic quantities to the background geometry and, in global equilibrium, can be
related to the inverse conformal   factor of the metric. The local thermodynamic
potential density is accordingly generalized to a function \(\chi=\chi(s,y)\), which
enters directly as the Lagrangian density of the zeroth-order fluid action.\footnote{Mancilla also allows for a nonvanishing chemical potential $\mu$ associated to electric charge, which we set to zero here.} Varying this
action and matching the resulting stress tensor to the perfect-fluid stress tensor then leads to a
generalized Euler identity,
\begin{equation}\label{eq:mancilla_geneuler}
\epsilon   \;=\; Ts -p + y\,\frac{\partial p}{\partial y}\,,
\end{equation}
where the additional term encodes the response of the fluid to the geometric scale $y$. For holographic fluids constructed from the boundary stress tensor of static AdS black holes, the correction term is nonzero only when the transverse space has nonvanishing curvature and precisely reproduces the final term in Eq.~\eqref{eq:mancilla_eulerlike}, with $\partial p/\partial y \propto k$. In global thermal equilibrium, the standard AdS Smarr relation may therefore be interpreted as the global thermodynamic dual, within the fluid/gravity correspondence, of the generalized local Euler equation for the boundary perfect fluid.

While this proposal is intriguing and internally consistent, its physical status remains
unclear. First, the variable \(y\) is introduced by enlarging the set of effective field theory scalars in a way
tailored to produce the extra term \(y\,\partial p/\partial y\) in \eqref{eq:mancilla_geneuler},
and its hydrodynamic interpretation is not as direct as that of \(s\) (or the chemical potential $\mu$).
Indeed, even though Mancilla provides a geometric motivation for \(y\) and discusses its
behavior under Weyl transformations in equilibrium, the identification
of this particular scalar as \emph{the} bookkeeping device for finite-volume/curvature effects
is not uniquely determined by standard thermodynamic or hydrodynamic reasoning. Second, the holographic application assumes from the outset a thermodynamic potential
of the form $\chi=\chi(s,y)$ obeying the generalized Euler relations
\eqref{eq:mancilla_geneuler}. The explicit form of $\chi$ is then fixed using the
black-hole thermodynamic dictionary, so that the construction presently offers a
reformulation and consistency check of the correction term in
\eqref{eq:mancilla_eulerlike}, rather than an independent derivation.
 Finally, it is not evident that the
effects of compactness/curvature in the dual CFT should generically be captured by a
single additional contribution of the form \(y\,\partial p/\partial y\) in an Euler
identity, as opposed to a more general pattern of subextensive corrections to thermodynamic
potentials. \\

\noindent\textbf{4. Failure of (quasi-)homogeneity.}
Finally, we present a new thermodynamic interpretation of the Killing-volume term in the
(quasi-local) AdS Smarr relation and of the corresponding correction in Eq.~\eqref{eq:mancilla_eulerlike}.
The three interpretations discussed above all assume that the additional term in the AdS
Smarr relation should be promoted to an independent thermodynamic or hydrodynamic
conjugate pair (e.g.\ $P_\Lambda V_\Lambda$, $\mu C$, or $y\partial p/\partial y$).
By contrast, we take $(E,T,S,P,V)$ to define a thermodynamic contact manifold with only
$n=2$ independent state variables, such as $(S,V)$ or $(T,V)$, and we do not introduce
any additional thermodynamic state variable, in particular none associated with
the cosmological constant.
We instead view the extra term  in the AdS
Smarr relation  as a manifestation of the failure of (quasi-)homogeneity
of the thermodynamic potentials, and therefore as a non-homogeneous correction
determined entirely by the existing thermodynamic variables.

The homogeneous   Euler relation $E = TS - PV$ follows from the degree-one homogeneity  of the energy
function $E(S,V)$, i.e.\ from the scaling relation
\(
E(\lambda S,\lambda V) = \lambda E(S,V).
\)
More generally, a quasi-homogeneous scaling relation of the form
\(
E(\lambda^{w_S} S,\lambda^{w_V} V) = \lambda^{w_E} E(S,V)
\)
yields a quasi-homogeneous Euler identity,
\(
w_E E = w_S T S - w_V P V .
\)
For asymptotically flat black holes, the quasi-local energy satisfies such a quasi-homogeneous
scaling relation, with $w_E = d-3$ and $w_S = w_V = d-2$, so that the quasi-local Smarr relation \eqref{eq:quasilocalsmarrflat}
maps directly to a quasi-homogeneous Euler identity.

For AdS black holes, however, the quasi-local Smarr relation
\eqref{eq:AdSsmarrrelation} contains an additional Killing-volume contribution that spoils
quasi-homogeneity. In static coordinates, $\tilde V_\xi$ depends explicitly on both the horizon
radius and the boundary radius, cf.~Eq.~\eqref{killingvolumeadsfinite}, and hence on the entropy
and thermodynamic volume that characterize the equilibrium configuration. This
  dependence, however, does not imply the existence of an additional state variable.

This viewpoint also clarifies the    asymptotic AdS Smarr formula. There are different ways to express the AdS Smarr formula, e.g. Eqs.~\eqref{smarrlimitnodim} and \eqref{standardadsasymptsmarr}, but they all express a failure of homogeneity or quasi-homogeneity. 
For instance, in our interpretation the correction term in \eqref{eq:mancilla_eulerlike}  reflects a
subextensive contribution induced by finite geometric scales. In an appropriate large-system limit  such subextensive terms may become negligible,
and the standard extensive Euler relation may emerge in this limit  (see Section \ref{sec:extadsblackholes}); however, away from this limit no generalized Euler identity need exist.

Furthermore, this interpretation is   closely aligned with Verlinde’s analysis \cite{Verlinde:2000wg} of CFT states  dual to AdS-Schwarzschild black holes, where the total energy is decomposed into an extensive thermal part and a subextensive Casimir contribution (see Section \ref{sec:cftextensivity}). This Casimir energy is due to the finite-temperature Casimir effect and is defined precisely as the part of the energy that violates the Euler relation. In this sense, the correction term in the AdS Smarr formula \eqref{eq:mancilla_eulerlike} plays the same role as Verlinde’s Casimir energy: it measures the failure of extensivity induced by finite geometric scales, and becomes negligible only in the appropriate large-system limit.

We develop this perspective further in the following Sections, in particular by exploring
the role of non-(quasi-)homogeneity and subextensive corrections in the thermodynamics of
AdS black holes.

\section{Extensivity in equilibrium thermodynamics} \label{sec:extensivityinthermo}

Having motivated a holographic interpretation of thermodynamic pressure and volume for black holes, and clarified the thermodynamic interpretation of the AdS Smarr relation, we can now ask whether black hole entropy is an extensive thermodynamic variable, perhaps in an appropriate large-system or ``thermodynamic'' limit. The distinction between extensive and intensive quantities is a central organizing principle of thermodynamics. Extensivity characterizes how thermodynamic variables scale with system size and underlies the existence of well-defined thermodynamic limits and  Euler relations.

Although often identified with linear scaling in the volume, extensivity admits several inequivalent but closely related formulations. 
Three such notions arise within standard thermodynamics,  whose logical relations are summarized in Appendix~\ref{appendixExtensivity}, and one notion arises in statistical mechanics~\cite{DunningDavies1983,DUNNINGDAVIES1985383,DUNNINGDAVIES1988705}. 
In this paper we adopt the definition of extensivity based on degree-one homogeneity of thermodynamic state functions. 
The present Section is purely thermodynamic; applications to (quasi-local) black hole thermodynamics are taken up later.

\subsection{Extensivity before   the large-system limit}\label{sec:extensivitythermolimit}

In equilibrium thermodynamics, the total state space may be described as a $(2n+1)$-dimensional manifold $M$ whose coordinates consist of a thermodynamic potential together with $n$ pairs of conjugate variables. At this level, the variables are a priori independent, and their variations are encoded in a differential contact one-form $\alpha$ on this space, turning it into a \textit{contact manifold}\footnote{More precisely, a contact manifold is a $2n+1$-dimensional manifold $M$ equipped with a \textit{contact structure}, which is a maximally non-integrable distribution $\mathcal{D}\subset TM$ of hyperplanes. These hyperplanes can be encoded as the kernel of a contact 1-form $\alpha$, i.e. $\mathcal{D}_p=\text{ker}(\alpha_p)$ for $p\in M$. The maximal non-integrability condition is then equivalent to $\alpha\wedge (d\alpha)^n\neq 0$. Note that $\alpha$ is not unique, since any $\xi\alpha$ with $\xi\in C^\infty(M)$ defines the same contact structure $\mathcal{D}$.} \cite{hermann1973geometry,MRUGALAgeometricformulation,Arnold:1989who,Bravetti:2018rts}. Concretely, if the thermodynamic potential is the internal energy $E$ and
the pairs of conjugate variables are $(T,S),(Y_i, \xi_i)$ and $(\mu_j, N_j)$, where $i=1,\dots, r$ and $j=1, \dots, s$ so that $1+ r +s =n$, then the contact one-form is
\begin{equation}
   \alpha =  dE  - T dS + \sum_{i=1}^{r}Y_i d \xi_i - \sum_{j=1}^{s}\mu_j dN_j =dE -  \sum_{i=1}^{n}y_i d x_i\,.
\end{equation}
Here, $T$ is the temperature; $S$ is the entropy; $\xi_i$ are deformation coordinates (such as the volume, surface area, elastic deformation parameters, or the magnetization in magnetic systems); $Y_i$ are generalized forces conjugate to the deformation coordinates (such as minus the pressure, surface tension, elastic stresses, or the magnetic field in the corresponding special cases);   $N_j$ are internal state variables (for instance, the number of molecules of component $j$); and   the $\mu_j$ are chemical potentials  \cite{landsberg1978thermodynamics}.

A \emph{fundamental relation} is the additional assumption that a chosen thermodynamic potential is a state function of $n$ independent thermodynamic variables. From the contact-geometric perspective, the fundamental relation corresponds to selecting a \textit{Legendre submanifold}\footnote{A Legendre submanifold is an integral submanifold of maximal dimension. An integral submanifold of a (contact) distribution $\mathcal{D}\subset TM$ is a submanifold $N\subset M$ such that $TN\subset\mathcal{D}$. Because of the maximal non-integrability of the contact distribution, the maximal dimension of an integral submanifold of a contact manifold is $n$. Since for a contact distribution $\mathcal{D}_p=\text{ker}(\alpha_p)$, it follows immediately that $\alpha$ vanishes identically on any integral submanifold.} $\mathscr{L} \subset M$ (which is automatically $n$-dimensional) of the $(2n+1)$-dimensional contact manifold. This $n$-dimensional submanifold of $M$ is identified with the space of physical equilibrium states.\footnote{The formalization of thermodynamics in the language of contact geometry has been extended to non-equilibrium configurations, beginning with \cite{1997haslachnonequil}.} On a Legendre submanifold   the contact one-form vanishes identically. That is, $\jmath^*\alpha=0$, where \(\jmath\colon \mathscr{L}\to M\) denotes the inclusion. This implies that the $n$ variables conjugate to the independent variables are partial derivatives of the potential with respect to the $n$ independent variables.

For example, for the internal energy $E$, the fundamental relation is    \begin{equation}\label{fundamentalrelationthis}E=E(S, \xi_1, \dots, \xi_r, N_1, \dots, N_s)\,,\end{equation} or, equivalently, $E=E(x_1, \dots, x_n)$.  The partial derivatives of the internal energy with respect to the independent variables are equal to their conjugate variables, i.e. $y_i= \partial E / \partial x_i$, or, equivalently,
\begin{equation} \label{conjugatevariables}
    T = \frac{\partial E}{\partial S}, \qquad Y_i = - \frac{\partial E}{\partial \xi_i} \,, \qquad \mu_j =  \frac{\partial E}{\partial N_j}\,.
\end{equation}
The differential of the fundamental relation is what is often called the ``first law'' in the   literature on black hole thermodynamics. We stick to this bad terminology, even though in standard thermodynamics the first law only expresses energy conservation. In terms of the internal energy, the first law reads
\begin{equation} \label{energyrepresentationfirstlaw}
    dE  = T dS - \sum_{i=1}^{r}Y_i d \xi_i + \sum_{j=1}^{s}\mu_j dN_j = \sum_{i=1}^{n}y_i d x_i\,,
\end{equation}
where the conjugate variables satisfy \eqref{conjugatevariables}.

We assume that the thermodynamic system is free of adiabatic partitions, i.e. internal constraints that prevent heat exchange between different parts of the system.\footnote{This assumption is weaker than Landsberg’s notion of a \emph{simple system} \cite{landsberg1978thermodynamics}. In addition to the absence of adiabatic partitions, Landsberg’s definition assumes extensivity of the entropy, \emph{i.e.}\ first-order homogeneity in the extensive variables. We do not impose this extensivity assumption, but only require the existence of a single global thermodynamic description.}
 The absence of such partitions implies that equilibrium states are characterized by a single entropy, a single internal energy, and hence a single temperature, rather than by separate thermodynamic descriptions for distinct subsystems. The stationary black hole geometries considered in this work satisfy this assumption.

Under these conditions, the equilibrium states admit a single global thermodynamic description in terms of a fixed set of independent variables and a single fundamental relation. In the contact-geometric formulation, this corresponds to a single globally defined Legendre submanifold of the thermodynamic contact manifold. A choice of \emph{thermodynamic representation} then amounts to selecting a thermodynamic potential and a corresponding set of independent variables parametrizing this submanifold \cite{callen1985thermodynamics}. Different representations lead to different forms of the   first law which are   equivalent, since they are all realized as the coordinate-free condition \(\jmath^*\alpha=0\). The potentials of different thermodynamic representations are related by Legendre transformations.   Legendre transformations act as contact transformations,\footnote{A contact transformation, or contactomorphism, on a contact manifold $(M,\alpha)$ is a diffeomorphism $f$ of $M$ preserving the contact structure. Since the contact structure is given by the kernel of $\alpha$, this is equivalent to $f^*\alpha=\xi \alpha$ for some $\xi\in C^\infty(M)$.} preserving the contact one-form up to multiplication by a smooth function.

In much of the thermodynamic literature it is then assumed that the equilibrium state coordinates $x_i$  in the energy representation are extensive, with their conjugate coordinates $y_i$
intensive. However, this does not follow from the construction of thermodynamic state space, but represents an additional assumption concerning
how coordinates on the equilibrium manifold behave under scaling and/or composition of
systems. Thus, Landsberg
\cite{landsberg1961thermodynamics,Landsberg1972FourthLaw}
referred to extensivity (understood as homogeneity of degree one) as the
\emph{fourth law of thermodynamics}. He emphasized that this property does not follow from the first three laws, but is instead an empirical characteristic of ``normal'' thermodynamic systems. Lieb and Yngvasson also note the independence of the extensivity property, though they do assume it for the entropy \cite{Lieb:1997mi}. Systems with long-range interactions typically violate this assumption,   highlighting the
limited domain of validity of the ``fourth law'' \cite{LandsbergTranah1980a,TranahLandsberg1980b,Landsberg1984,Landsberg1992bookchapter,Lieb:1997mi}. 

\paragraph{Extensivity as homogeneity of degree one.}
Now, extensivity can either be defined as a rather general property of thermodynamic state functions, which applies even before the thermodynamic limit, or as a property which in a sense defines the thermodynamic limit. It might seem paradoxical to speak of thermodynamic properties before the thermodynamic limit, but it is important to understand that there exists a thermodynamics of small systems, i.e.\ systems with a volume that has not (yet) been taken to infinity \cite{hillThermodynamicsSmallSystems1962,hillThermodynamicsSmallSystems1994} (see \cite{lavis2021,Jha2025} for philosophical literature on this topic). Since small systems may also be thermodynamic, we will instead refer to the thermodynamic limit (e.g.\ $V\to\infty$, with $N/V=$ constant) as the large-system limit. Small systems are generically non-extensive due to boundary effects. But in order to check this, one needs a notion of extensivity which applies to such systems. A simple illustration of subextensive corrections in a classical ideal gas at finite $N$ is reviewed in Appendix~\ref{idealgasappendix}.

There are three  notions of extensivity in thermodynamics that apply prior to the large-system limit:  additivity, doubling and homogeneity of degree one \cite{DunningDavies1983,DUNNINGDAVIES1985383,DUNNINGDAVIES1988705}. Additivity, together with continuity of the state function, implies homogeneity, while homogeneity in turn implies doubling. Although doubling is therefore the weakest of these conditions, we regard it as too weak to characterize extensivity, since it constrains the state function only at \textcolor{black}{integer scales and hence requires continuity to reach all scales.}
In the following we work with the homogeneity definition, since it is  the weakest physically reasonable  notion of extensivity. We refer to Appendix~\ref{appendixExtensivity} for a more detailed discussion of these definitions.   
\begin{description}
\item[(H) \emph{Homogeneity of degree one}.]
A thermodynamic state function $f(x_1, \dots, x_k, y_{k+1}, \dots, y_n)$, in shorthand notation $f(\textbf x, \textbf y)$, is said to be \emph{extensive in the homogeneous sense} if it is  homogeneous  of
degree one with respect to the $x_i$ variables, that is, if for all $\lambda\in\mathbb{R}_+$\footnote{
A function $f(\mathbf{x},\mathbf{y})$ is said to be homogeneous of degree $w$ with respect to the variables $\mathbf{x}$ if
\(
f(\lambda \mathbf{x},\mathbf{y})=\lambda^{w} f(\mathbf{x},\mathbf{y})
\)
for all $\lambda>0$, with the variables $\mathbf{y}$ held fixed.
}
\begin{equation} \label{eq:homogeneityoriginal}
f(\lambda\mathbf{x}, \textbf y)=\lambda f(\mathbf{x}, \textbf y)\,.
\end{equation}
\end{description}
We emphasize that the scaling transformation appearing in definition (H) must correspond to a change of system size in the thermodynamic sense. In addition to the requirement that all thermodynamic variables 
 $x_i$ scale linearly under this transformation, extensivity requires that the scaling does not merely change the state of a fixed-size system. 
 
 In the energy representation, thermodynamic state functions depend only on the variables $\{x_1,\dots,x_n\}$, in which case extensivity amounts to
 $
f(\lambda x_1,\dots,\lambda x_n)=\lambda f(x_1,\dots,x_n)\,.
$
A concrete example would be a system with fundamental relation $E(S,V,N)$. In other thermodynamic representations, some of the conjugate variables $y_i$ may instead be taken as independent, with the corresponding $x_i$ becoming dependent variables. The $y_i$ coordinates, however, are not scaled in the homogeneous scaling relation. For instance, in the canonical representation the temperature $T$ is an independent variable, but it is held fixed under scaling, so that \eqref{eq:homogeneityoriginal} amounts to
 $
f(\lambda x_1,\dots,\lambda x_{n-1},T)
=\lambda f(x_1,\dots,x_{n-1},T)\,.
 $
  
For completeness, we also define intensive thermodynamic quantities.
A thermodynamic state function \(g(x_1, \dots, x_k, y_{k+1}, \dots, y_n)\) is called \emph{intensive} if it is
homogeneous of degree zero with respect to the same scaling as in~(H), i.e.
\(
g(\lambda \textbf{x}, \textbf y)=g(\textbf{x}, \textbf y), 
\)
for all $\lambda \in \mathbb R_+$.
Equivalently, intensive quantities are invariant under uniform rescaling of the
$x_i$ variables. The class of intensive functions is not simply
the complement of the class of extensive ones, since there exist state functions
that are neither extensive nor intensive, such as $f(x)=x^{\gamma}$ for $\gamma \neq 0, 1$. For instance, a length scale $ V^{\frac{1}{3}}$ is neither extensive nor intensive.

Furthermore, following Section 7.1 of Landsberg \cite{landsberg1978thermodynamics}, we show that if the state function $f(x_1, \dots, x_k, y_{k+1}, \dots, y_n )$ is  extensive in the sense of (H), then all $(x_1, \dots, x_k)$ variables are extensive and all $ (y_{k+1}, \dots, y_n)$ variables are intensive. 

We   assume that the relation   $f(\mathbf x,\mathbf y)$
can be locally inverted for each $x_i$, so that
\begin{equation}
x_i = g_i(x_1,\dots,x_{i-1},f,x_{i+1},\dots, x_k, y_{k+1}, \dots, y_n)\,,
\label{eq:invertibility-system}
\end{equation}
where the functions $g_i$ are defined on an open neighborhood of the equilibrium state
under consideration.
By local invertibility, the definition \eqref{eq:homogeneityoriginal} in (H) can be solved for the
scaled variable $\lambda x_i$. Using the same inverse function $g_i$ as in
Eq.~\eqref{eq:invertibility-system}, one finds
\begin{equation}
\lambda x_i
= g_i(\lambda x_1,\dots,\lambda x_{i-1}, \lambda f,
\lambda x_{i+1},\dots, \lambda x_k, y_{k+1}, \dots, y_n)\,.
\end{equation}
Combining this relation with Eq.~\eqref{eq:invertibility-system} yields
\begin{equation}
\begin{aligned}
&g_i(\lambda x_1,\dots,\lambda x_{i-1}, \lambda f,
\lambda x_{i+1},\dots, \lambda x_k, y_{k+1}, \dots, y_n)
\\
&= \lambda\, g_i(x_1,\dots,x_{i-1},f,x_{i+1},\dots, x_k,
y_{k+1}, \dots, y_n)\,,
\end{aligned}
\end{equation}
showing that the inverse functions $g_i$ are homogeneous of degree one. Equivalently,
all variables $x_i$ scale linearly under the scaling transformation and are therefore
extensive. 

A completely analogous argument establishes intensivity of the variables $y_j$.
Assume in addition that the relation defined by $f(\mathbf x,\mathbf y)$ can be locally
inverted for each $y_j$, $j=k+1,\dots,n$, so that
\begin{equation}
y_j = h_j(x_1,\dots,x_k,f,y_{k+1},\dots,y_{j-1},y_{j+1},\dots,y_n)\,,
\label{eq:invertibility-y}
\end{equation}
with $h_j$ defined on an open neighborhood of the equilibrium state under consideration.
Using the homogeneity property \eqref{eq:homogeneityoriginal} and local invertibility,
one may solve again for the (unscaled) variable $y_j$ at the scaled point, obtaining
\begin{equation}
y_j
= h_j(\lambda x_1,\dots,\lambda x_k,\lambda f,y_{k+1},\dots,y_{j-1},y_{j+1},\dots,y_n)\,.
\end{equation}
Combining this relation with \eqref{eq:invertibility-y} shows that the functions $h_j$ are
homogeneous of degree zero, and therefore that the variables $y_j$ are invariant under the
scaling transformation, i.e.\ they are intensive.

\subsection{Quasi-homogeneous scaling and generalized Euler relations}
\label{sec:euler}

 When a thermodynamic potential exhibits a definite scaling behavior under rescalings of its arguments, this immediately implies algebraic relations among the conjugate pairs, known as Euler relations. The relevant mathematical input is Euler's theorem for generalized homogeneous functions. 
If a differentiable function $f(x_1,\ldots,x_n)$ satisfies the scaling relation
\begin{equation}
f(\lambda^{w_1}x_1,\ldots,\lambda^{w_n}x_n)
= \lambda^{w_f} f(x_1,\ldots,x_n)\,,
\end{equation}
then differentiation with respect to $\lambda$ and evaluation at $\lambda=1$ yields the identity
\begin{equation}
w_f f
= \sum_{i=1}^n w_i\, x_i \frac{\partial f}{\partial x_i}\,.
\end{equation}
This relation follows directly from the assumed scaling symmetry and does not rely on any thermodynamic interpretation.

In thermodynamics Euler's theorem can be applied to the scaling relation of  thermodynamic potentials. For instance, if the internal energy is a homogeneous function of degree one of its independent variables,
\begin{equation}
E(\lambda S,\lambda \xi_1, \dots , \lambda \xi_r, \lambda N_1, \dots , \lambda N_s) = \lambda E(S,\xi_1, \dots, \xi_r,N_1, \dots, N_s)\,,
\end{equation}
then Euler's theorem, with $w_E=w_S=w_{\xi_i}=w_{N_j}=1$, yields
\begin{equation}
E
= \frac{\partial E}{\partial S}S
+ \sum_{i=1}^{r}\frac{\partial E}{\partial \xi_i} \xi_i
+ \sum_{i=1}^{s} \frac{\partial E}{\partial N_j} N_j\,.
\end{equation}
Identifying the conjugate intensive variables \eqref{conjugatevariables}, one obtains the Euler relation
\begin{equation}
E = TS - \sum_{i=1}^{r}Y_i \xi_i + \sum_{j=1}^{s}\mu_j N_j\,.
\end{equation}
The Euler relation is therefore not an independent thermodynamic postulate, but a direct consequence of the extensivity of a thermodynamic potential. 

The form of the Euler relation is representation-dependent. For instance, in the Helmholtz representation the Euler relation takes the form
\begin{equation}
    F = - \sum_{i=1}^{r}Y_i \xi_i + \sum_{j=1}^{s}\mu_j N_j\,.
\end{equation}
Generically, if the thermodynamic potential $\Phi(\textbf x, \textbf y)= \Phi (x_1, \dots, x_k, y_{k+1}, \dots , y_n)$ satisfies the homogeneous-degree-one scaling relation $\Phi(\lambda \textbf x, \textbf y)=\lambda \Phi(\textbf x, \textbf y)  $, then the Euler relation takes the form $\Phi = \sum_{i=1}^{k}y_i x_i,$ where $y_i = \partial \Phi / \partial x_i.$

Since the Euler relation follows from extensivity, its violation implies that the system is not extensive. 
In the energy representation, the quantity~\cite{hillThermodynamicsSmallSystems1962,hillThermodynamicsSmallSystems1994}
\begin{equation}
  G \equiv E -T S + \sum_{i=1}^{r}Y_i \xi_i - \sum_{j=1}^{s}\mu_j N_j
\end{equation}
may be interpreted as a measure of non-extensive contributions to the energy. 
When extensivity holds, $ G=0$; when it does not, $G$ captures corrections arising from finite-size effects, 
boundary contributions, or interactions that spoil simple linear scaling. 

More generally, strict extensivity is not required for a scaling relation to exist. 
If the energy satisfies a generalized or  quasi-homogeneous scaling relation with nontrivial weights \cite{stanley1971introduction,Belgiorno2003,Quevedo:2018fzi},
\begin{equation}
E(\lambda^{w_S} S,\lambda^{w_{\xi_1}} \xi_1, \dots, \lambda^{w_{\xi_r}} \xi_r, \lambda^{w_{N_1}}N_1,\ldots, \lambda^{w_{N_s}}N_s)
= \lambda^{w_E} E(S,\xi_1,\ldots, N_s)\,,
\end{equation}
Euler's theorem instead implies a   quasi-homogeneous Euler identity, 
\begin{equation}
w_E E
=w_S TS - \sum_{i=1}^{r} w_{\xi_i}Y_i \xi_i + \sum_{j=1}^{s}w_{N_j}\mu_j N_j \,.
\end{equation}
While such relations reflect the presence of an underlying scaling property, 
they still correspond to non-extensive systems in the thermodynamic sense, 
since the energy no longer scales linearly with all independent variables. It is worth noting that in a quasi-homogeneous scaling relation the weights are not all independent. An overall rescaling of the weights can always be absorbed into a redefinition of the scaling parameter, so that only ratios of weights carry invariant meaning. Equivalently, one may, without loss of generality, fix one of the weights (for instance $w_E=1$), with the remaining weights characterizing the deviation from extensivity.

Finally, some thermodynamic systems admit no global scaling relation at all. 
In the absence of homogeneous or quasi-homogeneous behavior, Euler's theorem does not apply 
and no Euler-type relation can be derived.  From our perspective, this is precisely the situation encountered
for AdS black holes when the cosmological constant is not included as a thermodynamic
state variable: at finite volume and entropy, the equilibrium state space admits no global
scaling transformation, and correspondingly no   Euler relation exists.

\subsection{Extensivity in the large-system limit} In statistical mechanics, extensivity is usually formulated in terms of a thermodynamic limit. The precise definition of this limit depends on the choice of statistical ensemble, which in turn specifies which macroscopic variables are held fixed and which are allowed to vary. We apply an analogous limiting procedure to small thermodynamic systems, referring to it as the large-system limit. In this case, the definition of the limit depends on the choice of thermodynamic representation.

\begin{description}
\item[(L) \emph{Large-system limit}.]
The thermodynamic state function $f(x_1, \dots, x_k, y_{k+1}, \dots, y_n)$ is said to be \emph{extensive in the sense of the large-system limit} if, for every fixed vector of
ratios
\(
\boldsymbol{\rho}=(\rho_{1},\ldots,\rho_{k-1}) \equiv (x_1/ x_k, \dots, x_{k-1}/ x_k) 
\),
the limit
\begin{equation}
\lim_{x_k\to\infty}
\frac{1}{x_k}\,
f\!\left(
x_k\rho_{1},\ldots,x_k\rho_{k-1},x_k, y_{k+1},\ldots,y_n
\right)
\label{eq:L2_general}
\end{equation}
exists and does not vanish. If the limit goes to zero, then $f$ is said to be \emph{subextensive}, and if the limit diverges, then $f$ is \emph{superextensive}.
If the limit is finite and non-zero, its value defines a limiting function  denoted by
$f_\infty (\rho_{1},\ldots, \rho_{k-1},  y_{k+1},\ldots,y_n)= f_\infty (\boldsymbol \rho, \textbf y)$.
Here the variables $(x_a,y_a)$   form a thermodynamically conjugate pair in the energy representation~\eqref{energyrepresentationfirstlaw} of the first law.
 The ratios $ \boldsymbol \rho$ as well as the independent variables $\textbf y$ are held fixed as $x_k$ tends to infinity. 
\end{description}

Extensivity manifests itself in the large-system limit as linear growth of a state function with respect to the variable $x_k$. Typically, the variable $x_k$ taken to infinity is the volume. However, since all variables $x_i$ are sent to infinity simultaneously, because the ratios $\rho_i \equiv x_i/x_k$ are held fixed,  the limit may equivalently be defined by sending any other variable $x_i$, for $i = 1, \dots, k-1$, such as particle number (as is done in Appendix \ref{idealgasappendix} for the ideal gas) or entropy, to infinity. Crucially, in this limit the system size itself must grow large. Limits that only probe different states of a fixed system, such as high-energy or high-temperature limits, do not generally define a thermodynamic large-system limit.

In the thermodynamic representations  where the internal energy and the entropy are the thermodynamic potentials,  all  independent variables are given by ``extensive'' quantities  $\{x_i, \dots, x_n\}$, i.e. $k=n$, so no conjugate variables $y_i$ are held fixed in the large-system limit.  
If the limit exists,
intensive quantities arise as functions of the fixed ratios $\rho_i$, reflecting
the fact that the limit is taken at fixed densities. This agrees with the definition of
the thermodynamic limit in the microcanonical statistical ensemble, and it coincides with the thermodynamic limit discussed in~\cite{DunningDavies1983,DUNNINGDAVIES1985383,DUNNINGDAVIES1988705}.

In other thermodynamic representations, however, like the canonical and grand-canonical
representations, the independent variables include  parameters $y_i$, such as the
temperature or chemical potentials. These variables are held fixed in the large-system
limit, while all   variables $\{x_1,\dots,x_k\}$ are sent to infinity. This is
analogous to the thermodynamic limit of the (grand-)canonical ensemble, in which the
temperature is kept fixed as the system size is taken to infinity.

In statistical mechanics it is well known that different statistical ensembles are
thermodynamically equivalent in the thermodynamic limit, provided suitable stability
and convexity (or concavity) assumptions are satisfied \cite{Ruelle1969,Ellis1985}.
More precisely, thermodynamic equivalence holds whenever the limiting entropy is
concave  as a function of
the relevant densities, except possibly at points of phase coexistence where strict
concavity fails \cite{Touchette2005}. An analogous statement holds at the level of
thermodynamic representations: under   stability and convexity assumptions,
existence of the large-system limit in one   representation implies existence
of the large-system limit in a Legendre-related representation.
However, this   does not imply that extensive scaling properties are
preserved under the Legendre transformation: a thermodynamic potential may be extensive in the limit of one representation while
the   Legendre-related potential is subextensive in the large-system
limit of another representation, as we will see in the next Section on Schwarzschild black holes.

Finally, we establish the relation between the definitions (H) and (L) for the extensivity of state functions.  If the large-system limit exists, then the limiting function $f_\infty (\boldsymbol \rho, \textbf y )$ is automatically a homogeneous degree-\textcolor{black}{zero} state function.  
That is because if $f$ is extensive in the sense of~(L), then one may equivalently write
\begin{equation}
\label{asympexp}
f(x_1,\dots,x_k,y_{k+1},\dots,y_n)
=
x_k\,f_\infty(\boldsymbol{\rho},\mathbf{y})
+
o(x_k)\,,
\qquad
x_k\to\infty \ \text{at fixed } \boldsymbol{\rho}\, ,
\end{equation}
where $o(x_k)$ denotes a subextensive correction, i.e.\ $ o(x_k)/x_k\to 0$ as $x_k \to \infty$.
Such expansions are standard in statistical mechanics and asymptotic analysis
\cite{Ruelle1969,Ellis1985,Bruijn1981}.
Equation~\eqref{asympexp} implies asymptotic degree-one homogeneity along sequences of
fixed $\boldsymbol{\rho}$: for any fixed $\lambda>0$,
\[
f(\lambda\boldsymbol{x},\mathbf{y})
=
\lambda x_k\,f_\infty(\boldsymbol{\rho},\mathbf{y}) + o(x_k)
=
\lambda f(\boldsymbol{x},\mathbf{y}) + o(x_k)\,,
\qquad
x_k\to\infty\,.
\]
Thus, in the large-system limit, (L) $\Rightarrow$ (H).

Conversely, suppose that homogeneity (H) holds exactly, \emph{i.e.}\ as an identity prior to taking any limit. Fix \(\boldsymbol{\rho}\) and write \(\boldsymbol{x}=(X\boldsymbol{\rho},X)\) with \(X=x_k\). Degree-one homogeneity then gives
\[
\frac{f(X\boldsymbol{\rho},X,\mathbf{y})}{X}
=
f(\boldsymbol{\rho},1,\mathbf{y}),
\]
which is independent of \(X\). The large-system limit therefore exists trivially and is given by
\[
f_\infty(\boldsymbol{\rho},\mathbf{y}) = f(\boldsymbol{\rho},1,\mathbf{y}).
\]
Thus, exact homogeneity (H) implies the existence of the large-system limit (L), without any additional assumptions on system size.

\subsection{Extensivity of an equilibrium branch of a   thermodynamic representation}
\label{sec:extensivitysystem}

In   previous Sections we analyzed extensivity as a property of individual thermodynamic
state functions. Here we address the corresponding notion at the level of an equilibrium branch of a thermodynamic representation.  

Thermodynamic systems may admit multiple equilibrium solutions for the same values of the
control variables, for instance when the equations of state fail to be globally invertible.
Each such solution defines an \emph{equilibrium branch}. These branches potentially
differ  in stability properties  or scaling behavior.
In systems with multiple equilibrium branches, extensivity should therefore be formulated as a \emph{branchwise} property
within a given thermodynamic representation, rather than as a property of the thermodynamic
representation or of the system as a whole. We define extensivity      with respect to a fixed thermodynamic representation and
a fixed equilibrium branch within that representation. A central question of this Section
is whether extensivity so defined is preserved when the same equilibrium branch is investigated
  across Legendre-related representations.

Consider a general thermodynamic representation in which the equilibrium states are parametrized by a
complete set of independent variables $\{ \mathbf x,\mathbf y\}=\{ x_1, \dots, x_k, y_{k+1}, \dots, y_n\}$, obtained by selecting exactly one
variable from each conjugate pair $(x_a,y_a)$ appearing in the energy representation first law
\eqref{energyrepresentationfirstlaw}. This choice provides a non-redundant coordinate system
on the equilibrium manifold. The fundamental relation for the thermodynamic potential $\Phi$ in this representation is $\Phi = \Phi (\textbf x, \textbf y)$, and its first differential satisfies
\begin{equation}
    d \Phi = \sum_{i = 1}^k y_i dx_i - \sum_{j=k+1}^n x_j dy_j\,,
\end{equation}
with conjugate variables   defined by
\begin{equation}\begin{aligned}
     & y_i \equiv \frac{\partial \Phi}{\partial x_i}\,,\quad i = 1 , \dots, k,\\ &x_j \equiv - \frac{\partial \Phi}{\partial y_j} \, , \quad  j = k+1, \dots, n\,.
\end{aligned}\end{equation}
An equilibrium branch in this representation is said to be \emph{extensive} if there exists a
one-parameter family of scaling transformations $S_\lambda$, $\lambda>0$, interpreted as a
change of system size, such that on that branch the thermodynamic potential
$\Phi(\mathbf{x},\mathbf{y})$ and all
variables $x_a$ scale linearly and remain non-vanishing, while all conjugate variables $y_a$
remain invariant, for $a=1,\dots,n$.  Under these conditions the Euler relation in that representation, $\Phi = \sum_{i=1}^k y_i x_i$,  is satisfied
on the branch, which is   a  hallmark of extensivity.

Suppose in a given representation the potential is homogeneous of degree one with respect to the $(x_1, \dots, x_k)$ variables,   \begin{equation}\label{homogeneitypotential}\Phi(\lambda\mathbf x,\mathbf y)=\lambda\Phi(\mathbf x,\mathbf y)\,,\qquad \lambda>0\, . \end{equation} This implies that  all $x_1, \dots, x_k$ are extensive and all $y_{k+1}, \dots , y_n$ are intensive. Moreover,  differentiating Eq.~\eqref{homogeneitypotential} with respect to $x_i$ yields 
\begin{equation}
    \frac{\partial \Phi}{\partial x_i} (\lambda \textbf x, \textbf y ) = \frac{\partial \Phi}{\partial x_i} (\textbf x , \textbf y )\,,
\end{equation}
so that $y_i (\lambda \textbf x, \textbf y)= y_i ( \textbf x, \textbf y)$ for $i=1, \dots, k.$ Further, differentiating Eq.~\eqref{homogeneitypotential} with respect to $y_j$ yields
\begin{equation}
    \frac{\partial \Phi}{\partial y_j} (\lambda \textbf x, \textbf y ) =  \lambda \frac{\partial \Phi}{\partial y_j} (\textbf x, \textbf y)\,,
\end{equation}
so that $x_j (\lambda \textbf x, \textbf y ) = \lambda  x_j (\textbf x, \textbf y).$ Therefore, if the potential is extensive, then all $x_a$ variables  are extensive and all $y_a$ variables  are intensive, for $a=1, \dots, n $, and  the system admits a scaling transformation $(x_a, y_a)\mapsto (\lambda x_a, y_a)$.

We now show that if an equilibrium branch is extensive in one representation then typically it is also extensive in a Legendre-related representation. 
Let $I\subset\{1,\dots,n\}$ denote a subset of indices labeling
independent variables to be exchanged for their conjugates, then the
  Legendre-transformed potential is defined by
\cite{callen1985thermodynamics,Arnold:1989who,Zia2009}
\begin{equation} \label{generallegendretransf}
    \widetilde{\Phi}
    = \Phi(\mathbf x,\mathbf y)
      - \sum_{i\in I_x}  y_i x_i
      + \sum_{j\in I_y}  y_j x_j \,,
\end{equation}
where $I_x\subset I$ denotes indices originally associated with extensive variables and
$I_y\subset I$ those associated with intensive variables.  The transformed potential is intended to depend on
the complementary set of independent variables obtained by replacing each variable in $I$
by its conjugate variable.
This replacement is well defined precisely when the corresponding change of variables is
  invertible, i.e.\ when the Jacobian
$\det\!\left(\partial(y_i,x_j)/\partial(x_i,y_j)\right)$   is non-vanishing at the equilibrium configuration
under consideration.  Only under this condition can the   relations
$y_i=\partial\Phi/\partial x_i$ and $x_j=-\partial\Phi/\partial y_j$ be solved   to
express the transformed potential $\widetilde{\Phi}$ as a function of the new independent
variables.  When this condition fails, the same formal expression \eqref{generallegendretransf} might still define
$\widetilde{\Phi}$ branch-by-branch, but not as a
single-valued thermodynamic potential.

If the original thermodynamic potential $\Phi$ is \emph{exactly} homogeneous of degree one
under the scaling $(\mathbf x,\mathbf y)\mapsto(\lambda\mathbf x,\mathbf y)$ on a given
equilibrium branch, then Legendre transformations preserve extensivity. Indeed, each term
$ y_i x_i$ (and $  y_j x_j$) scales linearly under this transformation, so the
Legendre-transformed potential $\widetilde{\Phi}$ is likewise homogeneous of degree one
under the same scaling. In this case all Legendre-related thermodynamic potentials are
extensive and the extensivity of the branch is representation-independent.

However, when extensivity is realized only \emph{asymptotically} in a large-system limit,
this conclusion need not hold.  The essential mechanism is that, when extensivity is realized only asymptotically, the
leading extensive contribution to a thermodynamic potential can be removed by a Legendre
subtraction (on a given equilibrium branch), leading to a subextensive transformed
potential. 

Suppose that for large system size
$X \equiv x_k$ a thermodynamic potential admits an expansion of the form
\begin{equation}
\Phi(\boldsymbol{\rho}X, X, \mathbf y)
=
X\,\Phi_\infty(\boldsymbol{\rho},\mathbf y)
+
o(X)\,,
\end{equation}
where $\boldsymbol{\rho}=\mathbf x/X$ and the intensive variables $\mathbf y$ are held fixed.
To isolate the relevant mechanism, we consider a Legendre transformation in which only
intensive variables are exchanged for their conjugate extensive variables, so that the
transformation subtracts terms of the form $y_j x_j$. On the equilibrium branch, the
conjugate variables scale as
\begin{equation}
x_j
=
-\frac{\partial\Phi}{\partial y_j}
=
-X\,\frac{\partial\Phi_\infty}{\partial y_j}(\boldsymbol{\rho},\mathbf y)
+
o(X)\,,
\end{equation}
where the $y_j$ are intensive variables whose conjugates $x_j$ are extensive. Evaluating the
Legendre transformation~\eqref{generallegendretransf} with $I_x=\varnothing$ on the same
equilibrium branch then yields
\begin{equation}\label{dezevergelijkingisgoed}
\widetilde{\Phi}
=
X\!\left[
\Phi_\infty(\boldsymbol{\rho},\mathbf y)
-
\sum_{j\in I_y} y_j\,\frac{\partial\Phi_\infty}{\partial y_j}(\boldsymbol{\rho},\mathbf y)
\right]
+
o(X)\,.
\end{equation}
If $\Phi$ is exactly extensive, the bracketed combination approaches a finite, nonzero
function, so that $\widetilde{\Phi}\propto X$ and extensivity is preserved. If $\Phi$ is only
extensive asymptotically, however, the leading contribution need not survive on the
equilibrium branch. In particular, it may happen that the derivatives
$\partial_{y_j}\Phi_\infty(\boldsymbol{\rho},\mathbf y)$ are independent of $\mathbf y$, even though
$\Phi_\infty$ itself depends on the intensive variables. In this case the dependence of $\Phi_\infty$ on
$\mathbf y$ is   linear along the equilibrium branch, and the Legendre subtraction
removes the entire intensive-variable dependence of the leading term. The terms in square brackets  in \eqref{dezevergelijkingisgoed} can then cancel each other  on the branch,
leaving
\( 
\widetilde{\Phi}=o(X),
\)
despite the original potential $\Phi$ being extensive in the large-system limit.

The   Schwarzschild black hole   provides a concrete realization of this phenomenon: while the Helmholtz free energy becomes extensive on the large black hole branch in the canonical representation in the large-system limit, the corresponding internal energy remains subextensive (see Section~\ref{sec:limits_flat}).

\section{Non-extensivity of asymptotically flat black holes}
\label{nonextensivityflatblackholes}

Having developed a thorough understanding of extensivity, we now apply this notion to black hole thermodynamics. The key question is whether black holes admit a notion of system size that can be rescaled.
In gravitational thermodynamics the system is specified by a choice of boundary, and the system size is fixed by the geometry of that boundary. For spherically symmetric configurations this is captured by the boundary area, which, with the holographic identification developed in Section \ref{sec:quasilocalholographic}, defines the thermodynamic volume. Rescaling the system size therefore corresponds to rescaling this volume.

\subsection{Quasi-homogeneity of asymptotically flat black holes}
\label{sec:bh-quasihom}

Using quasi-local black hole thermodynamics in the sense of York, we analyze extensivity for asymptotically flat black holes enclosed by a finite boundary.
For  Schwarzschild black holes, one may formulate, using Eqs.~\eqref{eq:schwarzschildquantities}, a fundamental equation for the quasi-local energy as a function of entropy  and   volume,
\begin{equation} \label{esvschwarzschild}
E(S,V)
=
\frac{(d-2)\Omega_{d-2}}{8\pi G}
\left(\frac{V}{\Omega_{d-2}}\right)^{\!\frac{d-3}{d-2}}
\left[
1-
\sqrt{
1-
\left(
\frac{4G\,S}{V}
\right)^{\!\frac{d-3}{d-2}}
}
\right]\, .
\end{equation}
This energy satisfies the quasi-homogeneous scaling relation
\begin{equation}
E(\lambda^{d-2} S,\, \lambda^{d-2} V)
= \lambda^{d-3} E(S,V)\, ,
\end{equation}
from which the quasi-local Smarr relation \eqref{eq:quasilocalsmarrflat} follows by Euler’s theorem. Equivalently, introducing $\beta=\lambda^{d-2}$ shows that $E$ is homogeneous of degree $(d-3)/(d-2)$ rather than degree one. The quasi-local energy is therefore not extensive.
The deviation from extensivity may be quantified using the diagnostic introduced in Section~\ref{sec:euler}. Employing the quasi-local Smarr relation yields
\begin{equation}
G
\equiv E - TS + PV
= \frac{1}{d-3}\,(TS - PV)\, ,
\end{equation}
which is manifestly nonzero.

It is important to distinguish this result from the scaling properties of global black hole solutions. For a global stationary black hole the conserved charges defined at infinity are thermodynamic state variables, and they satisfy quasi-homogeneous scaling relations. In particular, for asymptotically flat rotating and charged black holes one has
\begin{equation}
S(\lambda^{d-3} M,\, \lambda^{d-2} J,\, \lambda^{d-3} Q)
= \lambda^{d-2} S(M,J,Q)\,,
\end{equation}
which implies the Smarr relation \cite{smarr1973}
\begin{equation}
(d-3)M
= (d-2)T S
+ (d-2)\Omega J
+ (d-3)\Phi Q \,.
\end{equation}
This is an Euler-type identity, but it does not arise from homogeneity of degree one.\footnote{
Equivalently, the quasi-homogeneity of $M(S,J,Q)$ may be rewritten as ordinary homogeneity by assigning equal scaling weight to all arguments. Since $S$ and $J$ scale as $[\text{length}]^{d-2}$ while $Q$ scales as $[\text{length}]^{d-3}$, introducing the variable $Q^{\frac{d-2}{d-3}}$ yields
\(
M(\beta S,\beta J,\beta Q^{\frac{d-2}{d-3}})
=\beta^{\frac{d-3}{d-2}} M(S,J,Q^{\frac{d-2}{d-3}}).
\)
Hence, the black hole mass is a homogeneous function of degree $(d-3)/(d-2)$. In $d=4$ this reproduces Smarr’s original statement \cite{smarr1973} that $M$ is homogeneous of degree $1/2$ in $(S,J,Q^2)$.}
 Rather, it reflects the intrinsic quasi-homogeneous structure of asymptotically flat black hole thermodynamics, as emphasized by Belgiorno~\cite{Belgiorno2003} and in subsequent work~\cite{Quevedo:2018fzi}.

The quasi-homogeneous scaling can be verified explicitly in four spacetime dimensions, where the Kerr-Newman entropy is given by
\begin{equation} \label{entropykerrnewman}
S(M,J,Q)=\pi  \left (2 G M^2 - Q^2 + 2 \sqrt{G^2 M^4 - G M^2 Q^2 - J^2} \right).
\end{equation}
Based on this non-linear scaling, Landsberg~\cite{Landsberg1984,Landsberg1992} ---  following earlier work with Tranah~\cite{LandsbergTranah1980a,TranahLandsberg1980b} that seems unavailable online --- argued that black holes are non-extensive. His argument relied on the doubling definition of extensivity \textcolor{black}{for $n=2$} (see Appendix~\ref{appendixExtensivity}).
In ordinary thermodynamic systems, such as gases, merging two identical subsystems doubles all extensive variables and yields an entropy satisfying the doubling condition
\(
S(2\mathbf{x}) = 2S(\mathbf{x}),
\)
which follows directly from definition (H) for $\lambda =2$.

For black holes, Landsberg identified the conserved charges $(M,J,Q)$ as the putatively extensive state variables and considered the transformation $(M,J,Q)\to(2M,2J,2Q)$, motivated by the merging of two identical black holes. While the entropy indeed fails to satisfy
\(
S(2M,2J,2Q)=2S(M,J,Q),
\)
this procedure does not test extensivity in the thermodynamic sense defined in Section~\eqref{sec:extensivitythermolimit}. Merging two black holes increases the total conserved charges, such as the energy, but it does not rescale the boundary with respect to which these charges are defined and therefore does not change the system size. Varying $(M,J,Q)$ only changes the thermodynamic state of the black hole system whose boundary is fixed at spatial infinity.   Consequently, quasi-homogeneity of the global  black hole variables does not, by itself, establish non-extensivity.

All in all, we conclude that asymptotically flat black holes define non-extensive thermodynamic systems in the quasi-local sense: their fundamental relations are intrinsically quasi-homogeneous rather than homogeneous of degree one with respect to the thermodynamic volume. A   fully conventional extensivity test based on a thermodynamic  fundamental equation $E(S,V,\ldots)$, involving the holographic volume, is available only in sufficiently symmetric settings (e.g.\ spherical cavities), while in the generic quasi-local formulation the work term is governed by the spatial stress tensor rather than a single scalar pressure-volume pair.

\subsection{Large-volume, high-energy, and high-temperature limits}
\label{sec:limits_flat}

\textbf{Large-volume limit.} Next we consider the large-system limit of black holes, which is defined   as the large-volume limit with appropriate
thermodynamic densities (the $\rho_i$ variables introduced in the previous section),
defined as ratios to the volume, and/or certain independent variables (the $y_i$ variables)
held fixed. For static, spherically symmetric black holes in a spherical cavity the large-volume limit in the energy representation, where $(S,V)$ are independent variables, is defined as
\begin{equation} \label{largesystemdeze}
    V \to \infty\,, \qquad \text{with} \quad s \equiv \frac{S}{V}  \quad \text{fixed}\,.
\end{equation}
Note this implies that also the entropy is taken to infinity. 

On the other hand,  in the canonical   representation the independent variables are $(T,V)$ and the large-system limit is defined as
\begin{equation} \label{largesystemdie}
    V \to \infty\,, \qquad \text{with} \quad T    \quad \text{fixed}\,. 
\end{equation}
It is useful to compare the leading large-volume behavior in these two distinct thermodynamic
representations.\footnote{For charged, static, spherically symmetric black holes the charge density $\rho = Q/V$ should also be kept fixed in both large-system limits.}

In the \emph{energy representation},  the quasi-local energy may be written as
$E(Vs,V)$. Using the explicit expression for this function in  Eq.~\eqref{esvschwarzschild}, one finds asymptotically at leading order
\begin{equation}
\frac{E(Vs,V)}{V}
\;\sim\;
V^{-\frac{1}{d-2}}\,
\mathcal{E}(s)
\;\xrightarrow{V\to\infty}\;0 \,,
\label{eq:largeV_energy_density}
\end{equation}
where $\mathcal{E}(s)$ is a positive, dimensionless function. Thus the energy is strictly subextensive with respect to the
holographic volume.

In the \emph{canonical  representation}, the thermodynamic potential is the Helmholtz free energy  $F(T,V)= E(T,V) - T S(T,V).$ Using the quasi-local expressions   in \eqref{eq:schwarzschildquantities}, the free energy may be written in four spacetime dimensions
as
\begin{equation}
\textcolor{black}{F(r_B, r_h)
=
\frac{r_B}{G}
\left[
1 - \sqrt{1 - \frac{r_h}{r_B}}
\right]
-
\frac{r_h}{4G}
\left( 1 - \frac{r_h}{r_B} \right)^{-1/2}\,.}
\end{equation}
At fixed boundary volume \(V\), thermal stability is determined by the local convexity
properties of \(F(T,V)\) with respect to~\(T\).
For the small black hole branch, \(F(T,V)\) is convex in \(T\), signaling a thermally
unstable phase with negative heat capacity,
\(C_V = -T\partial^2 F/\partial T^2 < 0\).
By contrast, for the large black hole branch \(F(T,V)\) is concave in \(T\), implying
thermal stability with
\(C_V = -T \partial^2 F/\partial T^2 > 0\).

For the small black hole branch \eqref{smallbhsch} in $d=4$ one finds the following expansion around $V=\infty$
\begin{equation}
    \textcolor{black}{F^{(1)} (T,V) =\frac{1}{16\pi GT}+\frac{1}{64\pi^{3/2} G T^{2}}\,V^{-1/2}
+\frac{1}{128\pi^{2} G T^{3}}\,V^{-1}
+\mathcal{O}\!\left(V^{-3/2}\right)\,,}
\end{equation}
where $V= 4\pi r_B^2.$
The entropy and energy follow from the standard thermodynamic relations
\begin{align}
S^{(1)}(T,V)
&=
- \frac{\partial F^{(1)}}{\partial T} 
=\textcolor{black}{\frac{1}{16\pi GT^2}
+\frac{1}{32\pi^{3/2}G\,T^{3}}\,V^{-1/2}
+\frac{3}{128\pi^{2}G\,T^{4}}\,V^{-1}
+\mathcal{O}\!\left(V^{-3/2}\right)\,,}
\\[1ex]
E^{(1)}(T,V)
&=
F^{(1)}+T S^{(1)}
=
\textcolor{black}{\frac{1}{8\pi GT}+\frac{3}{64\pi^{3/2}G\,T^{2}}\,V^{-1/2}
+\frac{1}{32\pi^{2}G\,T^{3}}\,V^{-1}
+\mathcal{O}\!\left(V^{-3/2}\right)\,.}
\end{align}
Now in the large-volume limit the free energy density vanishes
\begin{equation}
     \textcolor{black}{\frac{F^{(1)}(T,V)}{V}\sim V^{-1}  \;\xrightarrow{V\to\infty}\;0\,.}
\end{equation}
and the same applies to the entropy density and energy density. Hence, the free energy, entropy and internal energy of the small black hole are subextensive in the limit.

Moreover, for the large black hole branch \eqref{largebhsch} the free energy expansion around $V=\infty$ is
\begin{equation} \label{largebhfreeenergylargeV}
    F^{(2)}(T,V)
=
\textcolor{black}{-\frac{T}{4G}\,V
+\frac{1}{2\sqrt{\pi}\,G}\,V^{1/2}-\frac{1}{8\pi GT}
-\frac{1}{64\pi^{2} G T^{3}}\,V^{-1}
+\mathcal{O}\!\left(V^{-2}\right)\,,}
\end{equation}
and the entropy and energy expansions are
\begin{align}
S^{(2)}(T,V)
&=
\textcolor{black}{- \frac{\partial F^{(2)}}{\partial T} 
=
\frac{V}{4G}-\frac{1}{8\pi G T^2}
-\frac{3}{64\pi^{2}G\,T^{4}}\,V^{-1}
+\mathcal{O}\!\left(V^{-2}\right)\,,}
\\[1ex]
E^{(2)}(T,V)
&=
\textcolor{black}{F^{(2)}+T S^{(2)}
=
\frac{V^{1/2}}{2\sqrt{\pi}\,G}-\frac{1}{4\pi GT}
-\frac{1}{16\pi^{2}G\,T^{3}}\,V^{-1}
+\mathcal{O}\!\left(V^{-2}\right)\,.}
\end{align}
In this case the leading-order term scales differently for the energy compared to the free energy and entropy. In particular, the free energy density and entropy density remain finite in the large-volume limit
\begin{equation}
  \lim_{V \to \infty}  \frac{F^{(2)}(T,V)}{V}
= -\frac{T}{4G}\,, \qquad   \lim_{V \to \infty}  \frac{S^{(2)}(T,V)}{V}
=  \frac{1}{4G}\,.
\end{equation}
So the free energy and entropy   are extensive   functions of $(T,V)$ in the large-volume regime, while the energy density vanishes,
\begin{equation}
     \frac{E^{(2)}(T,V)}{V}\sim V^{-1/2}  \;\xrightarrow{V\to\infty}\;0\,,
\end{equation}
hence the energy is subextensive.\footnote{
It is also instructive to consider the pressure $P\equiv - \partial F / \partial V$.
In the large-volume limit the pressure of the small black  hole branch vanishes,
\(P^{(1)}\to 0\), whereas for the large black hole branch it approaches the finite value
\(P^{(2)}\to T/(4G)\).
This reflects the fact that only the large black hole branch carries a nontrivial
extensive free energy density in this limit.
 }
Since the Helmholtz free energy, entropy, and volume are extensive variables on the
large black hole branch in the large-volume regime of the canonical representation,  this equilibrium branch is   extensive at large volume
in the sense of Section~\ref{sec:extensivitysystem}. 
However, because the internal energy is not extensive in this representation nor in the energy representation \eqref{eq:largeV_energy_density}, the extensivity of the large black hole branch is representation-dependent, i.e. it only holds in the canonical representation.

This conclusion contradicts a recent claim in \cite{Banihashemi:2024yye}. They identified the same leading-order scaling
$F^{(2)} \sim S^{(2)} \sim r_B^{d-2}/G$ (in $d$ bulk spacetime dimensions) for the large black hole,
but the system was   classified as non-extensive based on this scaling. In
our view, however, $F^{(2)}$ and $S^{(2)}$ are extensive functions of $(T,V)$, since they
scale linearly with the holographic volume in the large-volume limit.

The subextensivity of the internal energy, despite the extensivity of the Helmholtz free
energy, can be traced to a cancellation of the leading extensive contributions in the
Legendre relation
\(
E = F + TS = F - T\,\partial_T F .
\)
On the large black hole branch the free energy admits the large-volume expansion
\(
F(T,V) = V f(T) + o(V),
\)
and the derivative of the leading free-energy density with respect to the intensive
variable $T$ is independent of $T$ along the equilibrium branch,
\(
\partial_T f(T) = \text{const}.
\)
As a result,  the leading $\mathcal O(V)$ terms in $F$ and $TS$ cancel, leaving
\( 
E = o(V).
\)
This cancellation realizes the mechanism described in
Section~\ref{sec:extensivitysystem}. Extensivity is therefore not preserved under the change of thermodynamic
representation on this branch.

\paragraph{High-energy and high-temperature limits.}
In addition to this large-system limit, one may consider other asymptotic limits that do
not correspond to increasing the system size. Both for black holes enclosed in a cavity
and for global black hole geometries, one may take the temperature or energy to be large at fixed
volume, depending on the thermodynamic representation. These limits probe
different asymptotic regions of the black hole state space and should be distinguished
from the large-system limit. 

Let us start with a global Kerr-Newman black hole system, for which the boundary has been taken to infinity, and take the high-energy limit by sending
\begin{equation}
     M \to \infty\,, \qquad \text{with} \quad \textcolor{black}{j \equiv \frac{J}{G M}}\  \quad \text{and} 
\quad
q \equiv \textcolor{black}{\frac{Q}{\sqrt{G}M} }\quad \text{fixed}\,.
\end{equation}
In terms of these   ratios the black hole entropy \eqref{entropykerrnewman} may  
be written in four dimensions  as
\begin{equation}
S(M,j,q)
=
\textcolor{black}{\pi G M^2
\left[
2 - q^2 + 2\sqrt{1 - q^2 - \frac{j^2}{M^2}}
\right]\,.}
\end{equation}
Taking $M \to \infty$ at fixed $(j,q)$ yields
\begin{equation}\label{superlinearentropy}
\frac{S(M,   j,  q)}{M} \sim M  \textcolor{black}{f(M,j,q) }\;\xrightarrow{M\to\infty}\;\infty\,,
\end{equation}
\textcolor{black}{with $f(M,j,q)$ a positive function that only depends on   $q$ in the limit}. Thus, the black hole entropy is superlinear (i.e. quadratic for $d=4$) in the energy. The same conclusion holds in higher spacetime dimensions, in which case $S  \sim M^{(d-2)/(d-3)}$ for large $M$. This is, however, an artifact of the infinite-boundary limit that was taken prior the infinite-mass limit. As we now explain, at a finite fixed boundary the entropy does not diverge for large energies.

At fixed boundary volume $V=\Omega_{d-2}r_B^{d-2}$ the quasi-local energy is bounded above,
since the horizon cannot extend beyond the cavity wall. Using Eq.~\eqref{esvschwarzschild}, the endpoint
$r_h\to r_B$ corresponds to $S\to V/4G$, so that the square root term vanishes and the
energy approaches a finite maximum,
\begin{equation}
E \xrightarrow{r_h\to r_B} E_{\max}(V)
=
\frac{(d-2)\Omega_{d-2}}{8\pi G}\left(\frac{V}{\Omega_{d-2}}\right)^{\frac{d-3}{d-2}}\,.
\end{equation}
In the same limit the entropy saturates at $S_{\max}(V)=V/(4G)$, and therefore
\begin{equation}
\lim_{E\to E_{\max}(V)}\frac{S(E,V)}{E}
=
\frac{S_{\max}(V)}{E_{\max}(V)}
=
\frac{2\pi}{d-2}\left(\frac{V}{\Omega_{d-2}}\right)^{\!\frac{1}{d-2}} .
\end{equation}
Thus, at fixed volume the large-energy limit does not correspond to an unbounded growth of the state space. Both the quasi-local energy and the entropy approach finite limiting values, and their ratio 
 $S/E$ tends to a constant proportional to the boundary radius. Increasing the energy beyond this point is impossible without enlarging the system itself. When $V$ is taken to infinity, $S/E$ diverges in agreement with the result \eqref{superlinearentropy} above. 

Finally, we consider the high-temperature limit at fixed boundary volume.  For
asymptotically flat Schwarzschild black holes in a cavity, this limit is controlled by the
single dimensionless parameter $r_B T$.  Indeed, by the   formula \eqref{tolmanflat}  for the boundary temperature,
taking $T\to\infty$ at fixed $r_B$ is equivalent to taking $r_B\to\infty$ at fixed $T$, in the
sense that both limits drive $r_B T\to\infty$.  Consequently, the same large-$r_B T$
expansions for the two York branches obtained in Eqs.~\eqref{smalllarger} and \eqref{largelarger} apply in either case: the small
black hole has $r_h$ asymptotically independent of $r_B$, while the large black hole has
$r_h\to r_B$.  In particular, at the
level of leading thermodynamic scaling, the high-temperature limit at fixed volume
reproduces the same asymptotic behavior as the large-volume limit at fixed temperature
discussed earlier \cite{york1986,Banihashemi:2024yye}. 

This coincidence is special to the asymptotically flat Schwarzschild family.  Once additional
scales are present, the large-volume and high-temperature limits generically probe distinct
asymptotic regions of state space and are no longer equivalent.  This is the case for charged
black holes, where the charge introduces an additional scale, and for
AdS black holes, where the AdS radius provides an intrinsic length scale (see Section \ref{sec:highTfixedVAdS}).

\section{Extensivity of AdS-Schwarzschild black holes}
\label{sec:extadsblackholes}

We now investigate whether AdS-Schwarzschild black holes admit an extensive thermodynamic description in an appropriate large-system limit. When enclosed by a spherical York boundary, AdS-Schwarzschild black holes are described by five thermodynamic state variables $(E,T,S,P,V)$. At finite boundary volume, these variables do not satisfy the homogeneous Euler relation, and the system is therefore non-extensive.

Unlike the asymptotically flat case, however, the quasi-local AdS Smarr relation \eqref{eq:AdSsmarrrelation} is not even quasi-homogeneous: it contains an additional contribution involving the cosmological constant and the Killing volume, where the latter depends explicitly on the thermodynamic state variables themselves. Rather than becoming negligible, this term plays a crucial role in the large-system limit. Specifically, it compensates the dimension-dependent coefficients appearing in the remaining terms of the Smarr relation, allowing an extensive Euler relation to emerge asymptotically. 

We address this question from two complementary perspectives. First, from the viewpoint of the holographically dual conformal field theory, extensivity emerges in the large-volume limit discussed above. Owing to scale invariance, the large-volume limit at fixed temperature   is equivalent to a high-temperature limit at fixed volume in the dual CFT.

Second, we show that even before taking the CFT limit, the quasi-local gravitational description already admits a large-volume limit in which AdS-Schwarzschild black holes become extensive. In this limit the quasi-local energy contains more extensive terms compared to the CFT large-volume limit. We analyze this limit explicitly in   the energy and canonical representation, and demonstrate that the Euler relation is recovered in the appropriate regimes.

\subsection{High-entropy and high-temperature limit in the dual CFT}
\label{sec:cftextensivity}

In what follows, we   evaluate the large-system limit   for the CFT dual of AdS-Schwarzschild black holes in both the energy and canonical thermodynamic representations.  

Let us begin by reviewing the holographic dictionary relating gravitational thermodynamic variables to those of the dual CFT for AdS-Schwarzschild black holes \cite{Witten:1998zw}. The $(d-1)$-dimensional CFT lives on the conformal completion of the $d$-dimensional asymptotically AdS spacetime, with the CFT metric identified with the boundary metric up to a Weyl rescaling~\cite{Gubser:1998bc,Witten:1998qj}. Taking $r\to\infty$ and rescaling by the Weyl factor $R/r$, the induced CFT metric becomes
\begin{equation}\label{cftmetricroverl}
ds^2 = -\frac{R^2}{L^2}dt^2 + R^2 d\Omega_{d-2}^2\,.
\end{equation}
Thus, the CFT lives on a spatial sphere of radius $R$, and the CFT time coordinate is rescaled by a factor $R/L$ relative to the bulk time coordinate. In the strict limit $r_B\to\infty$, the quasi-local energy, Tolman temperature, and surface pressure vanish. However, to obtain finite thermodynamic quantities in the dual CFT, these variables must be rescaled by the same Weyl factor as the boundary metric, leading to the definitions
\begin{equation}
E_{\text{CFT}}=\lim_{r_B\to\infty}E\frac{r_B}{R}\,,\qquad
T_{\text{CFT}}=\lim_{r_B\to\infty}T\frac{r_B}{R}\,,\qquad
P_{\text{CFT}} V_{\text{CFT}}=\lim_{r_B\to\infty}P V\frac{r_B}{R}\,.
\end{equation}
The entropy of the CFT thermal state is equal to the Bekenstein  entropy of the black hole, both denoted by $S$. Introducing the dimensionless horizon radius $x\equiv r_h/L$ and the standard holographic normalization of the central charge, which we keep fixed in scaling transformations,
\begin{equation}\label{eq:cftC}
C=\frac{\Omega_{d-2}L^{d-2}}{16\pi G}\,,
\end{equation}
the entropy, volume, internal energy, temperature, and pressure of the CFT thermal state dual to AdS-Schwarzschild can be written as a function of $x$ and $R$ as \cite{Cong:2021jgb,Lilani:2025gzt}
\begin{align}
&S = 4\pi C x^{d-2}\,,\qquad V_{\text{CFT}}=\Omega_{d-2}R^{d-2}\,,\\
&E_{\text{CFT}}=\frac{(d-2)Cx^{d-3}}{R}(x^2+1)\,, \label{cftenergy}\\
&T_{\text{CFT}}=\frac{1}{4\pi R x}\big[(d-1)x^2+(d-3)\big]\,,\label{cfttemperature}\\
&P_{\text{CFT}}=\frac{C x^{d-3}}{\Omega_{d-2}R^{d-1}}(x^2+1)\,. \label{cftpressure}
\end{align}
The energy is defined relative to the CFT vacuum on the sphere, so that the vacuum state dual to pure AdS has vanishing energy.

\paragraph{High-entropy limit.} In the energy representation  the fundamental thermodynamic relation is
\begin{equation} \label{cftenergyextsub}
E_{\text{CFT}}(S,V_{\text{CFT}})
=(d-2)C\left(\frac{\Omega_{d-2}}{V_{\text{CFT}}}\right)^{\!\frac{1}{d-2}}
\left[
\left(\frac{S}{4\pi C}\right)^{\!\frac{d-1}{d-2}}
+\left(\frac{S}{4\pi C}\right)^{\!\frac{d-3}{d-2}}
\right].
\end{equation}
This expression is dual to Eq.~\eqref{eq:AdSthermodynamicquantities} and  decomposes into two terms: the first is an extensive contribution,   and the second a subextensive correction,   
\begin{equation} \label{sumofcftextsubenergies}
E_{\text{CFT}}=E_{\text{CFT},\text{ext}}+E_{\text{CFT},\text{sub}}\,, 
\end{equation}
where the extensive part is homogeneous of degree one, while the subextensive term is homogeneous of degree $1-\frac{2}{d-2}=\frac{d-4}{d-2}$\,,
\begin{align}
&E_{\text{CFT},\text{ext}}(\lambda S,\lambda V_{\text{CFT}})=\lambda E_{\text{CFT},\text{ext}}(S,V_{\text{CFT}})\,,\\
&E_{\text{CFT},\text{sub}}(\lambda S,\lambda V_{\text{CFT}})=\lambda^{1-\frac{2}{d-2}}E_{\text{CFT},\text{sub}}(S,V_{\text{CFT}})\,.
\end{align}
The subextensive contribution is a Casimir-type energy \cite{Verlinde:2000wg}. While the Casimir effect is usually discussed at zero temperature as a consequence of vacuum quantum fluctuations, it generalizes at finite temperature to a finite-size correction to the thermal energy    \cite{Mehra:1967wf,AMBJORN19831,PLUNIEN198687,Kirsten1991,Nesterenko:2004gzu}.

In the energy representation, the large-system limit is $V_{\text{CFT}}
\to \infty$ at fixed entropy density $s=S/V_{\text{CFT}},$ or equivalently $S \to \infty$ at fixed entropy density $s$. In this limit, the energy in \eqref{cftenergyextsub} divided by the volume approaches a finite value,
\begin{equation}\label{eq:EdensityCFTlimit}
\lim_{V_{\text{CFT}}\to\infty}\frac{E_{\text{CFT}}(sV_{\text{CFT}},V_{\text{CFT}})}{V_{\text{CFT}}}
=(d-2)C(\Omega_{d-2})^{\frac{1}{d-2}}
\left(\frac{s}{4\pi C}\right)^{\frac{d-1}{d-2}}\,,
\end{equation}
demonstrating explicitly that  the extensive term contributes, while the subextensive correction   vanishes.

While a generic finite-volume CFT may exhibit several subextensive corrections, the AdS-Schwarzschild sector is special: in its energy representation $E_{\text{CFT}}(S,V_{\text{CFT}})$, in Eq.~\eqref{cftenergyextsub}, there is only one subextensive term beyond the extensive contribution. In the entropy representation, however, $S(E_{\text{CFT}},V_{\text{CFT}})$ contains an infinite sequence of subextensive corrections, arising from the asymptotic inversion at large $E_{\text{CFT}}$.

\paragraph{Cardy-Verlinde formula.}
The CFT entropy for thermal states dual to an AdS--Schwarzschild geometry admits particularly
simple expressions when written in terms of   the extensive or the sub-extensive part of
the energy, or both. Inverting the energy-entropy relations derived above yields
\begin{align}
S(E_{\text{CFT,ext}},R)
&=4\pi C\left(\frac{E_{\text{CFT,ext}}\,R}{(d-2)C}\right)^{\frac{d-2}{d-1}}, \label{cftentropyextensiveenergy} \\
S(E_{\text{CFT,sub}},R)
&=4\pi C\left(\frac{E_{\text{CFT,sub}}\,R}{(d-2)C}\right)^{\frac{d-2}{d-3}} .
\end{align}
These expressions describe the same entropy \(S\). For generic, non-holographic CFTs the scaling with the energies is identical, although the
overall proportionality coefficients depend on the matter content of the CFT.

Multiplying the corresponding energy-entropy relations prior to inversion, thereby
eliminating the central charge \(C\), one obtains an expression for the entropy   in
terms of both the extensive and sub-extensive energies
\begin{equation}
S(E_{\text{CFT,ext}},E_{\text{CFT,sub}},R)
=\frac{4\pi R}{d-2}\sqrt{E_{\text{CFT,ext}}\,E_{\text{CFT,sub}}}\,.
\end{equation}
Using Eq.~\eqref{sumofcftextsubenergies}, this may equivalently be written as
\begin{equation} \label{CVformula}
S(E_{\text{CFT}},E_{\text{CFT,sub}},R)
=\frac{4\pi R}{d-2}\sqrt{E_{\text{CFT,sub}}\bigl(E_{\text{CFT}}-E_{\text{CFT,sub}}\bigr)}\,.
\end{equation}
This reproduces the Cardy-Verlinde formula \cite{Verlinde:2000wg} for holographic CFTs upon identifying the sub-extensive \textcolor{black}{energy} with the thermal Casimir energy as
\(
 E_{\text{CFT,sub}} =E_C/2.
\)
The Cardy-Verlinde formula is the higher-dimensional generalization of the Cardy formula
\cite{Cardy:1986ie} for the microcanonical entropy of a two-dimensional CFT. In \(d=3\), the
sub-extensive energy is independent of the entropy and is related to (minus) the vacuum energy. Indeed,
setting \(E_{\text{CFT}}R=\textcolor{black}{2}L_0\) and \(E_{\text{CFT,sub}}R=\textcolor{black}{c/12}\), one recovers the Cardy formula
(with equal left- and right-moving energies \(\bar L_0=L_0\) and central charges \(\bar c=c\)):
\begin{equation}
S
=\textcolor{black}{4\pi}\sqrt{\frac{c}{6}\left(L_0-\frac{c}{24}\right)}\,.
\end{equation}

\paragraph{Euler relation from AdS Smarr formula.} Since the leading contribution to the energy in the large-entropy regime is extensive, it must satisfy the homogeneous Euler relation. This can be seen directly from the AdS Smarr formula \eqref{eq:AdSsmarrrelation}, which in the CFT variables takes the form
\begin{equation}
(d-3)E_{\text{CFT}}=(d-2)(T_{\text{CFT}}S-P_{\text{CFT}}V_{\text{CFT}})
-\lim_{r_B\to\infty}\frac{\Lambda\tilde V_\xi}{4\pi G N_{\text{bh}}(r_B)}\frac{r_B}{R}\,.
\end{equation}
The CFT limit of the Killing volume contribution \eqref{killingvolumeadsfinite} is
\begin{equation}
\lim_{r_B\to\infty}\frac{\Lambda\tilde V_\xi}{4\pi G N_{\text{bh}}(r_B)}\frac{r_B}{R}
=   \frac{(d-2)Cx^{d-3}}{R}(x^2-1)\,. 
\end{equation}
Note that this expression is a function of $x$ and $R$; in the energy representation it therefore depends on $S$ and $V_{\text{CFT}}$. In the high-entropy regime $x \gg 1$, the leading term of this expression coincides with the leading contribution to the internal energy, namely the extensive part $E_{\text{ext}}$. Consequently, in the large-entropy regime the Smarr relation reduces to the Euler relation
\begin{equation}
E_{\text{CFT},\text{ext}}
=
T_{\text{CFT},\text{int}}\,S
-
P_{\text{CFT},\text{int}}\,V_{\text{CFT}},
\qquad \text{for } S \gg 1,
\end{equation}
where only the extensive part of the CFT energy contributes.  Similarly, for the temperature and pressure only their intensive parts contribute, corresponding to the leading terms in the large-$x$ expansion of Eqs.~\eqref{cfttemperature} and~\eqref{cftpressure}. In the large-system limit, one should divide both sides of  this equation  by the volume, since the energy and entropy diverge in this limit, whereas their densities  remain finite. 
 
Away from the large-system limit, non-extensivity can be quantified by the violation of the Euler identity,
\begin{equation} \label{cftsubextensiveeuler}
    E_{\text{CFT}}- T_{\text{CFT}}S + P_{\text{CFT}}V_{\text{CFT}} = \frac{2 C x^{d-3}}{R}=2C\,
\left(\frac{\Omega_{d-2}}{V_{\text{CFT}}}\right)^{\frac{1}{d-2}}\left(\frac{S}{4\pi C}\right)^{\frac{d-3}{d-2}}
\,.
\end{equation}
This   is proportional to the subextensive contribution to the energy in~\eqref{cftenergyextsub}. Moreover, it matches with the AdS Smarr formula in \eqref{eq:mancilla_eulerlike}, where the final term should therefore be interpreted as a subextensive Casimir contribution.

\paragraph{High-temperature limit.} Finally, we   take the large-volume limit in the Helmholtz representation, where ($T_{\text{CFT}},V_{\text{CFT}}$) are the independent variables. Due to scale invariance of the CFT this is equivalent to the high-temperature limit at fixed volume. Inverting Eq.~\eqref{cfttemperature} for the temperature, the $x$ variable can be expressed as a function of $T_{\text{CFT}}$ and $R$,
\begin{equation}
x(T_{\text{CFT}},R)
=\frac{1}{d-1}\left(2\pi T_{\text{CFT}}R\pm\sqrt{4\pi^2T_{\text{CFT}}^2R^2-(d-1)(d-3)}\right),
\end{equation}
where the plus sign corresponds to the large black hole branch and the minus sign to the small black hole branch. Expanding in the high-temperature or large-volume regime $T_{\text{CFT}}R=\infty$ yields  for the small and large black holes, respectively,
\begin{equation}
\begin{aligned}
x^{(1)} &= \frac{d-3}{4\pi T_{\text{CFT}}R}
+ \frac{(d-1)(d-3)^2}{64\pi^3(T_{\text{CFT}}R)^3}
+ \mathcal O((T_{\text{CFT}}R)^{-5})\,,\\
x^{(2)} &= \frac{4\pi T_{\text{CFT}}R}{d-1}
-\frac{d-3}{4\pi T_{\text{CFT}}R}
+ \mathcal O((T_{\text{CFT}}R)^{-3})\,.
\end{aligned}
\end{equation}
Note that for $d=3$ there is no small black hole solution. Substituting these expressions into the entropy, internal energy and Helmholtz free energy we find for the small  black hole,
\begin{equation}
\begin{aligned} \label{cftsmallblackhole}
    S^{(1)}(T_{\text{CFT}},R) &= 4\pi C\left(\frac{d-3}{4\pi T_{\text{CFT}}R}\right)^{d-2}
    \left[
    1+\frac{(d-2)(d-1)(d-3)}{16\pi^2 (T_{\text{CFT}}R)^2}
    +\mathcal O\!\left((T_{\text{CFT}}R)^{-4}\right)
    \right] \\
    E^{(1)}(T_{\text{CFT}},R) &= \frac{(d-2)\,C}{R}\left(\frac{d-3}{4\pi T_{\text{CFT}}R}\right)^{d-3}
    \left[
    1+\frac{d(d-3)^2}{16\pi^2 (T_{\text{CFT}}R)^2}
    +\mathcal O\!\left((T_{\text{CFT}}R)^{-4}\right)
    \right] \\
    F^{(1)}(T_{\text{CFT}},R)&= \frac{C}{R}\left(\frac{d-3}{4\pi T_{\text{CFT}}R}\right)^{d-3}
    \left[
    1+\frac{(d-2)(d-3)^2}{16\pi^2 (T_{\text{CFT}}R)^2}
    +\mathcal O\!\left((T_{\text{CFT}}R)^{-4}\right)
    \right] \,.
\end{aligned}
\end{equation}
and for the   large black hole,
\begin{equation}
\begin{aligned}
    S^{(2)}(T_{\text{CFT}},R) &= 4\pi C\left(\frac{4\pi T_{\text{CFT}}R}{d-1}\right)^{d-2}
    \left[
    1-\frac{(d-2)(d-3)(d-1)}{16\pi^2 (T_{\text{CFT}}R)^2}
    +\mathcal O\!\left((T_{\text{CFT}}R)^{-4}\right)
    \right] \\
    E^{(2)}(T_{\text{CFT}},R) &= \frac{(d-2)\,C}{R}\left(\frac{4\pi T_{\text{CFT}}R}{d-1}\right)^{d-1}
    \left[
    1-\frac{(d-4)(d-1)^2}{16\pi^2 (T_{\text{CFT}}R)^2}
    +\mathcal O\!\left((T_{\text{CFT}}R)^{-4}\right)
    \right] \\
    F^{(2)}(T_{\text{CFT}},R) &= -\,\frac{C}{R}\left(\frac{4\pi T_{\text{CFT}}R}{d-1}\right)^{d-1}
    \left[
    1-\frac{(d-2)(d-1)^2}{16\pi^2 (T_{\text{CFT}}R)^2}
    +\mathcal O\!\left((T_{\text{CFT}}R)^{-4}\right)
    \right] \,.
\end{aligned}
\end{equation}
 This structure matches the general perturbative expansion of finite-temperature CFTs on a sphere around $T_{\text{CFT}}R=\infty$ \cite{Kutasov:2000td}. While the scaling of each term in  the expansion is universal, the coefficients encode dynamical information determined by the theory; in holographic CFTs they are fixed. In the large-volume limit at fixed temperature, the entropy density, energy density, and free energy density of the small black hole vanish, whereas they remain finite   for the large black hole. Thus, the small black hole does not define an extensive equilibrium branch, while the large black hole does.

We emphasize that two distinct limits have been taken in this section. First, we took the CFT limit of quasi-local AdS thermodynamics, retaining only the leading term in a $1/r$ expansion. Second, within the resulting CFT, we took the high-entropy and high-temperature limit, retaining only the leading terms in $x=r_h/L$. Accordingly, there are two distinct sources of corrections to the extensive CFT limit: subextensive corrections intrinsic to the CFT, controlled by $r_h/L$, and corrections to the CFT itself, controlled by inverse powers of the bulk radius $r$.

However, taking these limits in this order misses an important structural feature. The quasi-local energy  of AdS-Schwarzschild black holes, viewed as a function $E(S,V)$ of entropy and volume, contains an infinite set of extensive contributions beyond the leading CFT term. Therefore, the procedure of first taking $r\to\infty$ to project onto the CFT, and then going to a regime in which the subextensive CFT energy is negligible, is too restrictive to capture the full large-system limit. The appropriate large-volume limit $V\to\infty$ at fixed $S/V$,  where $V$ is not the CFT volume but rather the volume of a holographic theory at a finite cutoff,   should reproduce all extensive contributions to the quasi-local energy. Understanding how these additional extensive terms arise, despite being subleading in $1/r$, requires a careful analysis of the finite-$r$ corrections to the quasi-local energy, which we turn to next.

\subsection{Large-volume limit in the energy representation}
\label{sec:energycorrectionsstructure}

We will now show that the quasi-local energy of AdS-Schwarzschild black holes, with entropy and volume as independent variables, becomes extensive in the limit $V \to \infty$ with $S/V$ held fixed. In terms of static coordinates, this corresponds to the limit $r_B \to \infty$ with $r_h/r_B$ kept constant. In this regime, the leading term reproduces the extensive CFT energy identified in the previous section, but, remarkably, there exist infinitely many additional extensive contributions in a $1/r_B$ expansion.

To elucidate the origin of this infinite tower of extensive terms, we recall the explicit expression for the quasi-local energy as a function of the boundary radius $r_B$ and the horizon radius $r_h$,
\begin{align}
E
=
\frac{(d-2)\Omega_{d-2} r_B^{d-3}}{8\pi G}
\left(
\sqrt{1+\frac{r_B^2}{L^2}}
-
\sqrt{1+\frac{r_B^2}{L^2}
-
\frac{r_h^{d-3}}{r_B^{d-3}}
\left(1+\frac{r_h^2}{L^2}\right)}
\right)\,.
\end{align}
We factor out $\sqrt{1+r_B^2/L^2}$ and expand the remaining square root around $r_B=\infty$, obtaining
\begin{align} \label{expandingenergy1}
E
&=
\frac{(d-2)\Omega_{d-2} r_B^{d-3}}{8\pi G}
\left(
\frac{\upsilon}{2L(L^2+r_B^2)^{1/2}}
+
\frac{\upsilon^2}{8L(L^2+r_B^2)^{3/2}}
+
\frac{\upsilon^3}{16L(L^2+r_B^2)^{5/2}}
+
\cdots
\right)\,,
\end{align}
where we have introduced
\begin{equation}
\upsilon \equiv \frac{r_h^{d-3}}{r_B^{d-3}}(L^2+r_h^2)\,.
\end{equation}
At this stage the denominators in \eqref{expandingenergy1} have not yet been expanded. To illustrate the emergence of additional extensive contributions beyond the leading CFT term, we explicitly expand the first two terms in the large-$r_B$ limit,
\begin{align}
E(r_B,r_h)
&=
\frac{(d-2)\Omega_{d-2}}{16\pi G}
\frac{(L^2+r_h^2)\,r_h^{d-3}}{L r_B}
\left[
1-\frac{L^2}{2r_B^2}
+\frac{3L^4}{8r_B^4}
+\mathcal{O}(r_B^{-6})
\right]
\nonumber\\[6pt]
&\quad+
\frac{(d-2)\Omega_{d-2}}{64\pi G}
\frac{(L^2+r_h^2)^2\,r_h^{2(d-3)}}{L r_B^{d}}
\left[
1-\frac{3L^2}{2r_B^2}
+\frac{15L^4}{8r_B^4}
+\mathcal{O}(r_B^{-6})
\right]
+\cdots .
\label{eq:E_large_r_two_terms}
\end{align}
In the first line, the leading term in the $1/r_B$ expansion and leading in $r_h$ constitutes an extensive contribution. After an appropriate Weyl rescaling, this term matches the extensive piece of the CFT energy \eqref{cftenergy}. It scales as $r_h^{d-1}/r_B$ and is therefore homogeneous of degree one under the scaling $(V,S)\to(\lambda V,\lambda S)$, or equivalently $(r_B,r_h)\to(\lambda^{1/(d-2)}r_B,\lambda^{1/(d-2)}r_h)$.

Crucially, the second line in \eqref{eq:E_large_r_two_terms} also contains an extensive contribution, again given by the leading term in the $1/r_B$ expansion and in $r_h$. This term scales as $r_h^{2(d-1)}/r_B^{d}$ and is likewise homogeneous of degree one under the same scaling transformation. Dividing both contributions by the volume $V\sim r_B^{d-2}$ yields finite energy densities in the large-volume limit at fixed $r_h/r_B$, scaling respectively as $(r_h/r_B)^{d-1}$ and $(r_h/r_B)^{2(d-1)}$. All remaining terms contain lower powers (either in $r_h$ or $r_B$ or both) and are therefore subextensive.

Next, we investigate the general structure of the extensive contributions to the energy.
Using the binomial expansion of the square root for $|z|<1$
\begin{align}
    \sqrt{1+z}=1+\frac{z}{2}-\frac{z^2}{8}+\frac{z^3}{16}-\cdots=\sum_{k=0}^\infty\binom{1/2}{k}z^k\,,
\end{align}
we see that the terms in the expansion~\eqref{expandingenergy1} of the quasi-local energy are of the form
\begin{align}\label{eq:energytermsgeneralform}
    -\frac{(d-2)\Omega_{d-2}}{8\pi G}\binom{1/2}{k}\frac{r_B^{d-3}}{L}\frac{(-\upsilon)^k}{(L^2+r_B^2)^{(2k-1)/2}}\,,\;\;\;\;\; k\in\mathbb{N}_{>0}\,.
\end{align}
At this stage the factor $(L^2+r_B^2)^{-(2k-1)/2}$ has not yet been expanded around $r_B=\infty$.
Its large-$r_B$ expansion is
\begin{equation}
\begin{aligned}
    (L^2+r_B^2)^{-(2k-1)/2}
     &=\frac{1}{r_B^{2k-1}}\left(1-\frac{(2k-1)L^2}{2r_B^{2}}+\frac{(2k-1)(2k+1)L^4}{8r_B^{4}}-\cdots\right) \\
    &=\sum_{l=0}^\infty\binom{-(2k-1)/2}{l}L^{2l}r_B^{-(2k-1+2l)}\,. \label{expansionofLrBetc}
\end{aligned}
\end{equation}
This implies that the $(k,l)$ term in the expansion of the energy is proportional to
\begin{align}
    \frac{r_B^{d-3}}{L}L^{2l}\frac{\upsilon^k}{r_B^{2k-1+2l}}=\frac{r_B^{d-3}L^{2l-1}}{r_B^{2k-1+2l}}\left(\frac{r_h^{d-3}}{r_B^{d-3}}\right)^{k}(L^2+r_h^2)^k=\frac{L^{2l-1}\left(r_h^{(d-3)}(L^2+r_h^2)\right)^k}{r_B^{(k-1)(d-3)+2k-1+2l}}\,.
\end{align}
From this expression it is clear that, for each fixed $k$, the $l=0$ term contains precisely one extensive contribution to the energy, namely the term that is leading in $r_h$, obtained by keeping only the $r_h^{2k}$ part of $(L^2+r_h^2)^k$. Indeed, this contribution is proportional to
\begin{align}
    \frac{r_h^{k(d-3)}r_h^{2k}}{r_B^{(k-1)(d-3)+2k-1}}=r_h^{d-3}r_B\left(\frac{r_h}{r_B}\right)^{(k-1)(d-3)+2k}.
\end{align}
For every $k\in \mathbb{N}_{>0}$, all terms with $l\geq1$ scale with fewer powers of $r_B$ than the leading $l=0$ contribution,  and are therefore subextensive. Additionally, for each $l$, terms in the expansion of $(L^2+r_h^2)^k$ other than $r_h^{2k}$, e.g. $kL^2r_h^{2(k-1)}$, contain fewer powers of $r_h$ than the extensive term, resulting in further subextensive scaling. So among the higher corrections one finds terms with lower powers of $r_h, r_B$ or both.

The total extensive energy is thus given by keeping only the $r_h^{2k}$ terms in all $l=0$ contributions:
\begin{equation}
E_{\text{ext}}
=
-\sum_{k=1}^{\infty}
\frac{(d-2)\Omega_{d-2}}{8\pi G}
\binom{1/2}{k}
\frac{(-1)^k\, r_h^{k(d-1)}}{L\, r_B^{(k-1)(d-1)+1}} \,.
\label{eq:Eext}
\end{equation}
Each term in Eq. \eqref{eq:Eext} is extensive, since under the scaling transformation
$(r_B,r_h)\to\lambda^{1/(d-2)} (r_B, r_h)$ the $k$-th term scales with $\lambda$, as the combined power of $r_h$ and $r_B$ is $d-2$,
so that each contribution is homogeneous of degree one. By consequence, dividing \eqref{eq:Eext} by the volume   yields a finite energy density in the large-system limit.

The above analysis reveals an infinite tower of extensive contributions to the energy. 
The leading term  reproduces the extensive CFT energy, scaling as
$$
E_{k=1}\sim r_h^{d-1} r_B^{-1}
\sim S^{\frac{d-1}{d-2}} V^{-\frac{1}{d-2}} .
$$
The next contribution  scales as
$$
E_{k=2}\sim r_h^{2(d-1)} r_B^{-d}
\sim S^{2\frac{d-1}{d-2}} V^{-\frac{d}{d-2}} ,
$$
and is likewise extensive. More generally, successive terms in the $k$-expansion differ by an additional factor
$r_h^{d-1} r_B^{-(d-1)}\sim S^{\frac{d-1}{d-2}} V^{-\frac{d-1}{d-2}}$, which does not affect extensivity since the
$\lambda$-scaling of $S$ and $V$ cancels.

Accordingly, the large-system limit must be taken so as to retain this entire tower of extensive contributions. Taking $V\to\infty$ at fixed entropy, as in Section \ref{sec:cftextensivity},  suppresses all higher-$k$ contributions relative to the $k=1$ term and is therefore too restrictive. Instead, the appropriate limit in the energy representation is
$V\to\infty$ with $S/V$ held fixed,  allowing entropy to compensate for the growing volume. In this limit, all extensive contributions scale linearly with $V$ times a function of the entropy density, whereas
subextensive corrections grow with lower powers of $V$ and hence drop out of
the energy density.

\subsection{General formula for entropy and extensive energy}

 The contributions to the extensive energy \(E_{\text{ext}}\) in
Eq.~\eqref{eq:Eext} can be resummed explicitly. Writing the series as a
binomial expansion yields
\begin{align}
    E_\text{ext}&=-\frac{(d-2)\Omega_{d-2}r_B^{d-2}}{8\pi GL }\sum_{k=1}^\infty\binom{1/2}{k}\left(-\frac{r_h^{d-1}}{r_B^{d-1}}\right)^k\nonumber
    \\
    &=\frac{(d-2)\Omega_{d-2}r_B^{d-2}}{8\pi GL }\left(1-\sqrt{1-\frac{r_h^{d-1}}{r_B^{d-1}}}\right)\,.
\end{align}
Expressing this result in terms of the entropy \(S\) and holographic
volume \(V=\Omega_{d-2}r_B^{d-2}\), one finds
\begin{align} \label{extensiveenergyclosedform}
    E_\text{ext} (S,V)=\frac{(d-2)V}{8\pi GL }\left(1-\sqrt{1-\left(\frac{4GS}{V}\right)^{\frac{d-1}{d-2}}}\right)\,.
\end{align}
This yields a remarkably simple closed-form expression for the total
extensive energy in a spherical cavity in terms of the
thermodynamic variables \((S,V)\). This expression is indeed extensive, since it is a homogeneous function of degree one in its independent variables. Note that $E_{\text{ext}}$ vanishes in the   limit $L \to \infty$, reproducing the   non-extensivity  of asymptotically flat Schwarzschild black holes in the   energy representation, see \eqref{eq:largeV_energy_density}.

The relation \eqref{extensiveenergyclosedform} can be inverted straightforwardly to obtain the entropy as
a function of the extensive energy and the volume,
\begin{align} \label{entropyextensiveenergy}
    S (E_{\text{ext}},V)=\frac{V}{4G}\left(1-\left(1-\frac{8\pi GL E_\text{ext}}{(d-2)V}\right)^2\right)^{\frac{d-2}{d-1}}\,.
\end{align}
This provides a general expression for the entropy of an AdS-Schwarzschild black hole in arbitrary spacetime dimension and at arbitrary finite
boundary radius \(r_B\).

The extensive energy can also be used to calculate the intensive part of the temperature,
\begin{align}
    T_\text{int}(S,V)=\frac{\partial E_\text{ext}}{\partial S}=\frac{d-1}{4\pi L} \left( \frac{4GS}{V} \right)^{\frac{1}{d-2}} \left( 1 - \left( \frac{4GS}{V} \right)^{\frac{d-1}{d-2}} \right)^{-1/2}\,,
\end{align}
or in terms of $r_h$ and $r_B$,
\begin{align}
    T_{\text{int}}(r_h,r_B) = \frac{(d-1) r_h}{4\pi L r_B \sqrt{1 - \left( \frac{r_h}{r_B} \right)^{d-1}}}\,.
\end{align}
Similarly we find the intensive pressure
\begin{align}
    P_{\text{int}}(S,V) = \frac{d-2}{8\pi GL} \left( \sqrt{1 - \left( \frac{4GS}{V} \right)^{\frac{d-1}{d-2}}} - 1 \right) + \frac{(d-1) \left( \frac{4GS}{V} \right)^{\frac{d-1}{d-2}}}{16\pi GL \sqrt{1 - \left( \frac{4GS}{V} \right)^{\frac{d-1}{d-2}}}}\,,
\end{align}
or in terms of $r_h$ and $r_B,$
\begin{align}\label{PintSV}
    P_{\text{int}}(r_h,r_B) = \frac{1}{8\pi GL} \left[ (d-2) \left( \sqrt{1 - \frac{r_h^{d-1}}{r_B^{d-1}}} - 1 \right) + \frac{(d-1) r_h^{d-1}}{2 r_B^{d-1} \sqrt{1 - \frac{r_h^{d-1}}{r_B^{d-1}}}} \right].
\end{align}
These expressions for the temperature and pressure depend only on the ratio \(S/V\) and are therefore homogeneous functions of degree zero, i.e. they are intensive quantities. In the large-system limit  the energy density is given by $E_\text{ext}/V$ and the temperature and pressure become $T_\text{int}$ and $P_\text{int}$. These variables satisfy the Euler relation
\begin{equation}
    E_{\text{ext}} = T_{\text{int}} S - P_{\text{int}} V\,.
\end{equation}

\paragraph{CFT limit.}
We also recover the CFT relation between the entropy and the extensive energy by expanding the former at 
large volume. Starting from Eq.~\eqref{entropyextensiveenergy}, we rewrite the entropy
in terms of the central charge \(C\) (defined in Eq.~\eqref{eq:cftC}),  the volume of a sphere whose radius is the AdS radius
\(V_L=\Omega_{d-2}L^{d-2}\), and the dimensionless extensive energy
\(\mathcal E_{\text{ext}} = E_{\text{ext}}L\):
\begin{equation} \label{entropyfintiecutoff}
S
= 4\pi C \,\frac{V}{V_L}
\left[
1-\left(
1-\frac{\mathcal E_{\rm ext}V_L}{(d-2)2C V}\,
\right)^2
\right]^{\frac{d-2}{d-1}}\,.
\end{equation}
It is convenient to introduce the dimensionless expansion parameter
\begin{equation}\label{dimensionlessexpansionparameter}\varepsilon = \frac{\mathcal E_{\rm ext}V_L}{(d-2)2C V}\,,\end{equation}
which is small in the large-volume regime. Expanding the entropy around small $\varepsilon$ gives
\begin{equation}S = \frac{4\pi C}{V_L}V (2\varepsilon)^{\frac{d-2}{d-1}} \left[ 1 - \left( \frac{d-2}{d-1} \right) \frac{\varepsilon}{2} - \frac{d-2}{8(d-1)^2} \varepsilon^2 + \dots \right],\end{equation}

The leading term reproduces the CFT entropy-energy relation. Substituting
Eq.~\eqref{dimensionlessexpansionparameter} into the leading piece     yields
\begin{equation}
S
=\frac{4\pi C}{V_L}\,V^{\frac{1}{d-1}}
\left(\frac{\mathcal E_{\rm ext}V_L}{(d-2)C}\right)^{\frac{d-2}{d-1}}
\,+\,\mathcal O\!\left(V^{-\frac{d-2}{d-1}}\right).
\end{equation}
When expressing this entropy in terms of the dimensionful   energy $E_{\text{ext}}$ we note that the dependence on $V_L$ cancels, 
\begin{equation} \label{cftentropyextensive}
S
=4\pi C\left[
\frac{E_{\text{ext}}}{(d-2)C}
\left(\frac{V}{\Omega_{d-2}}\right)^{\frac{1}{d-2}}
\right]^{\frac{d-2}{d-1}} +\,\mathcal O\!\left(V^{-\frac{d-2}{d-1}}\right)\,,
\end{equation}
in agreement with Eq.~\eqref{cftentropyextensiveenergy}. 

Finally, we emphasize that this  differs from the Cardy-Verlinde formula
\eqref{CVformula}, which depends explicitly on both the extensive and sub-extensive parts of
the energy. By contrast, our formulas \textcolor{black}{\eqref{entropyextensiveenergy} and \eqref{entropyfintiecutoff}}   express  the entropy entirely in terms of the
extensive contribution. Remarkably, these expressions are nonetheless exact, even at finite $r_B$: the full entropy
can be recovered without explicit reference to the sub-extensive energy. In this sense, the
entropy is already fully encoded in the extensive part of the energy.

\subsection{Large-volume limit at fixed temperature}

Above we showed that AdS-Schwarzschild black holes become extensive in the large-system limit when described in the energy representation. We now address the same question in the canonical representation. In this representation, fixing \(T\) and \(V\) admits two AdS-Schwarzschild solutions, corresponding to the small  and large black hole branches. As already indicated by the analysis of the dual CFT, only the large black hole branch is expected to admit an extensive large-volume limit, while the small black hole branch is not. In the following, we show that this expectation is borne out in quasi-local AdS thermodynamics: at fixed temperature, extensivity arises exclusively for the large black hole branch as \(V \to \infty\).

\paragraph{Small black hole.} 
A closed-form inversion of the temperature relation $T = T(r_h,r_B)$ in Eq.~\eqref{eq:AdSthermodynamicquantities} for $r_h$ is not available in general for AdS-Schwarzschild black holes.
 To proceed, we use \texttt{AsymptoticSolve} in \textit{Mathematica} to obtain an asymptotic expansion for $r_h(T,r_B)$ about $r_B=\infty$. In four spacetime dimensions, there are five solutions for $r_h$ but only two of them are physical. The resulting expression for the small black hole is
\begin{align} \label{fourdimensionsexpansionsmallbh}
    r_h(T,r_B)=\frac{L}{4 \pi r_B T} + \frac{3 L - 8  \pi^2 L^3  T^2}{64 \pi^3 r_B^3 T^3}+\mathcal{O}\left(r_B^{-5}\right).
\end{align}
Substituting back into the thermodynamic quantities \eqref{eq:AdSthermodynamicquantities}  gives
\begin{equation}
\begin{aligned}
    E(T,r_B)&=\frac{L^2}{8 \pi G T} \frac{1}{r_B^2} + \frac{L^2 - 4 \pi^2 L^4 T^2}{32 \pi^3 G T^3} \frac{1}{r_B^4} + \mathcal{O}\left(\frac{1}{r_B^6}\right),
    \\
    S(T,r_B)&=\frac{L^2}{16 \pi G T^2} \frac{1}{r_B^2} + \frac{3L^2 - 8 \pi^2 L^4 T^2}{128 \pi^3 G T^4} \frac{1}{r_B^4} + \mathcal{O}\left(\frac{1}{r_B^6}\right),
    \\
    P(T,r_B)&=\frac{L^2}{64 \textcolor{black}{\pi^2} G T} \frac{1}{r_B^4} + \frac{L^2 - 8 \pi^2 L^4 T^2}{256 \textcolor{black}{\pi^4} G T^3} \frac{1}{r_B^6} + \mathcal{O}\left(\frac{1}{r_B^8}\right),
    \\
    F(T,r_B)&=\frac{L^2}{16 \pi G T} \frac{1}{r_B^2} + \frac{L^2 - 8 \pi^2 L^4 T^2}{128 \pi^3 G T^3} \frac{1}{r_B^4} + \mathcal{O}\left(\frac{1}{r_B^6}\right).
\end{aligned}
\end{equation}
It is immediately evident that the small black hole is non-extensive, since the (free) energy scales with $r_B^{-2}\sim V^{-1}$. One readily verifies that the entropy and pressure satisfy the   thermodynamic relations $S = - \partial F / \partial T$ and $P = - \partial F / \partial V$. At leading order, the thermodynamic quantities coincide with those of the dual CFT, Eq. \eqref{cftsmallblackhole}, upon identifying the dimensionless combination \(T r_B = T_{\mathrm{CFT}} R\); however, this agreement breaks down at subleading order, reflecting finite-cutoff effects that are not captured by the undeformed CFT.

\paragraph{Large black hole.}

For the large black hole in $d=4$, the expansion of the horizon radius around $r_B=\infty$ at fixed temperature $T$ is
\begin{align} \label{eq:expansionforlargeradiuslargebh}
r_h(T,r_B) = \frac{r_B}{\rho } - \frac{L^2}{r_B} {\color{black}
\frac{\rho(\rho-1)(2\rho^2+2\rho-1)}
     {3(2\rho^3+1)}
} + \mathcal{O}\left( r_B^{-3}\right)\,,
\end{align}
where $\rho=\rho(T)$ is defined as the root of the depressed cubic equation
\begin{align} \label{eq:rootcubicequation}
    16 L^2 \pi^2 T^2 \rho^3 - 9\rho - 16 L^2 \pi^2 T^2 = 0\,.
\end{align}
By Descartes' rule of signs, this equation admits exactly one positive real root. Depending on the value of 
 $TL$, the cubic may have either one or three real roots in total. The general depressed cubic $t^3+pt+q=0$ has one real root if $q^2/4+p^3/27>0$ and three distinct real roots if $q^2/4+p^3/27<0$. In the present  case
\begin{align}
    p=\frac{-9}{16 L^2\pi^2 T^2}\,,\;\;\;\;\; q=-1\,,
\end{align}
so the condition for   a unique real root is  
\begin{align}
    \frac{1}{4}>\frac{27}{(4\pi  T L)^6}\,, \qquad
\text{or} \qquad
    T L>\frac{\sqrt{3}}{2^{5/3}\pi }\,.
\end{align}
If we assume that we are in this regime of sufficiently high temperature, then the cubic admits a unique real root. By Descartes’ rule of signs, this root is necessarily the unique positive root. In that case, the unique real solution is given by Cardano's formula
\begin{align}
    \left(\frac{-q}{2}+\sqrt{\frac{q^2}{4}+\frac{p^3}{27}}\right)^{1/3}+\left(\frac{-q}{2}-\sqrt{\frac{q^2}{4}+\frac{p^3}{27}}\right)^{1/3},
\end{align}
which in the present case yields
\begin{align}
    \rho(T)=\left(\frac{1}{2}+\sqrt{\frac{1}{4}-\frac{27}{(4\pi LT)^6}}\right)^{1/3}+\left(\frac{1}{2}-\sqrt{\frac{1}{4}-\frac{27}{(4\pi LT)^6}}\right)^{1/3}.
\end{align}
In the low-$T$ regime,     the cubic admits the following three real roots
\begin{align}
    2\sqrt{-p/3}\cos\left(\frac{1}{3}\arccos\left(\frac{3q}{2p}\sqrt{-3/p}\right)-\frac{2\pi k}{3}\right),\;\;\;\;\; k=0,1,2\,.
\end{align}
The unique positive root corresponds to $k=0$.

Next, we use the expansion \eqref{eq:expansionforlargeradiuslargebh} for $r_h(T,r_B)$, expressed in terms of the general root $\rho=\rho(T)$, to compute the leading large-$r_B$ behavior of the thermodynamic quantities
\begin{equation}
\begin{aligned}  \label{largebhexpansionatlargeradius}
S(T, r_B) &= \frac{\pi r_B^2}{G \rho^2} + \mathcal{O}\left(r_B^{0}\right),
\\
E(T, r_B) &= \frac{r_B^2}{GL}(1-y) + \mathcal{O}\left(r_B^{0}\right),
\\
P(T, r_B) &= \frac{1}{4\pi G L} \left(\frac{1-y}{y}-\frac{1}{4\rho^3 y}\right)+ \mathcal{O}\left(r_B^{-2}\right),
\\
F(T, r_B) &= \frac{r_B^2}{GL}\left((1-y)-\frac{3}{4\rho^3 y}\right)+\mathcal{O}\left(r_B^{0}\right)\,.
\end{aligned}
\end{equation} 
Here $y = \sqrt{1 - \rho^{-3}}$  is a convenient quantity since it satisfies $T=3/(4\pi L\rho y)$. The (free) energy scales with $r_B^2\sim V$ to leading order, showcasing extensivity. We emphasize that the temperature dependence here is more general than in the CFT, where thermodynamic quantities  in the canonical representation depend only on the product $T_{\text{CFT}} R.$ Moreover, in contrast to the large-volume limit of asymptotically flat large black holes, the internal energy here is also linear in the volume, indicating that the large AdS black hole branch is extensive across representations.

Finally, we check the homogeneous Euler relation explicitly. We have
\begin{equation}
\begin{aligned} 
TS &= \frac{3r_B^2}{4GL\rho^3 y}  + \mathcal{O}\left(r_B^{0}\right),
\\
PV &= \frac{r_B^2}{ G L} \left(\frac{1-y}{y}-\frac{1}{4\rho^3 y}\right)+ \mathcal{O}\left(r_B^{0}\right),
\end{aligned}
\end{equation}
so that to leading order 
\begin{align}
    TS-PV\approx\frac{r_B^2}{GL}\left(\frac{1}{\rho^3 y}+\frac{y-1}{y}\right) 
    =\frac{r_B^2}{GL}(-y+1)\approx E\,.
\end{align}
This is again a clear indication that the large black hole system is extensive, even though we are in a finite $T$ regime.

\subsection{High-temperature limit at fixed volume}
\label{sec:highTfixedVAdS}

We now consider the high-temperature limit $T\to\infty$ at fixed volume. 
This limit is generally distinct from the large-volume limit $V\to\infty$ at fixed temperature studied in the previous section. 
While the two limits coincide for CFTs and for  asymptotically flat Schwarzschild black holes, we demonstrate that they differ for AdS-Schwarzschild black holes.

\paragraph{Small black hole.}

For the small black hole branch, we invert the Tolman temperature relation perturbatively. 
Specifically, we expand $T$ as a series in $r_h$ about $r_h=0$ and use  \texttt{InverseSeries}  in \textit{Mathematica} to obtain the corresponding expansion of $r_h$ in terms of $T$ at fixed boundary data. 
This yields in general dimensions
\begin{align}
    r_h(T,r_B)&=\frac{L(d-3)}{4\pi(L^2+r_B^2)^{1/2}T}+\frac{L(d-3)^2(d-1)}{64\pi^3 (L^2+r_B^2)^{3/2}T^3}+\frac{L(d-3)^3(d-1)^2}{512\textcolor{black}{\pi^5}(L^2+r_B^2)^{5/2}T^5}+\mathcal{O}(T^{-7}) \nonumber
    \\
    &+\frac{L^{d}(d-3)^{d-2}}{2^{2d-3}\pi^{d-2}r_B^{d-3}(L^2+r_B^2)^{d/2}T^{d-2}}+\mathcal{O}(T^{-d})\,. \label{eq:smallBHhighTexpansion}
\end{align}
Note that for $d=4$ the $T^{-(d-2)}$ correction gives rise to a $T^{-2}$ term, while for $d=5$ it combines with the $T^{-3}$ contribution. 
For $d\geq 6$, the leading behavior is governed by the $T^{-1}$ and $T^{-3}$ terms. 
The $T^{-(d-2)}$ term represents a genuine non-CFT correction, while the remaining terms may be interpreted, at least for large $r_B$ where $L^2+r_B^2\approx r_B^2$, as finite-size corrections within the CFT, since they depend only on the combination $r_B T$.

We can then insert this expansion for $r_h(T,r_B)$ to obtain the high-temperature expansions of the energy, entropy, and pressure \textcolor{black}{for $d\geq 5$:}
\begin{equation}
\begin{aligned}
    E(T,r_B)&=\frac{(d-2)\Omega_{d-2}L^{d-2}(d-3)^{d-3}}{2^{2d-2}\pi^{d-2} G\left(L^2+r_B^2\right)^{(d-2)/2}T^{d-3}}+\mathcal{O}(T^{-(d-1)})\,,
    \\
    S(T,r_B)&=\frac{(d-3)\Omega_{d-2}L^{d-2}(d-3)^{d-3}}{2^{2d-2}\pi^{d-2} G\left(L^2+r_B^2\right)^{(d-2)/2}T^{d-2}}+\mathcal{O}(T^{-d})\,,
    \\
    P(T,r_B)&=\frac{L^{d-2}(d-3)^{d-3}}{2^{2d-2}\pi^{d-2} Gr_B^{d-4}\left(L^2+r_B^2\right)^{d/2}T^{d-3}}+\mathcal{O}(T^{-(d-1)})\,.\end{aligned}
\end{equation}
\textcolor{black}{For $d=4$ the leading terms are the same, but the remainders scale as $\mathcal{O}(T^{-2})$ for $E$ and $P$ and as $\mathcal{O}(T^{-3})$ for $S$.} We emphasize that this expansion does not coincide with the expansion taken at fixed $T$ around $r_B=\infty$, due to the presence of the additional scale set by the AdS radius $L$. Nevertheless, in the large-$r_B$ limit the two expansions agree at leading order, and this leading behavior also matches that of the dual CFT.

As an aside, the high-temperature expansion in Eq.~\eqref{eq:smallBHhighTexpansion} can be used to obtain a large-$r_B$ expansion of the horizon radius $r_h(T,r_B)$. Using Eq.~\eqref{expansionofLrBetc}, one finds that the structure of this expansion is highly constrained. In particular, the leading $T^{-1}$ contribution generates terms proportional to $r_B^{-1}$ together with higher odd inverse powers $r_B^{-3}, r_B^{-5},\ldots$, while the $T^{-3}$ contribution starts only at order $r_B^{-3}$. There are also contributions with powers containing $d$, starting with $T^{-(d-2)}$ at order $r_B^{-(2d-3)}$, which themselves receive further $r_B^{-2}$ corrections. The resulting expansion therefore takes the form
\begin{equation}\begin{aligned}
    r_h(T,r_B)&=\frac{(d-3)L}{4\pi Tr_B}+\frac{1}{r_B^{3}}\left(-\frac{(d-3)L^3}{8\pi T}+\frac{(d-3)^2(d-1)L}{64\pi^3 T^3}\right)+\mathcal{O}\left(r_B^{-5}\right)
    \\
    &+\frac{(d-3)^{d-2}L^d}{2^{2d-3}\pi^{d-2}T^{d-2}r_B^{2d-3}}+\mathcal{O}\left(r_B^{-(2d-1)}\right).
\end{aligned}
\end{equation}
In $d=4$, this expansion agrees with the earlier result in Eq.~\eqref{fourdimensionsexpansionsmallbh}.

\paragraph{Large black hole.}
For the large black hole branch in $d=4$, solving for $r_h(T,r_B)$ in the high-temperature regime yields
\begin{align}
    r_h(T, r_B) = r_B - \frac{3r_B^2 + L^2}{16 \pi^2 L^2 r_B} \frac{1}{T^2} \textcolor{black}{+} \frac{(3r_B^2 - 2L^2)(3r_B^2 + L^2)}{256 \pi^4 L^4 r_B^3} \frac{1}{T^4} 
    + \mathcal{O}\left(T^{-6}\right).
\end{align}
Substituting this expansion into the quasi-local thermodynamic quantities leads to
\begin{equation}\begin{aligned}
    S(T, r_B) &= \frac{\pi r_B^2}{G} - \frac{3r_B^2 + L^2}{8\pi G L^2 T^2} + \mathcal{O}(T^{-4})\,,
    \\
    E(T, r_B) &= \frac{r_B}{G} \sqrt{1 + \frac{r_B^2}{L^2}} - \frac{3r_B^2 + L^2}{4\pi G L^2 T} + \mathcal{O}(T^{-2})\,,
    \\
    P(T, r_B) &= \frac{T}{4G} - \frac{1 + \frac{2r_B^2}{L^2}}{8\pi G r_B \sqrt{1 + \frac{r_B^2}{L^2}}} + \mathcal{O}(T^{-1})\,,
    \\
    F(T, r_B) &= -\frac{\pi r_B^2 T}{G} + \frac{r_B}{G} \sqrt{1 + \frac{r_B^2}{L^2}} - \frac{3r_B^2 + L^2}{8\pi G L^2 T} + \mathcal{O}(T^{-2})\,.
\end{aligned}
\end{equation}
This high-temperature expansion at fixed $r_B$ is qualitatively different from the large-volume expansion \eqref{largebhexpansionatlargeradius} performed at fixed temperature. In particular, the leading terms obtained in these two limits do not coincide, reflecting the fact that the limits $T\to\infty$ at fixed $r_B$ and $r_B\to\infty$ at fixed $T$ do not commute in the presence of a finite AdS radius.

Moreover, the structure of the present expansion closely parallels that of the high-$T$ expansion of asymptotically flat Schwarzschild black holes, cf.\ Eq.~\eqref{largebhfreeenergylargeV}. As $T\to\infty$, the horizon radius approaches the cavity wall, $r_h\to r_B$, so that the thermodynamics probes the near-horizon Rindler geometry rather than a genuine large-volume regime. In this limit, both the Helmholtz free energy and the entropy scale linearly with the holographic volume $V\sim r_B^2$ and therefore are extensive in the canonical representation. However, \textcolor{black}{the internal energy is only extensive for $r_B \gg L $, but is otherwise subextensive in the high-temperature expansion.}


\section{Discussion}

In this paper we argued that quasi-local gravitational thermodynamics, when combined with the holographic principle, provides a   physically meaningful notion of thermodynamic pressure and volume for black holes. A central outcome of this analysis is that the appropriate thermodynamic volume is not a bulk quantity in the naive geometric sense, but instead a codimension-2 object from the bulk spacetime perspective. This feature is not accidental: it is a direct consequence of diffeomorphism invariance, which implies that all conserved and thermodynamic quantities in gravity are necessarily defined on boundaries rather than in the bulk. From this viewpoint, it is the boundary area that plays the role of volume in the dual thermodynamic description, while the conjugate surface pressure arises from the quasi-local gravitational stress tensor.

Using this holographic identification, we examined the issue of extensivity for Schwarzschild black holes in both asymptotically flat and asymptotically AdS spacetimes. For asymptotically flat black holes, we found that the entropy as a function of energy and volume is non-extensive, even in the large-volume   limit. This is not unexpected: gravity is a long-range interaction, and self-gravitating systems are well known to violate extensivity in ordinary Newtonian settings \cite{Katz2003,Campa2009,Callender2011}. From this perspective, the non-extensive character of asymptotically flat black hole thermodynamics, in the energy or entropy representations, is  natural.

What is more striking is the contrasting behavior in the canonical representation, in which the large black hole branch becomes extensive in the appropriate large-system limit. In particular, the free energy scales linearly with the holographic volume (while the internal energy does not). The small black hole branch, on the other hand, is non-extensive, showing that extensivity is a representation- and branch-dependent property. 

For asymptotically AdS-Schwarzschild black holes the situation is again different, since the system is already extensive in the large-system limit of the entropy and energy representations. This shows that the asymptotic structure of black hole spacetimes can modify their extensivity. The canonical representation is similar to that of asymptotically flat black holes: the large black hole branch is extensive, whereas the small branch is not. 
This behavior mirrors that of local   quantum field theories at finite temperature, whose canonical free energy is extensive in the thermodynamic limit. The extensive limit for large AdS black holes across different representations is therefore consistent with the locality of the dual conformal field theory. 

Conversely, the   non-extensivity of asymptotically flat black holes (in the energy or entropy representation) suggests an obstruction to realizing a holographic dual of flat-space gravity as a   local quantum field theory. The notion of locality probed here is the many-body locality underlying ordinary equilibrium thermodynamics, namely the suppression of correlations between sufficiently separated subsystems that leads to additive entropy and free energy in the large-system limit. Our analysis is carried out within a framework in which the holographic dual is assumed to live on a finite timelike boundary. From this perspective, thermodynamic non-extensivity provides a concrete constraint on any holographic description of asymptotically flat gravity formulated on such a boundary. It would be interesting to compare this conclusion with other approaches to flat-space holography and investigate whether they violate locality in a similar   sense.

Since our analysis is formulated entirely at the level of equilibrium thermodynamics, we have not addressed the underlying statistical mechanical description. In the case of AdS black holes, however, the dual conformal field theory provides a well-defined microscopic framework. In that context, the large-system limit considered should agree with  the thermodynamic limit of the dual theory. It would therefore be interesting to analyze this limit directly at the level of the density of states, and to determine whether it reproduces the extensive behavior   for large AdS black holes.

Related work \cite{Anninos:2023epi,Anninos:2024wpy,Banihashemi:2024yye}  on quasi-local thermodynamics with conformal boundary conditions has shown that high-temperature gravitational thermodynamics can be extensive, even at finite cutoff and for vanishing or positive cosmological constant. In this framework, the boundary data held fixed consist of the conformal class of the induced metric on the boundary together with the trace of the extrinsic curvature of the boundary. While a first-law-type relation can be derived, involving variations of the trace of the extrinsic curvature, the precise thermodynamic interpretation of this term remains unclear. In particular, it is not obvious how to identify a corresponding notion of pressure-volume work or other standard thermodynamic conjugate pairs. By contrast, the present work employs a finite cutoff with Dirichlet boundary conditions, allowing for a more standard identification of quasi-local variables with thermodynamic quantities.

Several extensions of our analysis would be worth pursuing. One natural direction is to generalize our discussion of extensivity to charged and rotating black holes, both in asymptotically flat and   AdS spacetimes. In such cases, additional charges and chemical potentials   enter the thermodynamic description, and it would be interesting to determine whether an extensive equilibrium branch persists in the AdS case and, if so, what the corresponding extensive energy is. For rotating black holes the additional property arises that the surface pressure is  defined only locally as the conjugate to the area element, see Eq.~\eqref{reflocalsurfacepressure}. Accordingly, extensivity should be understood in terms of local additivity, in the sense that the thermodynamic variations are additive over patches of the boundary, rather than extensivity being characterized by a single global variable such as 
 $V=A$. This is consistent with a continuum, local-equilibrium description of the boundary system. For a fluid description of large rotating AdS black holes in AdS/CFT we refer to \cite{Bhattacharyya:2007vs}. 

Another important direction is to explore finite-size and subextensive corrections in more detail. In local conformal field theories, such corrections admit an interpretation in terms of a thermal effective action and encode theory-specific data beyond the leading extensive contribution~\cite{Allameh:2024qqp}. It would be interesting to investigate whether an analogous (non-local) effective description exists in holographic settings with finite  Dirichlet boundary or in asymptotically flat gravity, or whether the absence of a leading extensive term in the latter case obstructs such an effective description altogether.

Further, our results are closely related to recent developments on holography at finite cutoff \cite{Banihashemi:2024yye,Zhang:2025dgm} and $T\bar T$ deformations of conformal field theories \cite{McGough:2016lol}. In particular, Wall and Soni \cite{Soni:2024aop} formulated a   holographic covariant entropy bound within the framework of Cauchy slice holography \cite{Araujo-Regado:2022gvw}, which bounds the logarithm of the number of states of the dual theory that can pass through a given codimension-$2$ cut of a spatial hypersurface. In their construction, the bound is determined entirely by geometric data intrinsic to the cut, such as the induced metric and extrinsic curvature; an explicit closed-form expression is obtained in three-dimensional gravity. By contrast, our approach employs a finite timelike boundary and characterizes gravitational equilibrium in terms of quasi-local thermodynamic quantities. It would be useful to clarify the relationship between the codimension-$2$ geometric data appearing in their covariant entropy bound and the boundary thermodynamic variables in our framework.

Finally, our analysis assumes that a holographic dual description exists for gravitational systems enclosed by a finite boundary, extending  AdS/CFT   beyond asymptotic infinity. While this idea is supported by developments such as $T \bar T$ deformations and related finite-cutoff constructions, these are best understood for two-dimensional CFTs, and their generalization to higher dimensions remains less well established. More generally, the structure of the dual theory at a finite boundary in AdS is not fully understood, particularly at scales below the AdS radius. The dual system may not correspond to a conventional local quantum field theory, but rather to a nonlocal or effective description. It is also not entirely clear whether quasi-local thermodynamic quantities are uniquely selected by such a dual description, or whether their identification depends on choices of background subtraction or boundary counterterms. These issues are especially significant in asymptotically flat and de Sitter spacetimes, where no precise dual description is known, let alone at finite boundaries (see, however, \cite{Leuven:2018ejp,Coleman:2021nor} for some proposals). Understanding whether holography can be extended to spacetimes with non-AdS asymptotics and to finite boundaries, and if so in what precise sense, remains an important direction for future work.


\subsection*{Acknowledgements}

SB thanks the Julian Schwinger Foundation for financial support at the
2025 Peyresq Spacetime Meeting, where useful discussions contributed to this work. MRV thanks the audiences of the 2025 Peyresq Spacetime Meeting, the National Seminar Theoretical High Energy Physics at NIKHEF and  the IMAPP meeting at Radboud University, where this work was presented, for their interesting questions. This project is supported by the Spinoza Grant of the
Dutch Science Organization (NWO) awarded to Klaas Landsman.

\appendix

\section{Definitions of extensivity and their logical relations}\label{appendixExtensivity}

Dunning-Davies \cite{DunningDavies1983} distinguished three inequivalent notions of extensivity within standard thermodynamics, none of which coincides with the notion of extensivity commonly employed in statistical mechanics. To formulate these definitions precisely, we fix a thermodynamic representation in which the \(n\)-dimensional equilibrium (Legendre) submanifold \(\mathscr{L}\) is coordinatized by
\(
(\mathbf{x},\mathbf{y})=(x_1,\ldots,x_k,\, y_{k+1},\ldots,y_n),
\)
representing a complete set of independent thermodynamic variables. Each coordinate is chosen from the full collection of conjugate pairs \((x_a,y_a)\) appearing in the energy representation, such that no conjugate pair contributes more than one independent variable. Dunning-Davies restricted attention to the energy and entropy representations, which are precisely the representations in which all independent variables are drawn from the \(x_a\). Here we allow for more general thermodynamic representations obtained by Legendre transformations, in which variables of either type may be taken as independent, including, for example, the canonical representation.

Thermodynamic quantities are smooth real-valued state functions
$f:\mathscr{L}\to\mathbb{R}$, $\mathbf{(x,y)}\mapsto f(\mathbf{x,y})$. Examples of such state functions depend on the chosen thermodynamic representation. 
They include thermodynamic potentials, such as the internal energy \(E(S,V,N)\), the 
Helmholtz free energy \(F(T,V,N)\), or the Gibbs free energy \(G(T,P,N)\). 
They also include derived thermal quantities defined on \textcolor{black}{\(\mathscr L\)}, such as 
the heat capacities \(C_V(T,V,N)\) and \(C_P(T,P,N)\), compressibilities, thermal expansion coefficients, 
and other response functions obtained from second derivatives of the thermodynamic potentials. 
All such quantities are functions on the equilibrium manifold and are uniquely determined 
once a thermodynamic representation is fixed.

Extensivity is then formulated as a property of a state function $f$ relative to a chosen
coordinate chart and to certain operations (e.g. adding, doubling or scaling) on $\mathscr{L}$. 
We now state
three standard  definitions of extensivity in equilibrium thermodynamics \cite{DunningDavies1983}:

\begin{description}
\item[(A) \emph{Additivity}.]
The function $f$ is said to be extensive in the additive sense if its value in a composite
system --- formed from two independent subsystems with state coordinates $(\mathbf{x}_1, \textbf y)$ and
$(\mathbf{x}_2, \textbf y)$ ---  equals the sum of its values in each of the subsystems,
\begin{equation}
f(\mathbf{x}_1+\mathbf{x}_2, \textbf y )=f(\mathbf{x}_1, \textbf y )+f(\mathbf{x}_2, \textbf y )\,,
\label{eq:additivity}
\end{equation}
where addition is understood componentwise.

\item[(D) \textcolor{black}{\emph{Doubling}}.]
The function $f$ is said to be extensive in the sense of doubling if its value is \textcolor{black}{scaled by $n$ when
the $x_i$ variables are simultaneously scaled by $n$,}
\begin{equation}
\textcolor{black}{f(n\mathbf{x,y})=nf(\mathbf{x,y}),\qquad n\in\mathbb{N}_{>0}\,.}
\label{eq:doubling}
\end{equation}
\textcolor{black}{The term \textit{doubling} comes from the special case $n=2$, but we use this term to refer to integer scaling for all $n$.}

\item[(H) \emph{Homogeneity}.]
The function $f$ is said to be extensive in the homogeneous sense if it is homogeneous of
degree one with respect to the $x_i$ variables, that is, if for all $\lambda\in\mathbb{R}_+$,
\begin{equation}
f(\lambda\mathbf{x,y})=\lambda f(\mathbf{x,y})\,.
\label{eq:homogeneity}
\end{equation}
\end{description}

\noindent These three definitions of extensivity   are not equivalent in general~\cite{DunningDavies1983,DUNNINGDAVIES1985383,DUNNINGDAVIES1988705}.
The logical relations between the different definitions depend sensitively on additional
assumptions imposed on the state function $f$: continuity of the state function~$f$ (denoted by C);  and, in certain cases, the restriction that state vectors
appearing in composition laws be parallel (denoted by~P).
In the remainder of this section we review the relations between the thermodynamic definitions (A), (D), and (H). 

The implication (H) $\Rightarrow$ (D)
is immediate by setting \textcolor{black}{$\lambda=n$} in \eqref{eq:homogeneity}. Likewise (A) $\Rightarrow$ (D)
follows from~\eqref{eq:additivity} by taking $n$ identical subsystems:
$f(2\mathbf{x,y})=f(\mathbf{x}+\mathbf{x,y})=2f(\mathbf{x,y})$.
The converse implications require continuity.   Under the additional assumption that $f$ is
continuous, both (A) and (D) imply homogeneity of degree one (H)
\cite{DunningDavies1983}. 

For completeness, we prove   (A) $+$ C $\Rightarrow$ (H).
Fix $(\mathbf{x,y})$ and apply (A) repeatedly to a composite system consisting
of $m$ identical subsystems to obtain $f(m\mathbf{x},\textbf y )=m f(\mathbf{x},\textbf y)$ for all
$m\in\mathbb{N}$.
Writing $\mathbf{x}=q(\mathbf{x}/q)$, additivity yields   $f(\mathbf{x}/q,\textbf y)=f(\mathbf{x},\textbf y)/q$,
and hence for any rational $\lambda=p/q>0$,
\[
f(\lambda\mathbf{x}, \textbf y )
= f\!\left(p\frac{\mathbf{x}}{q},\textbf y\right)
= p f\!\left(\frac{\mathbf{x}}{q},\textbf y\right)
= \frac{p}{q} f(\mathbf{x},\textbf y)
= \lambda f(\mathbf{x},\textbf y)\,.
\]
Thus \eqref{eq:homogeneity} holds for all $\lambda\in\mathbb{Q}_+$. By continuity of $(\textbf{x},\textbf y) \mapsto f(\textbf{x,y})$ and $(\lambda, ( \textbf x,\textbf y)) \mapsto (\lambda \textbf{x},\textbf y)$, the composition $(\lambda, (\textbf x,\textbf y)) \mapsto   f(\lambda \textbf{x},\textbf y)$ is also continuous, hence at fixed $(\textbf{x},\textbf y)$ the function $\lambda \mapsto f(\lambda\textbf{x},\textbf y)$ is  continuous. Since for $\lambda \in \mathbb {Q}_+$ this function is equal to $\lambda f(\textbf{x}\textcolor{black}{,\textbf{y}})$, by continuity this should hold for all positive real $\lambda$, proving (H).

It is important to note that additivity is, \emph{a priori}, a much stronger condition than
homogeneity. Homogeneity constrains the behavior of $f$ only along rays
$\{\lambda\mathbf{x}:\lambda>0\}$ in the space of additive variables (with $\mathbf{y}$ held fixed), whereas additivity involves the composition
of arbitrary vectors $\mathbf{x}_1+\mathbf{x}_2$, which need not be parallel. Homogeneity therefore
implies additivity only in the special case where the $\textbf x $ vectors appearing in
\eqref{eq:additivity} are assumed to be parallel, so that composition reduces to scaling along a
single ray.

We now prove (H) $+$ P $\Rightarrow$ (A) $+$ P. If composition is restricted to parallel vectors in the
space of additive variables, i.e.\ $\mathbf{x}\parallel\mathbf{z}$ so that
$\mathbf{z}=\mu\mathbf{x}$ with $\mu\ge 0$, then $\mathbf{x}+\mathbf{z}=(1+\mu)\mathbf{x}$ lies on the
same ray. Homogeneity therefore gives
$f(\mathbf{x}+\mathbf{z},\mathbf{y})=f((1+\mu)\mathbf{x},\mathbf{y})=(1+\mu)f(\mathbf{x},\mathbf{y})$,
while also $f(\mathbf{z},\mathbf{y})=f(\mu\mathbf{x},\mathbf{y})=\mu f(\mathbf{x},\mathbf{y})$.
Combining these yields
$f(\mathbf{x}+\mathbf{z},\mathbf{y})=f(\mathbf{x},\mathbf{y})+f(\mathbf{z},\mathbf{y})$, hence
additivity holds when the $\textbf x$ vectors in \eqref{eq:additivity} are parallel. Since (D) $+$ C
$\Rightarrow$ (H), we also have (D) $+$ C $+$ P $\Rightarrow$ (A) $+$ P.

Exact additivity is not expected to hold for generic thermodynamic systems,
because composition need not preserve independence; a more natural expectation
is superadditivity, obtained by replacing the equality in \eqref{eq:additivity}
by an inequality:
\begin{description}
\item[(S) \emph{Superadditivity}.]
The function $f$ is called superadditive if, for any two independent
subsystems with state vectors $(\mathbf{x}_1, \textbf y)$ and $(\mathbf{x}_2, \textbf y )$, the
value of $f$ for the composite system satisfies
\begin{equation}
f(\mathbf{x}_1+\mathbf{x}_2, \textbf y )\;\ge\; f(\mathbf{x}_1, \textbf y )+f(\mathbf{x}_2, \textbf y )\,.
\label{eq:superadditivity}
\end{equation}
\end{description}
Additivity corresponds to saturation of this inequality; more generally,
superadditivity allows for the presence of (non-local) correlations or interactions between subsystems
that invalidate a simple decomposition into independent parts. Superadditivity is therefore expected to be relevant
in black-hole or more generally gravitational thermodynamics. In particular, it has been shown to hold for the
microcanonical entropy $S(M,J,Q)$ of asymptotically flat Kerr-Newman black holes, where
long-range gravitational interactions render strict additivity implausible and classical
merger processes are governed by an area-increase inequality
\cite{LandsbergTranah1980a,TranahLandsberg1980b,Landsberg1984,Cvetic:2018dqf}.

These relations may be summarized as follows. For finite systems, the thermodynamic
definitions (A), (D), and (H) are logically distinct and become equivalent only under additional
assumptions. Among them, the doubling property (D)
is formally the weakest condition, but by itself is too weak to characterize extensivity, as
it constrains the state function only \textcolor{black}{at integer scales}. Under a continuity   assumption, however, doubling extends to full homogeneity of degree one,
so that (D) and (H) become equivalent.

Homogeneity of degree one (H) therefore provides the weakest physically meaningful notion of
extensivity: it imposes a uniform scaling constraint without assuming any particular rule for
the composition of subsystems or the existence of a thermodynamic limit. Assuming continuity,
the violation of (H) necessarily implies the failure of both doubling and additivity.

\section{Subextensive corrections for a classical ideal gas }\label{idealgasappendix}

In this Appendix we derive the subextensive corrections to the entropy and Euler relation for a
$d$-dimensional, monatomic, classical ideal gas, i.e., a non-interacting   gas obeying Maxwell-Boltzmann statistics.  The
system is treated in the (semi)classical approximation, with quantum mechanics entering only
through the phase-space cell size $h^d$, and thermodynamic quantities are computed from
classical phase-space integrals with the Gibbs $1/N!$ factor.  This description applies in
the dilute, high-temperature regime in which quantum degeneracy is negligible and  Bose-Einstein or Fermi-Dirac  particle
statistics reduce to Maxwell-Boltzmann counting.  The subextensive corrections obtained
below arise entirely from the finite-$N$ combinatorics encoded in the Stirling expansion of
$\ln N!$ and quantify deviations from exact extensivity at finite $N$, i.e.,  from the
thermodynamic limit $N,V\to\infty$ with $N/V$ held fixed.

\paragraph{Canonical partition function at finite $N$.} The   partition function in the canonical ensemble for an ideal gas of $N$ identical particles each of mass $m $ is:
\begin{align}
Z(T,V,N) = \frac{1}{N! h^{dN}} \int \prod_{i=1}^{N} d^dq_i d^dp_i e^{-\beta \frac{p_i^2}{2m}}\,,
\end{align}
where $h$ is Planck's constant, $V$ is the $d$-dimensional volume and $\beta = 1/(k_B T)$ is the inverse temperature. This expression is valid in the dilute,
non-degenerate regime $(N/V) \lambda^d \ll 1$, where $\lambda = \sqrt{h^2 \beta / (2\pi m)}$ is the thermal de Broglie wavelength. For
distinguishable particles the Gibbs factor $1/N!$ is absent.

Evaluating the Gaussian integrals exactly yields
\begin{align}Z(T,V,N) = \frac{V^N}{N!} \left( \frac{2\pi m}{h^2 \beta} \right)^{dN/2} = \frac{1}{N!} \left( \frac{V}{\lambda^d} \right)^N\,.\end{align}
 Taking the logarithm gives
\begin{align} \label{lnZ1}
\ln Z = N \ln V - dN \ln \lambda - \ln N!\,.\end{align}
We approximate the factorial using the higher-order Stirling series:
\begin{align}
\ln N! = N\ln N - N + \frac{1}{2}\ln(2\pi N) + \frac{1}{12N} - \frac{1}{360N^3} + \dots
\end{align}
Substituting this into \eqref{lnZ1}, we obtain:
\begin{align}
\ln Z = N \left[ \ln \left( \frac{V}{N \lambda^d} \right) + 1 \right] - \frac{1}{2}\ln(2\pi N) - \frac{1}{12N} + \frac{1}{360N^3} - \dots
\end{align}
The term proportional to $N$ in square brackets reproduces the standard extensive
Sackur--Tetrode contribution.  All remaining terms grow more slowly than linearly with $N$
and therefore represent subextensive corrections, beginning with a logarithmic correction
 and followed by inverse powers of $N$.
Since the Helmholtz free energy is given by $F=-k_B T \ln Z$, the leading term defines the
extensive part of the free energy, while the remaining contributions are subextensive and
vanish relative to the extensive term in the thermodynamic limit \begin{equation}
\lim_{N\to\infty}\frac{F(T,\textcolor{black}{N/n},N)}{N}
= -k_B T\left[\ln\!\left(\frac{1}{n\lambda^d}\right)+1\right],
\end{equation}  where $n \equiv N/V$ is the number density.

\paragraph{Finite-$N$ corrections to the entropy.}
The entropy is obtained from the Helmholtz free energy via
\begin{align}
S = - \frac{\partial F}{\partial T} 
= k_B\!\left(\ln Z + T\frac{\partial \ln Z}{\partial T}\right).
\end{align}
Since the thermal de Broglie wavelength scales as $\lambda \propto T^{-1/2}$, the temperature
derivative acts only on the $\lambda$-dependent term in $\ln Z$, yielding
\begin{align}
T\frac{\partial \ln Z}{\partial T}
= T\frac{\partial}{\partial T}\!\left(-\frac{Nd}{2}\ln \lambda^2\right)
= \frac{Nd}{2}.
\end{align}
Crucially, the subextensive terms in $\ln Z$ (such as $-\tfrac12\ln(2\pi N)$ and the inverse
powers of $N$) are independent of $T$ and therefore pass unchanged into the entropy.

Substituting the expression for $\ln Z$, we obtain
\begin{equation}
\begin{aligned}\label{eq:entropyfiniteN}
S(T,V,N)
&= k_B\!\left[\ln\!\left(\frac{V^N}{N!\lambda^{dN}}\right)+\frac{dN}{2}\right]
\\
&= Nk_B\!\left[\ln\!\left(\frac{V}{N\lambda^d}\right)+1+\frac{d}{2}\right]
-\frac{k_B}{2}\ln(2\pi N)
-\frac{k_B}{12N}
+\cdots .
\end{aligned}
\end{equation}
Keeping only the term proportional to $N$ in the second line reproduces the
Sackur-Tetrode entropy of a classical ideal gas in the thermodynamic  
limit, while the remaining terms represent finite-$N$ subextensive corrections.

\paragraph{Pressure and chemical potential.} The pressure is obtained from the volume derivative of the Helmholtz free energy.  Since all
subextensive corrections to $F$ depend only on $N$, they do not contribute to this derivative.
One therefore finds
\begin{align}
P(T,V,N) = - \frac{\partial F}{\partial V}   = k_B T  \frac{\partial \ln Z}{\partial V}  = \frac{N k_B T}{V}.
\end{align}
Thus the ideal gas law $PV = Nk_B T$ remains exact even when finite-$N$ corrections are
included.
The chemical potential, by contrast, is sensitive to subextensive corrections.  Using
$\mu = \partial F/\partial N $ and differentiating the expansion of $\ln Z$ with
respect to $N$ yields
\begin{align}
\mu(T,V,N) = -k_B T  \frac{\partial \ln Z}{\partial N}   = -k_B T \left[ \ln \left( \frac{V}{N \lambda^d} \right) - \frac{1}{2N} + \frac{1}{12N^2} + \dots \right].
\end{align}
The leading logarithmic term reproduces the standard intensive chemical potential in the
thermodynamic limit, while the remaining terms constitute finite-$N$ corrections that vanish
as $N\to\infty$ at fixed density.

\paragraph{Finite-$N$ corrections to the Euler relation.}
Using $E=\tfrac{d}{2}Nk_B T$ and $PV = Nk_B T$, together with the finite-$N$ expression for
the entropy derived above, we find
\begin{align}
E - TS + PV - \mu N
= \frac{k_B T}{2}\ln(2\pi N)
- \frac{k_B T}{2}
+ \frac{k_B T}{6N}
+ \cdots .
\end{align}
This expression shows explicitly that, at finite $N$, the Euler relation is violated.  The
leading deviation is logarithmic in $N$ and originates from the subextensive combinatorial
contribution to the entropy.  In the thermodynamic limit, the Euler relation is recovered in
the sense that
\begin{equation}
\frac{1}{N}\bigl(E - TS + PV - \mu N\bigr) \;\longrightarrow\; 0
\qquad \text{as } N\to\infty
\end{equation}
at fixed density.
Thus an ideal gas at finite $N$ is a non-extensive thermodynamic system.  This non-extensivity arises from the entropy and chemical potential,
which receive finite-$N$ corrections, while the internal energy and pressure retain their
extensive and intensive character, respectively, even at finite $N$. The ideal gas therefore provides a simple example in which extensivity is violated
at finite system size but recovered asymptotically in the large-system limit.

\bibliographystyle{jhep}
\bibliography{bibliography}

\end{document}